\documentclass[12pt,twoside, eps]{article}

\usepackage{color,graphicx,times}

\makeatother

\makeatletter
\long\def\M#1{\leavevmode\setbox\@tempboxa\hbox{#1}\@tempdima\fboxrule
    \advance\@tempdima \fboxsep \advance\@tempdima \dp\@tempboxa
   \hbox{\lower \@tempdima\hbox
  {\vbox{\hrule \@height \fboxrule
          \hbox{  \hskip\fboxsep
          \vbox{\vskip\fboxsep \box\@tempboxa\vskip\fboxsep}\hskip
                 \fboxsep\vrule \@width \fboxrule}%
                  }}}}
\makeatother

\let \ttorg \tt \def \tt{\ttorg \obeyspaces}

\begin{document}

\date{}

\title{\Large\bf Knot Logic and  Topological Quantum Computing with Majorana Fermions}

\author{Louis
H. Kauffman\\ Department of Mathematics, Statistics \\ and Computer Science (m/c
249)    \\ 851 South Morgan Street   \\ University of Illinois at Chicago\\
Chicago, Illinois 60607-7045\\ $<$kauffman@uic.edu$>$}

\maketitle

\thispagestyle{empty}

\subsection*{\centering Abstract}

{\em This paper is an introduction to relationships between quantum topology and quantum computing. 
We show how knots are related not just to braiding and quantum operators, but to quantum set theoretical foundations, algebras of fermions, and we show how the operation of negation in logic, seen as both a value and an operator, can generate the fusion algebra for a Majorana fermion. We call negation in this mode the mark, as it operates on itself to change from marked to unmarked states. The mark viewed recursively as a simplest discrete dynamical system naturally generates the fermion algebra, the quaternions and the braid group representations related to Majorana fermions. The paper begins with these fundamentals. It then discusses unitary
solutions to the Yang-Baxter equation that are universal quantum gates, quantum entanglement and topological entanglement, and gives an exposition of
knot-theoretic recoupling theory, its relationship with topological quantum field theory and applies these methods to produce unitary representations of the
braid groups that are dense in the unitary groups. These methods are rooted in the bracket state sum model for the Jones polynomial. A  self-contained study of the quantum
universal Fibonacci model is given. Results are applied to give quantum algorithms for the computation of the colored Jones polynomials for knots
and links, and the Witten-Reshetikhin-Turaev invariant of three manifolds. Two constructions are given for the Fibonacci model, one based in Temperley-Lieb recoupling theory, the other quite elementary and also based on the Temperley-Lieb algebra.  This paper begins an exploration of 
quantum epistemology in relation to the structure of discrimination as the underpinning of basic logic, perception and measurement.}
\bigbreak

\section{Introduction}
This paper is an introduction to relationships between quantum topology and quantum computing. 
We take a foundational approach, showing how knots are related not just to braiding and quantum operators, but to quantum set theoretical foundations and algebras of fermions. We show how the operation of negation in logic, seen as both a value and an operator, can generate the fusion algebra for a Majorana fermion, a particle that is its own anti-particle and interacts with itself either to annihilate itself or to produce itself. We call negation in this mode {\em the mark}, as it operates on itself to change from marked to unmarked states. The mark viewed recursively as a simplest discrete dynamical system naturally generates the fermion algebra, the quaternions and the braid group representations related to Majorana fermions. The paper begins with these fundmentals. They provide a conceptual key to many of the models that appear later in the paper. In particular, the Fibonacci model for topological quantum computing is seen to be based on the fusion rules for a Majorana fermion and these in turn are the interaction rules for the mark seen as a logical particle. It requires a shift in viewpoint  to see that the operator of negation can also be seen as a logical value. This is explained in Sections 3, 4 and 5. The quaternions emerge naturally from the reentering mark. All these models have their roots in unitary representations of the Artin braid group to the quaternions.  
\bigbreak

An outline of the parts of this paper is given below.
\bigbreak

\begin{enumerate}
\item Introduction
\item Knots and Braids
\item Knot Logic
\item Fermions, Majorana Fermions and Algebraic Knot Sets
\item Laws of Form
\item Quantum Mechanics and Quantum Computation
\item Braiding Operators and Universal Quantum Gates
\item A Remark about $EPR,$ Entanglement and Bell's Inequality
\item The Aravind Hypothesis
\item $SU(2)$ Representations of the Artin Braid Group
\item The Bracket Polynomial and the Jones Polynomial
\item Quantum Topology, Cobordism Categories, Temperley-Lieb Algebra and Topological Quantum Field Theory
\item Braiding and Topological Quantum Field Theory
\item Spin Networks and Temperley-Lieb Recoupling Theory
\item Fibonacci Particles
\item The Fibonacci Recoupling Model
\item Quantum Computation of Colored Jones Polynomials and the Witten-Reshetikhin-Turaev Invariant
\item A Direct Construction of the Fibonacci Model
\end{enumerate}
 
Much of what is new in this paper proceeds from thinking about knots and sets and distinctions.
The Sections 3, 4 and 5 are self-contained and self-explanatory. These sections show how a formal system discovered by Spencer-Brown  \cite{LOF}, underlying Boolean logic, is composed of a ``logical particle", the mark $~\M{~}~$, that interacts with itself to either produce itself or to cancel itself.
$$\M{~}\,\,\M{~} \longrightarrow \,  \M{~}$$
$$\M{\M{~}} \longrightarrow ~~$$
In this sense the mark is a formal model of a Majorana fermion.
The oirginal formal structure of the mark gives the fusion algebra for the Majorana fermion.
In Section 5 we show that this iconic representation of the particle is directly related to modeling with surface cobordisms and this theme occurs throughout the paper.
In Section 5 we also show that the mark, viewed as a generator of a discrete dynamical system, generates the Clifford algebra associated with a Majorana fermion and we end this section by showing 
how this iterant viewpoint leads naturally to the Dirac equation using the approach of \cite{Rowlands}.
This is part of the contents of the Sections 3, 4, 5. In these sections we examine relationships with knots as models of non-standard set theory. The algebra of fermions is directly relevant to this
knot set theory and can be formulated in terms of the Clifford algebra of Majorana fermions.
\bigbreak

We weave this material with 
the emergence of unitary braid group representations that are significant for quantum information theory.
In particular we weave the topology with the algebra of fermions and in order to clarify this development, we give a quick summary of that algebra and a quick summary of topological quantum computing in the rest of this introduction.
\bigbreak

\noindent {\bf Fermion Algebra.}
Recall fermion algebra. One has fermion creation operators $\psi$ and their
conjugate annihilation operators $\psi^{\dagger}.$
One has $\psi^{2} = 0 = (\psi^{\dagger})^{2.}$
There is a fundamental commutation relation
$$\psi \psi^{\dagger} + \psi^{\dagger} \psi = 1.$$
If you have more than one of them say $\psi$ and $\phi$,
then they anti-commute:
$$\psi \phi = - \phi \psi.$$
The Majorana fermions $c$ that satisfy $c^{\dagger} = c$ so that they
are their own anti-particles. There is a lot of interest in these as
quasi-particles and they are related to braiding and to topological
quantum computing.  A group of researchers  \cite{Kouwenhouven} claims, at this writing, to
have found quasiparticle Majorana fermions in edge effects in nano-wires.
(A line of fermions could have a Majorana fermion happen non-locally from
one end of the line to the other.) The Fibonacci model that we discuss is also based on
Majorana particles, possibly related to collecctive electronic excitations. 
If $P$ is a Majorana fermion particle, then $P$ can interact with itself to either produce itself or to annihilate itself. This is the simple ``fusion algebra" for this particle. One can write
$P^2 = P + 1$ to denote the two possible self-interactions the particle $P.$
The patterns of interaction and braiding of such a particle $P$ give
rise to the Fibonacci model.\
\bigbreak

Majoranas are related to standard fermions as follows:
The algebra for Majoranas is $c = c^{\dagger}$ and $cc' = -c'c$ if $c$ and $c'$ are
distinct Majorana fermions with  $c^{2}= 1$ and  $c'^{2}= 1.$
One can make a standard fermion from two Majoranas via
$$\psi = (c + ic')/2,$$
$$\psi^{\dagger} = (c -ic')/2.$$
Similarly one can
mathematically make two Majoranas from any single fermion.
Now if you take a set of Majoranas
$$\{ c_1, c_2, c_3, \cdots , c_n \}$$
then there are natural braiding operators that act on the vector space with
these $c_k$ as the basis. The operators are mediated by algebra elements
$$\tau_{k} =(1 + c_{k+1} c_{k})/2,$$
$$\tau_{k}^{-1} = (1 - c_{k+1} c_{k})/2.$$
Then the braiding operators are
$$T_{k}: Span \{c_1,c_2,\cdots, ,c_n \} \longrightarrow Span \{c_1,c_2,\cdots, ,c_n \}$$
via 
$$T_{k}(x) = \tau_{k} x \tau_{k}^{-1}.$$
The braiding is simply:
$$T_{k}(c_{k}) = c_{k+1},$$
$$T_{k}(c_{k+1}) = - c_{k},$$
and $T_{k}$ is the identity otherwise.
This gives a very nice unitary representaton of the Artin braid group and
it deserves better understanding.
\bigbreak

It is worth noting that a triple of Majorana fermions say $a,b,c$ gives rise
to a representation of the quaternion group. This is a generalization of
the well-known association of Pauli matrices and quaternions.
We have $a^2 = b^2 = c^2 = 1$ and they anticommute.
Let $I = ba, J = cb, K = ac.$
Then $$I^2 = J^2 = K^2 = IJK = -1,$$ giving the quaternions.
The operators
$$A = (1/\sqrt{2})(1 + I)$$
$$B = (1/\sqrt{2})(1 + J)$$
$$C = (1/\sqrt{2})(1 + K)$$
braid one another: $$ABA = BAB,BCB =CBC, ACA = CAC.$$
This is a special case of the braid group representation described above
for an arbitrary list of Majorana fermions.
These braiding operators are entangling and so can be used for universal
quantum computation, but they give only partial topological quantum
computation due to the interaction with single qubit operators not
generated by them.
\bigbreak

In Section 5 we show how the dynamics of the reentering mark leads to two (algebraic) Majorana fermions $e$ and $\eta$ that correspond to the spatial and temporal aspects of this recursive process.
The corresponding standard fermion creation and annihilation operators are then given by the formulas below.
$$\psi = (e + i\eta)/2$$ and
$$\psi^{\dagger} = (e - i\eta)/2.$$
This gives a model of a fermion creation operator as a point in a non-commutative spacetime.
This suggestive point of view, based on knot logic and Laws of Form, will be explored in subsequent publications.
\bigbreak

\noindent{\bf Topological quantum computing.}
This paper describes relationships between quantum topology and quantum computing as  a modified version of
Chapter 14 of the book \cite{CKL} and an expanded version of \cite{SpinTop} and an expanded version of a chapter in \cite{AMS}.  Quantum topology is, roughly speaking, that part of
low-dimensional topology that interacts with statistical and quantum physics. Many invariants of knots, links and three dimensional manifolds have been born
of this  interaction, and the form of the invariants is closely related to the form of the computation of amplitudes in quantum mechanics. Consequently, it is
fruitful to move back and forth between quantum topological methods and the techniques of quantum information theory.
\bigbreak

We sketch the background topology, discuss analogies (such as topological entanglement and quantum entanglement),
show direct correspondences between certain topological operators (solutions to the Yang-Baxter equation) and universal quantum gates. We 
describe the background for topological quantum computing in terms of Temperley--Lieb (we will sometimes abbreviate this to $TL$) recoupling theory.
This is a recoupling theory that generalizes standard angular momentum recoupling theory, generalizes the Penrose theory of spin networks and is inherently
topological.  Temperley--Lieb recoupling Theory is based on the bracket polynomial model \cite{KA87,KP} for the Jones polynomial. It is built in terms of
diagrammatic combinatorial topology. The same structure can be explained in terms of the $SU(2)_{q}$ quantum group, and has relationships with functional
integration and Witten's approach to topological quantum field theory. Nevertheless, the approach given here is elementary.   The structure  is built from simple beginnings and this structure and its recoupling language can be
applied to many things including colored Jones polynomials, Witten--Reshetikhin--Turaev invariants of three manifolds, topological quantum field theory
and quantum computing.
\bigbreak

In quantum computing, the simplest non-trivial example of the Temperley--Lieb recoupling Theory gives
the so-called Fibonacci model. The recoupling theory yields representations of the Artin braid group into unitary groups $U(n)$ where $n$ is a Fibonacci number. These representations are {\em dense} in the unitary group, and can be used to model quantum computation
universally in terms of representations of the braid group. Hence the term: topological quantum computation. In this paper, we outline the basics of the Temperely--Lieb Recoupling Theory, and show explicitly how the Fibonacci model arises from it.
The diagrammatic computations in the sections 12 to 18 are completely self-contained and can be used by a reader who has just learned the bracket polynomial,
and wants to see how these dense unitary braid group representations arise from it. In the final section of the paper we give a separate construction for the Fibnacci model that is based on $2 \times 2$ 
complex matrix representations of the Temperley--Lieb algebra. This is a completely elementary construction independent of the recoupling theory of the previous sections. We studied this construction 
in \cite{Fibonacci} and a version of it has been used in \cite{Shor}.
\bigbreak 

The relationship of the work here with the mathematics of Chern-Simons theory and conformal field theory occurs through the work of Witten, Moore and Seiberg and Moore and Read \cite{MooreRead}. One can compare the mathematical techniques of the present paper with 
the physics of the quantum Hall effect and its possibilities for topological quantum computing.
This part of the story will await a sequel to the present exposition.
\bigbreak

\begin{figure}
     \begin{center}
     \begin{tabular}{c}
     \includegraphics[height=6cm]{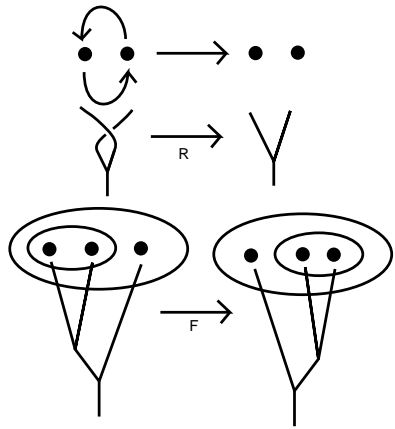}
     \end{tabular}
     \end{center}
     \caption{\bf Braiding Anyons }
     \label{Figure 1 }
     \end{figure}

Here is a very condensed presentation of how unitary representations of the braid group are constructed via topological quantum field theoretic methods, leading to the Fibonacci model and its generalizations.
These representations are more powerful, in principle, than the representations we have just given, because they encompass a dense collection of all unitary transformations, including single qubit transformations needed for universal quantum computing.
One has a mathematical particle with label $P$
that can interact with itself to produce either itself labeled $P$ or itself
with the null label $*.$ We shall denote the interaction of two particles $P$ and $Q$ by the expression
$PQ,$ but it is understood that the ``value" of $PQ$ is the result of the interaction, and this may partake of a number of possibilities.
Thus for our particle $P$, we have that $PP$ may be equal to $P$ or to $*$ in a given situation.
When $*$ interacts with $P$ the result is always $P. $ When $*$ interacts with $*$ the result is always $*.$ One considers
process spaces where a row of particles labeled $P$ can successively
interact, subject to the restriction that the end result is $P.$ For example
the space $V[(ab)c]$ denotes the space of interactions of three particles
labeled $P.$ The particles are placed in the positions $a,b,c.$ Thus we
begin with $(PP)P.$ In a typical sequence of interactions, the first two $P$%
's interact to produce a $*,$ and the $*$ interacts with $P$ to produce $P.$ 
\[
(PP)P \longrightarrow (*)P \longrightarrow P. 
\]
\noindent In another possibility, the first two $P$'s interact to produce a $%
P,$ and the $P$ interacts with $P$ to produce $P.$ 
\[
(PP)P \longrightarrow (P)P \longrightarrow P. 
\]
It follows from this analysis that the space of linear combinations of
processes $V[(ab)c]$ is two dimensional. The two processes we have just
described can be taken to be the qubit basis for this space. One obtains
a representation of the three strand Artin braid group on $V[(ab)c]$ by
assigning appropriate phase changes to each of the generating processes. One
can think of these phases as corresponding to the interchange of the
particles labeled $a$ and $b$ in the association $(ab)c.$ The other operator
for this representation corresponds to the interchange of $b$ and $c.$ This
interchange is accomplished by a {\it unitary change of basis mapping} 
\[
F:V[(ab)c] \longrightarrow V[a(bc)]. 
\]
\noindent If 
\[
A:V[(ab)c] \longrightarrow V[(ba)c] 
\]
is the first braiding operator (corresponding to an interchange of the first
two particles in the association) then the second operator 
\[
B:V[(ab)c] \longrightarrow V[(ac)b] 
\]
is accomplished via the formula $B = F^{-1}RF$ where the $R$ in this formula
acts in the second vector space $V[a(bc)]$ to apply the phases for the
interchange of $b$ and $c.$ These issues are illustrated in Figure~\ref{Figure 1 }, where the parenthesization of the particles 
is indicated by circles and by also by trees. The trees can be taken to indicate patterns of particle interaction, where
two particles interact at the branch of a binary tree to produce the particle product at the root. See also Figure~\ref{Figure 28 } for an illustration
of the braiding $B = F^{-1}RF$
\bigbreak

     \bigbreak

In this scheme, vector spaces corresponding to associated strings of
particle interactions are interrelated by {\it recoupling transformations}
that generalize the mapping $F$ indicated above. A full representation of
the Artin braid group on each space is defined in terms of the local
interchange phase gates and the recoupling transformations. These gates and
transformations have to satisfy a number of identities in order to produce a
well-defined representation of the braid group. These identities were
discovered originally in relation to topological quantum field theory. In
our approach the structure of phase gates and recoupling
transformations arise naturally from the structure of the bracket model for
the Jones polynomial. Thus we obtain a knot-theoretic basis for topological
quantum computing. \bigbreak

In modeling the quantum Hall effect \cite{Wilczek,Fradkin,B1,B2}, the braiding of  quasi-particles (collective excitations) leads to non-trival
representations of the Artin braid group. Such particles are called {\it Anyons}. The braiding in these models is related to 
topological quantum field theory. It is hoped that
the mathematics we explain here will form a bridge between theoretical models of anyons and their applications to quantum computing.
\bigbreak

\noindent {\bf Acknowledgement.}  
Much of this paper is based upon joint work with Samuel J.  Lomonaco in the papers \cite{TEQE,Spie,QK1,QK2,QK3,QK4,QK5,BG,AnyonicTop,QCJP2,Fibonacci,AMS}. I have woven  this work into the present paper in a form that is coupled with recent and previous work on relations with logic and with Majorana fermions. The relations with logic stem from the following previous papers of the author
\cite{FormDynamics,SignSpace1,SignSpace2,SignSpace3,KnotLogic,Eigenform,Reflexivity,Reflexphys,SpaceTime,KN1,KN2,NonCom1,NonCom2,Emerg}. These previous papers are an exploration of the foundations of knot theory in relation to Laws of Form, non-standard set theory, recursion and discrete dynamical systems. At the level of discrete dynamical systems the papers are related to foundations of physics. More work needs to be done in all these domains. 

Two recent books contain material relevant to the context of this paper.
They are \cite{Bernd} and \cite{Rowlands}. The interested reader should examine these approaches to fundamental physics. It is planned to use this paper and other joint work as a springboard for a book 
\cite{TQIT} on topological quantum information theory and for a book that expands on the foundational issues raised in this paper and the previous papers of the author.
\bigbreak

\section{Knots and Braids}
The purpose of this section is to give a quick introduction to the diagrammatic theory of knots,
links and braids. A {\it knot} is an embedding of a circle in three-dimensional space, taken up to ambient 
isotopy. The problem of deciding whether two knots are isotopic is an example of a {\it placement problem}, a problem of studying the topological forms that
can be made by placing one space inside another. In the case of knot theory we consider the placements of a circle inside three dimensional space.
There are many applications of the theory of knots. Topology is a background for the physical structure of real
knots made from rope of cable. As a result, the field of practical knot tying is a field of applied topology that existed well before the 
mathematical discipline of topology arose. Then again long molecules such as rubber molecules and DNA molecules can be knotted and linked. There have been
a number of intense applications of knot theory to the study of $DNA$ \cite{Sumners1} and to polymer physics \cite{KA}. Knot theory is closely related to
theoretical physics as well with applications in quantum gravity \cite{Smolin,CRLS,KaufLiko} and many applications of ideas in physics to the topological
structure of  knots themselves \cite{KP}. 
\bigbreak

{\it Quantum topology} is the study and invention of topological invariants via the use
of analogies and techniques from mathematical physics. Many invariants such as the Jones polynomial are 
constructed via partition functions and generalized quantum amplitudes. As a result, one expects to see relationships between knot theory
and physics. In this paper we will study how knot theory can be used to produce unitary representations of the braid group. Such representations
can play a fundamental role in quantum computing.
\bigbreak

\begin{figure}
     \begin{center}
     \begin{tabular}{c}
     \includegraphics[height=4cm]{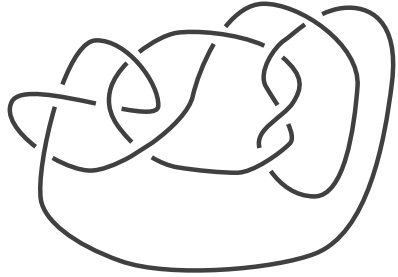}
     \end{tabular}
     \end{center}
     \caption{\bf A knot diagram}
     \label{Figure 2 }
     \end{figure} 
     \bigbreak

\begin{figure}
     \begin{center}
     \begin{tabular}{c}
     \includegraphics[height=6cm]{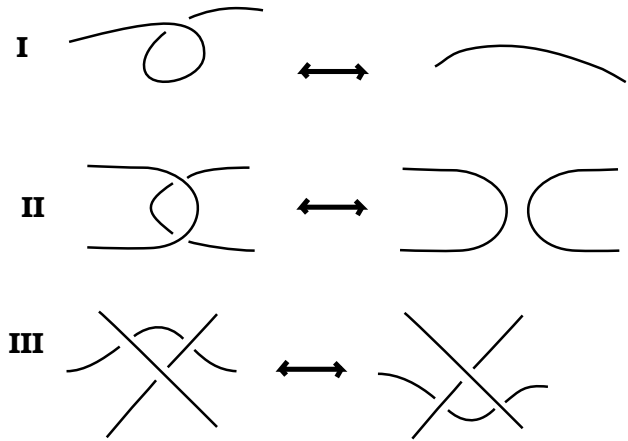}
     \end{tabular}
     \end{center}
     \caption{\bf The Reidemeister Moves}
     \label{Figure 3 }
     \end{figure} 
     \bigbreak

\noindent That is, two knots are regarded as equivalent if one embedding can be obtained from the other
through a continuous family of embeddings of circles in three-space. A {\it link} is an embedding of a disjoint
collection of circles, taken up to ambient isotopy. Figure~\ref{Figure 2 } illustrates a diagram for a knot. The diagram is regarded
both as a schematic picture of the knot, and as a plane graph with extra structure at the nodes (indicating how the curve of 
the knot passes over or under itself by standard pictorial conventions).

\begin{figure}
     \begin{center}
     \begin{tabular}{c}
     \includegraphics[height=6cm]{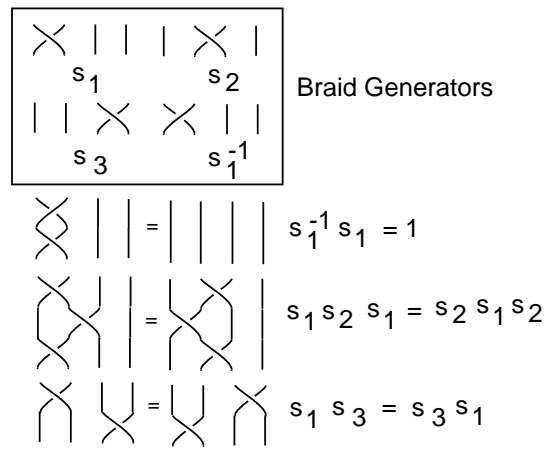}
     \end{tabular}
     \end{center}
     \caption{\bf Braid Generators  }
     \label{Figure 4 }
     \end{figure} 
     \bigbreak

Ambient isotopy is mathematically the same as the equivalence relation generated on diagrams by the {\it Reidemeister moves}. These moves are
illustrated in Figure~\ref{Figure 3 }. Each move is performed on a local part of the diagram that is topologically identical to the part of the diagram illustrated
in this figure (these figures are representative examples of the types of Reidemeister moves) without changing the rest of the diagram. The Reidemeister
moves are useful in doing combinatorial topology with knots and links, notably in working out the behaviour of knot invariants. A {\it knot invariant}
is a function defined from knots and links to some other mathematical object (such as groups or polynomials or numbers) such that equivalent diagrams are
mapped to equivalent objects (isomorphic groups, identical polynomials, identical numbers). The Reidemeister moves are of great use for analyzing the 
structure of knot invariants and they are closely related to the {\it Artin braid group}, which we discuss below.

\begin{figure}
     \begin{center}
     \begin{tabular}{c}
     \includegraphics[height=6cm]{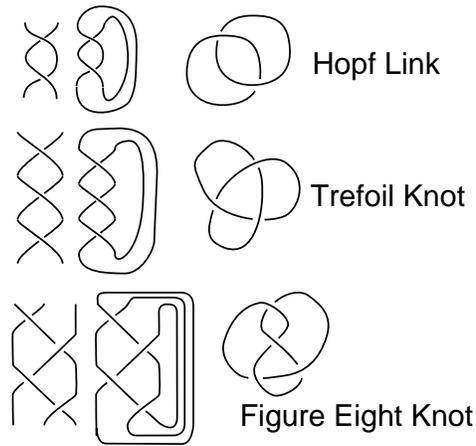}
     \end{tabular}
     \end{center}
     \caption{\bf Closing Braids to form knots and links. }
     \label{Figure 5 }
     \end{figure} 
     \bigbreak

\begin{figure}
     \begin{center}
     \begin{tabular}{c}
     \includegraphics[height=6cm]{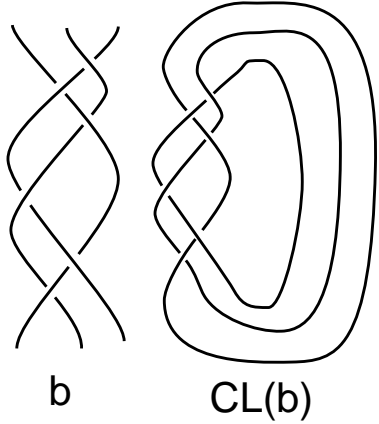}
     \end{tabular}
     \end{center}
     \caption{\bf Borromean Rings as a Braid Closure  }
     \label{Figure 6 }
     \end{figure} 
     \bigbreak

A {\it braid} is an embedding of a collection of strands that have 
their ends in two rows of points that are set one above the other with respect to a choice of vertical. The strands are not
individually knotted and they are disjoint from one another. See Figure~\ref{Figure 4 }, Figure~\ref{Figure 5 } and Figure~\ref{Figure 6 } for illustrations of braids and moves on braids. Braids can be 
multiplied by attaching the bottom row of one braid to the top row of the other braid. Taken up to ambient isotopy, fixing the endpoints, the braids form
a group under this notion of multiplication. In Figure~\ref{Figure 4 } we illustrate the form of the basic generators of the braid group, and the form of the
relations among these generators. Figure~\ref{Figure 5 } illustrates how to close a braid by attaching the top strands to the bottom strands by a collection of 
parallel arcs. A key theorem of Alexander states that every knot or link can be represented as a closed braid. Thus the theory of braids is critical to the 
theory of knots and links. Figure~\ref{Figure 6 } illustrates the famous Borromean Rings (a link of three unknotted loops such that any two of the loops are unlinked)
as the closure of a braid.
\bigbreak

Let $B_{n}$ denote the Artin braid group on $n$ strands.
We recall here that $B_{n}$ is generated by elementary braids $\{ s_{1}, \cdots ,s_{n-1} \}$
with relations 

\begin{enumerate}
\item $s_{i} s_{j} = s_{j} s_{i}$ for $|i-j| > 1$, 
\item $s_{i} s_{i+1} s_{i} = s_{i+1} s_{i} s_{i+1}$ for $i= 1, \cdots n-2.$
\end{enumerate}

\noindent See Figure~\ref{Figure 4 } for an illustration of the elementary braids and their relations. Note that the braid group has a diagrammatic
topological interpretation, where a braid is an intertwining of strands that lead from one set of $n$ points to another set of $n$ points.
The braid generators $s_i$ are represented by diagrams where the $i$-th and $(i + 1)$-th strands wind around one another by a single 
half-twist (the sense of this turn is shown in Figure~\ref{Figure 4 }) and all other strands drop straight to the bottom. Braids are diagrammed
vertically as in Figure~\ref{Figure 4 }, and the products are taken in order from top to bottom. The product of two braid diagrams is accomplished by
adjoining the top strands of one braid to the bottom strands of the other braid. 
\bigbreak 

In Figure~\ref{Figure 4 } we have restricted the illustration to the
four-stranded braid group $B_4.$ In that figure the three braid generators of $B_4$ are shown, and then the inverse of the
first generator is drawn. Following this, one sees the identities $s_{1} s_{1}^{-1} = 1$ 
(where the identity element in $B_{4}$ consists in  four vertical strands), 
$s_{1} s_{2} s_{1} = s_{2} s_{1}s_{2},$ and finally
$s_1 s_3 = s_3 s_1.$ 
\bigbreak

Braids are a key structure in mathematics. It is not just that they are a collection of groups with a vivid topological interpretation.
From the algebraic point of view the braid groups $B_{n}$ are important extensions of the symmetric groups $S_{n}.$ Recall that the 
symmetric group $S_{n}$ of all permutations of $n$ distinct objects has presentation as shown below.
\begin{enumerate}
\item $s_{i}^{2} = 1$ for  $i= 1, \cdots n-1,$
\item $s_{i} s_{j} = s_{j} s_{i}$ for $|i-j| > 1$, 
\item $s_{i} s_{i+1} s_{i} = s_{i+1} s_{i} s_{i+1}$ for $i= 1, \cdots n-2.$
\end{enumerate}
Thus $S_{n}$ is obtained from $B_{n}$ by setting the square of each braiding generator equal to one. We have an exact sequence of groups
$${1} \longrightarrow B_{n} \longrightarrow S_{n} \longrightarrow {1}$$ exhibiting the Artin braid group as an extension of the symmetric group.
\bigbreak

In the next sections we shall show how representations of the Artin braid group are rich enough to provide a dense set of transformations in the 
unitary groups. Thus the braid groups are {\it in principle} fundamental to quantum computation and quantum information theory.
\bigbreak

\section{Knot Logic}

We shall use knot and link diagrams to represent sets. More about this point of view can be found in  the author's paper "Knot Logic" \cite{KL}.
\bigbreak

\begin{figure}
     \begin{center}
     \begin{tabular}{c}
     \includegraphics[height=4cm]{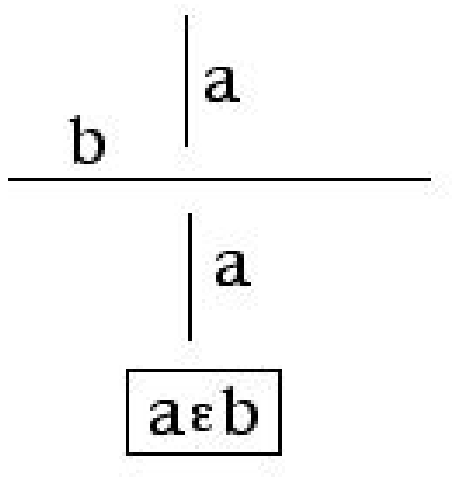}
     \end{tabular}
     \end{center}
     \caption{\bf Membership }
     \label{member}
     \end{figure}

Set theory is about an asymmetric relation called membership. 
We write $a \in S$ to say that $a$ is a member of the set $S.$  In this section we shall diagram the membership relation as in Figure~ \ref{member}.

     \bigbreak

The entities $a$ and $b$ that are in the relation $a \in b$ are diagrammed as segments of lines or curves, with the $a$-curve passing underneath the $b$-curve.  Membership is represented by under-passage of curve segments.  A curve or segment with no curves passing underneath it
is the empty set.

\begin{figure}
     \begin{center}
     \begin{tabular}{c}
     \includegraphics[height=4cm]{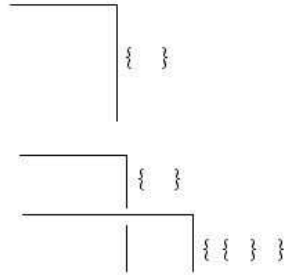}
     \end{tabular}
     \end{center}
     \caption{\bf Von Neumann 1 }
     \label{von1}
     \end{figure} 
     \bigbreak

\begin{figure}
     \begin{center}
     \begin{tabular}{c}
     \includegraphics[height=4cm]{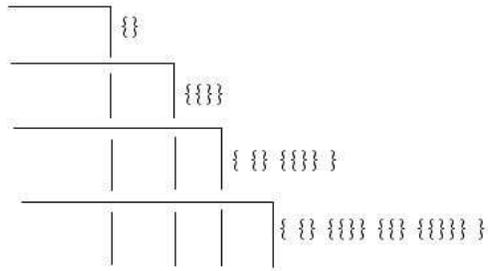}
     \end{tabular}
     \end{center}
     \caption{\bf Von Neumann 2 }
     \label{von2}
     \end{figure}

In the Figure~\ref{von1}, we indicate two sets. The first (looking like a right-angle bracket that we refer to as the {\em mark}) is the empty set. The second, consisting of a mark crossing over another mark, is the set whose only member is the empty set.
We can continue this construction, building the von Neumann construction of the natural numbers in this notation as in Figure~\ref{von2}

     \bigbreak

\begin{figure}
     \begin{center}
     \begin{tabular}{c}
     \includegraphics[height=4cm]{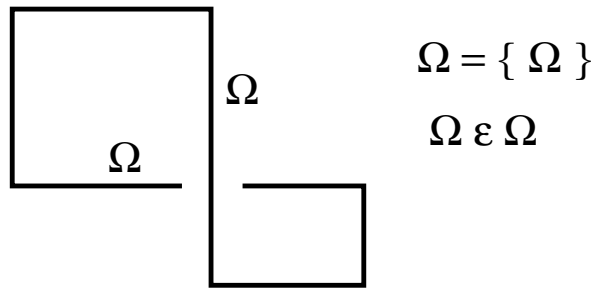}
     \end{tabular}
     \end{center}
     \caption{\bf Omega is a member of Omega.}
     \label{omega}
     \end{figure}

\begin{figure}
     \begin{center}
     \begin{tabular}{c}
     \includegraphics[height=4cm]{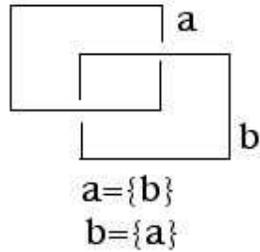}
     \end{tabular}
     \end{center}
     \caption{\bf Mutual Membership}
     \label{mutual}
     \end{figure}

This notation allows us to also have sets that are members of themselves as in Figure~\ref{omega},
and and sets can be members of each other as in Figure~\ref{mutual}. This mutuality is diagrammed as topological linking. This leads to the question beyond flatland: Is there a topological interpretation for this way of looking at set-membership? 

     \bigbreak

     \bigbreak

\begin{figure}
     \begin{center}
     \begin{tabular}{c}
     \includegraphics[height=6cm]{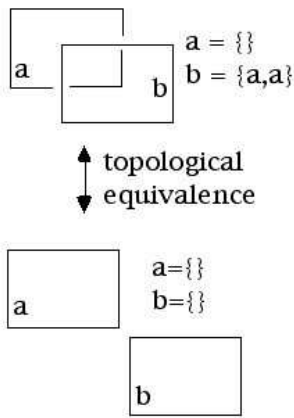}
     \end{tabular}
     \end{center}
     \caption{\bf Cancellation }
     \label{cancel}
     \end{figure} 
     \bigbreak

Consider the example in Figure~\ref{cancel}, modified from the previous one.
The link consisting of $a$ and $b$ in this example is not topologically linked. The two components slide over one another and come apart.
The set a remains empty, but the set $b$ changes from $b = \{a,a\}$ to
empty. This example suggests the following interpretation.

{\it  Regard each diagram as specifying a multi-set
(where more than one instance of an element can occur), and the rule for reducing to a set with one representative for each element is:
Elements of knot sets cancel in pairs.
Two knot sets are said to be equivalent if one can be obtained from the other by a finite sequence of pair cancellations.}

\begin{figure}
     \begin{center}
     \begin{tabular}{c}
     \includegraphics[height=4cm]{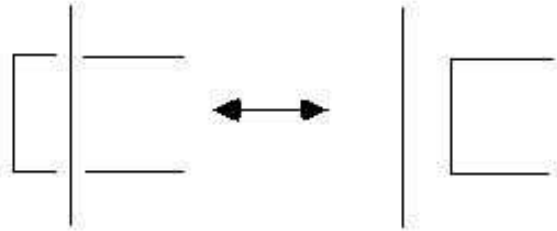}
     \end{tabular}
     \end{center}
     \caption{\bf Reidemeister 2 }
     \label{r2}
     \end{figure}

This equivalence relation on knot sets is in exact accord with the first Reidemeister move as shown in Figure~\ref{r2}.

     \bigbreak

\begin{figure}
     \begin{center}
     \begin{tabular}{c}
     \includegraphics[height=6cm]{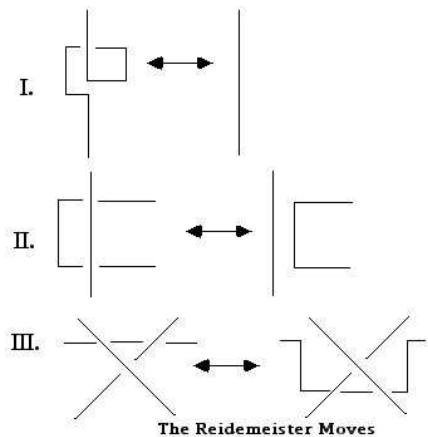}
     \end{tabular}
     \end{center}
     \caption{\bf Reidemeister Moves}
     \label{reid}
     \end{figure}

There are other topological moves, and we must examine them as well.  In fact, it is well-known that topological equivalence of knots (single circle embeddings), links (mutltiple circle embeddings) and tangles (arbitrary diagrammatic embeddings with end points fixed and the rule that you are not allowed to move strings over endpoints) is generated by three basic moves (the Reidemeister moves) as shown in Figure~\ref{reid}. See \cite{KP}.

     \bigbreak

\begin{figure}
     \begin{center}
     \begin{tabular}{c}
     \includegraphics[height=4cm]{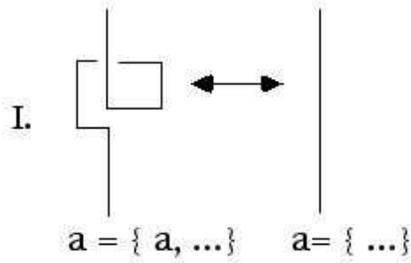}
     \end{tabular}
     \end{center}
     \caption{\bf Reidemeister I: Replacing Self-Membership with No Self-Membership }
     \label{reid1}
     \end{figure}

It is apparent that move III does not change any of the relationships in the knot multi-sets. The line that moves just shifts and remains underneath the other two lines. On the other hand move number one can change the self-referential nature of the corresponding knot-set.
One goes, in the first move, between a set that indicates self-membership to a set that does not indicate self-membership (at the site in question). See Figure~\ref{reid1}
This means that in knot-set theory every set has representatives
(the diagrams are the representatives of the sets) that are members of themselves, and it has representatives that are not members of themselves. In this domain, self-membership does not mean infinite descent. We do not insist that $$a = \{a\}$$ implies that 
$$a = \{ \{ \{ \{  \cdots \} \} \} \}. $$ Rather, $a = \{ a \}$ just means that $a$ has a little curl in its diagram. The Russell set of all sets that are not members of themselves is meaningless in this domain.
\bigbreak

     \bigbreak

     \begin{figure}
     \begin{center}
     \begin{tabular}{c}
     \includegraphics[height=4cm]{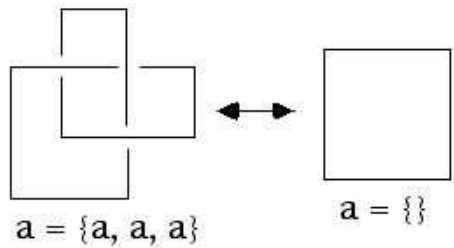}
     \end{tabular}
     \end{center}
     \caption{\bf  Trefoil is an empty knotset.}
     \label{sorrow}
     \end{figure} 
     \bigbreak

     \begin{figure}
     \begin{center}
     \begin{tabular}{c}
     \includegraphics[height=4cm]{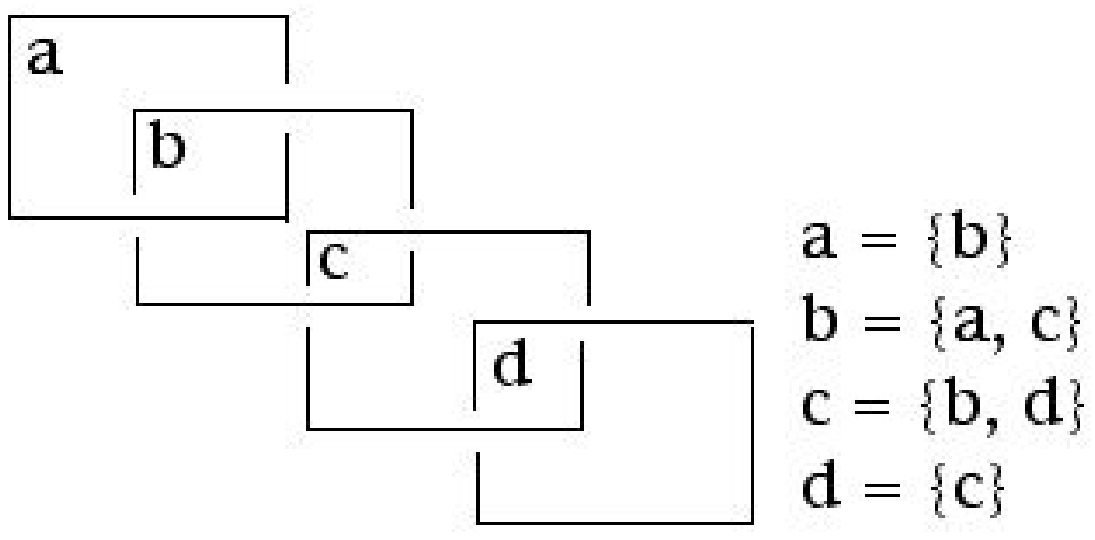}
     \end{tabular}
     \end{center}
     \caption{\bf Chain }
     \label{chain}
     \end{figure} 
     \bigbreak

\begin{figure}
     \begin{center}
     \begin{tabular}{c}
     \includegraphics[height=4cm]{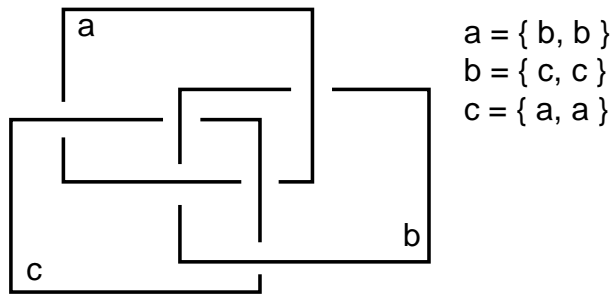}
     \end{tabular}
     \end{center}
     \caption{\bf Borromean Rings }
     \label{boro}
     \end{figure} 
     \bigbreak

We can summarize this first level of knot-set theory in the following two equivalences:
\begin{enumerate}
\item Self-Reference:   $$a = \{b,c, \cdots \} \Longleftrightarrow a = \{a,b,c, \cdots \}$$
\item Pair Cancellation: $$ S= \{a,a, b, c, \cdots \} \Longleftrightarrow  S = \{b,c, \cdots \}$$
\end{enumerate}
With this mode of dealing with self-reference and multiplicity, knot-set theory has the interpretation in terms of topological classes of diagrams. We could imagine that the flatlanders felt the need to invent three dimensional space and topology, just so their set theory would have such an elegant interpretation.
\bigbreak

But how elegant is this interpretation, from the point of view of topology? Are we happy that knots are equivalent to the empty knot-set as shown in Figure~\ref{sorrow}? For this, an extension of the theory is clearly in the waiting. 
We are happy that many topologically non-trivial links correspond to non-trivial knot-sets.
In the Figure~\ref{chain} , a chain link becomes a linked chain of knot-sets. But consider the link shown in Figure~\ref{boro}. These rings are commonly called the Borromean Rings. The Rings have the property that if you remove any one of them, then the other two are topologically unlinked. They form a topological tripartite relation. Their knot-set is described by the three equations

$$a = \{b,b \}$$
$$b = \{c,c \}$$
$$c = \{a,a \}.$$

Thus we see that this representative knot-set is a "scissors-paper-stone" pattern. Each component of the Rings lies over one other component, in a cyclic pattern. But in terms of the equivalence relation on knot sets that we have used, the knot set for the Rings is empty (by pair cancellation). 
\bigbreak

In order to go further in the direction of topological invariants for knots and links it is necessary to use more structure than the simple membership relation that motivates the knots-sets. Viewed from the point of view of the diagrams for knots and links there are a number of possible directions. For example, one can label all the arcs of the diagram and introduce algebraic relations at each crossing. This leads to the fundamental group and the quandle \cite{KP}. One can also label all the arcs of the diagram from an index set and view this labeling as a state in analogous to a state of a physical system in statistical mechanics. Then evaluations of these states and summations of the evaluations over all the states give the class of knot invariants called {\it quantum invariants} for knots and links \cite{KP}. These include the Jones polynomial and its generalizations. In this paper we will explain and use the Jones polynomial and the so-called colored Jones polynomials. See Section 17 for this development. The purpose of this section has been to introduce the subject of knot and link diagrams in the context of thinking about foundations of mathematics. However, it is worthwhile adding structure to the knot set theory so that it can at least see the higher order linking of the Borommean rings. We do this in the next subsection by keeping track of the order in which sets are encountered along the arc of a given component, and by keeping track of both membership and {\it co-membership} where we shall say that $A$ is  co-member of $B$ if $B$ is a member of $A.$ As one moves along an arc one sequentially encounters members and co-members.
\bigbreak

\subsection {Ordered Knot Sets}
\begin{figure}
     \begin{center}
     \begin{tabular}{c}
     \includegraphics[height=4cm]{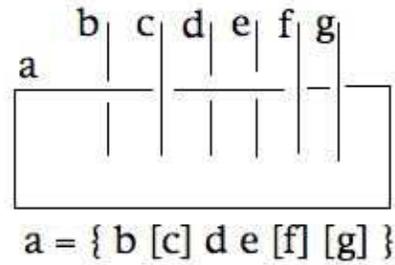}
     \end{tabular}
     \end{center}
     \caption{\bf  An ordered knot set.}
     \label{order}
     \end{figure}

Take a walk along a given component.
Write down the sequence of memberships and belongings that you encounter on the walk  as shown in 
Figure~\ref{order}.

     \bigbreak

\begin{figure}
     \begin{center}
     \begin{tabular}{c}
     \includegraphics[height=4cm]{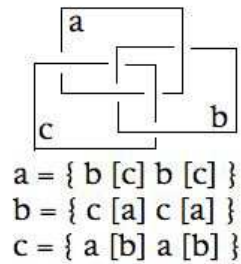}
     \end{tabular}
     \end{center}
     \caption{\bf  Borromean Rings as Ordered Knot Set.}
     \label{bororder}
     \end{figure}

In this notation, we record the order in which memberships and ``co-memberships" ( $a$ is a co-member of $b$ if and only if $b$ is a member of $a$) occur along the strand of a given component of the knot-set. 
We do not choose a direction of traverse, so  it is ok to reverse the total order of the contents of a given component and to take this order up to cyclic permutation. Thus we now have the representation of the Borromean Rings as shown in Figure~\ref{bororder}.

     \bigbreak

\begin{figure}
     \begin{center}
     \begin{tabular}{c}
     \includegraphics[height=9cm]{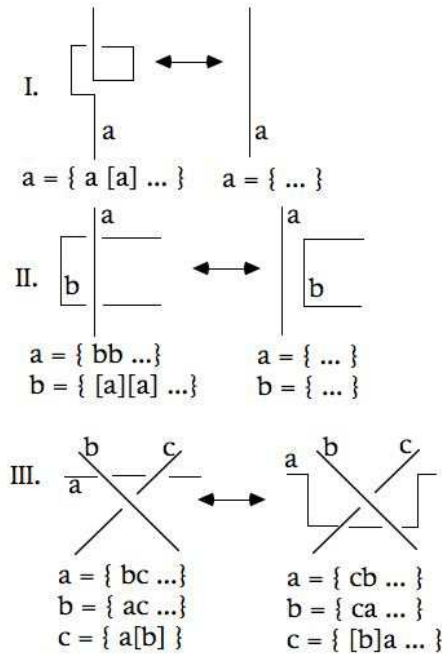}
     \end{tabular}
     \end{center}
     \caption{\bf  Reidemeister Moves for Ordered Knot Sets.}
     \label{rmovesordered}
     \end{figure}

With this extra information in front of us, it is clear that we should not allow the pair cancellations unless they occur in direct order, with no intervening co-memberships. Lets look at the revised Reidemeister moves as in Figure~\ref{rmovesordered}.

     \bigbreak

As is clear from the above diagrams, the Reidemeister moves tell us
that we should impose some specific equivalences on these ordered knot sets:
\begin{enumerate}
\item We can erase any appearance of $a[a]$ or of $[a]a$ inside the set for $a.$
\item If $bb$ occurs in $a$ and $[a][a]$ occurs in $b$, then they can both be erased.
\item  If $bc$ is in $a$, $ac$ is in $b$ and$ a[b]$ is in $c$, then we can reverse the order of each of these two element strings.  
\end{enumerate}

We take these three rules (and a couple of variants suggested by the diagrams) as the notion of equivalence of ordered knot-sets.
One can see that the ordered knot-set for the Borromean rings is non-trivial in this equivalence relation. In this sense we have a a proof that the Borromean rings are linked, based on their scissors, paper, stone structure. The only proof that I know for the non-triviality of the Borommean ordered knot set uses the concept of coloring discussed in the next subsection. 
\bigbreak

Knots and links are represented by the diagrams themselves, taken up the equivalence relation generated by the Reidemeister moves. This calculus of diagrams is quite complex and it is remarkable, the number and depth of different mathematical approaches that are used to study this calculus and its properties. Studying knots and links is rather like studying number theory. The objects of study themselves can be constructed directly, and form a countable set. The problems that seem to emanate naturally from these objects are challenging and fascinating.
For more about knot-sets, see \cite{KnotLogic}

\subsection {Quandles and Colorings of Knot Diagrams}
\begin{figure}
     \begin{center}
     \begin{tabular}{c}
     \includegraphics[height=4cm]{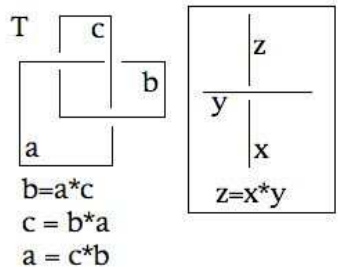}
     \end{tabular}
     \end{center}
     \caption{\bf  The Quandle for the Trefoil Knot.}
     \label{quandle}
     \end{figure}

There is an approach to studying knots and links that is very close to our ordered knot sets, but starts from a rather different premise.
In this approach each arc of the diagram receives a label or ``color".
An arc of the diagram is a continuous curve in the diagram that starts at one under crossing and ends at another under crossing. 
For example, the trefoil diagram is related to this algebra as shown in Figure~\ref{quandle}.

     \bigbreak

Each arc corresponds to an element of a ``color algebra"
$IQ(T)$ where $T$ denotes the trefoil knot. We have that$IQ(T)$ is generated by colors $a$,$b$ and $c$ with the relations
$c*b=a, a*c=b, b*a = c, a*a = a$. Each of these relations is a description of one of the crossings in $T.$ These relations are specific to the trefoil knot. If we take on an algebra of this sort, we want its coloring structure to be invariant under the Reidemeister moves. This implies the following global relations:
$$x*x = x$$
$$(x*y)*y= x$$
$$(x*y)*z = (x*z)*(y*z)$$
for any $x$, $y$ and $z$ in the algebra (set of colors) $IQ(T).$
An algebra that satisfies these rules is called an Involutory Quandle (See \cite{KP}),  hence the initials $IQ.$ These global relations are really expressions of the concept of self-crossing and iterated crossing in the multiplicity of crossings that are available in a calculus of boundaries where the notation
indicates the choice of interpretation, where one boundary is seen to cross (over) the other boundary.  
If we adopt these global relations for the algebra $IQ(K)$ for any knot or link diagram $K$, then two diagrams that are related by the Reidemeister moves will have isomorphic algebras. They will also inherit colorings of their arcs from one another. Thus the calculation of the algebra $IQ(K)$ for a knot or link $K$ has the potentiality for bringing forth deep topological structure from the diagram.
\bigbreak

In the case of the trefoil, one can see that the algebra actually closes at the set of elements $a,b,c.$
We have the complete set of relations $$c*b=a, a*c=b, b*a = c, a*a = a, a*a = a, b*b=b, c*c = c,$$ forming 
the {\em three-color quandle}. Three-coloring turns out to be quite useful for many knots and links.
Thus we have seen that the trefoil knot is knotted due to its having a non-trivial three-coloring.
By the same token, one can see that the Borommean rings are linked by checking that they do {\em not} have a non-trivial three-coloring! This fact is easy to check by directly trying to color the rings. That uncolorability implies that the rings are linked follows from the fact that there is a non-trivial coloring of three unlinked rings (color each ring by a separate color). This coloring of the unlinked rings would then induce a coloring of the Borommean rings. Since there is no such coloring, the Borommean rings must be linked.
\bigbreak

The ordered knot set corresponding to a link can be colored or not colored in the same manner as a link diagram. The spaces between the letters in the ordered code of the knot set can be assigned colors in the same way as the arcs of a link diagram. In this way, the coloring proofs can be transferred to ordered knot sets in the case of links. We leave the details of this analysis of link sets to another paper.
\bigbreak

Knot theory can be seen as a natural articulation not of three dimensional space (a perfectly good interpretation) but of the properties of interactions of boundaries. Each boundary can be regarded as that boundary transgressed by another boundary. The choice of who is the transgressed and who transgresses is the choice of a crossing, the choice of membership in the context of knot-set theory.  \bigbreak

\section {\bf Fermions, Majorana Fermions and Algebraic Knot Sets}
In the last part of our discussion of knot sets we added order and co-membership to the structure.
In this way of thinking, the knot set is an ordered sequence of memberships and co-memberships that are encountered as one moves along the strand of that part of the weave. Lets take this view, but go back to the ordinary knot sets that just catalog memberships. Then the knot set is a ordered list of the 
memberships that are encountered along the weave. For example, in Figure~\ref{chain} we have
$a = \{b \}, b = \{ a,c \}, c= \{ b,d \}, c = \{c\},$ and this would become the algebraic statements
$a = \{b\}, b = \{ac\}, c= \{bd\}, c = \{c\},$ where we remove the parentheses and write the contents of each set as a algebraic product. We retain the brackets in order to continue to differentiate the set from its contents. Then we would have that  $ \{ bccd \} = \{ bd \}$ since repetitions are eliminated, and we see that the rule $x^2 = 1$ should be obeyed by this algebra of products of set members.
\bigbreak

What shall we do about $a = \{ bcdc \}?$ we could decide that $xy = yx$ for all $x$ and $y$ in a given knot set. This commutative law would disregard the ordering, and we would have 
$\{ bcdc \} = \{ bccd \} \{ bd \}.$ The simplest algebraic version of the knot sets is to have a commutative algebra with $x^2 = 1$ for all members. Then we can define $X \Delta Y$ for sets
$X =\{ \alpha \}$ and $Y = \{ \beta \}$ by the equation $$X \Delta Y = \{ \alpha \beta \}$$ where 
$\alpha \beta$ represents the product of the members of $X$ and $Y$ taken together.
The operation $X \Delta Y$ represents the union  of knot sets and corresponds to exclusive or in standard set theory. 
\bigbreak

For example, suppose $$A = \{ yx \}, B= \{ zy \}, C = \{ xz \}. $$ Then we have
$A^2 = B^2 = C^2 = \{ 1 \}$ where it is understood that $\{ 1 \} = \{  \}$ represents the empty set.
(That is, in the algebra $1$ represents the empty word.) Furthermore we have
$AB = C, BC = A, CA =B.$ The relations in this example are very close to the quaternions. This example suggests that we could change the algebraic structure so that members satisfy $xy = - yx,$ adding
a notion of {\it sign } to the algebraic representation of the knot sets.  We then get the pattern of the quaternion group: $A^2 = B^2 = C^2 = ABC = -1$ where $-1$ denotes the ``negative" empty set.
\bigbreak

By introducing the {\it Clifford algebra} with $x^2 = 1$ and $xy = -yx$ for generators, we bring the knot sets into direct correspondence with an algebra of {\it Majorana fermions}. The generators of this Clifford algebra represent fermions that are their own anti-particles.  
For a long time it has been conjectured that neutrinos may be Majorana fermions. More recently, it has been suggested that Majorana fermions may occur in collective electronic phenomena \cite{Majorana,Ivanov,Beenakker,Georgiev,Topical}.
\bigbreak

There is a natural association of fermion algebra to knot sets. In order to explain this association, we first give a short exposition of the algebra of fermion operators. In a standard collection of fermion operators
$m_{1},\cdots , m_{k}$ one has that each $m_{i}$ is a linear operator on a Hilbert space with an adjoint
operator $m_{i}^{\dagger}$ (corresponding to the anti-particle for the particle created by $m_{i}$) and 
relations
$$m_{i}^2 = 0,$$
$$m_{i}m_{i}^{\dagger} + m_{i}^{\dagger}m_{i} = 1,$$
$$m_{i}m_{j} + m_{j}m_{i} = 0$$
when $i \ne j.$
\bigbreak

There is another  brand of Fermion algebra where we have
generators $c_{1}, \cdots c_{k}$ and $c_{i}^2 = 1$ while $c_{i}c_{j} = -c_{j}c_{i}$ for all $i \ne j.$
These are the {\em Majorana fermions}. There is a algebraic translation between the fermion algebra and Majorana fermion algebra. Given two Majorana fermions $a$ and $b$ with $a^2 = b^2 = 1$ and
$ab = -ba,$ define $$m = (a + ib)/2$$ and
$$m^{\dagger} = (a - ib)/2.$$ It is then easy to see that $a^2 = b^2 = 1$ and $ab = -ba$ imply that
$m$ and $m^{ \dagger }$ form a fermion in the sense that $m^2 = (m^{\dagger})^{2} = 0$ and 
$m m^{\dagger} + m^{\dagger} m = 1.$ Thus pairs of Majorana fermions can be construed as ordinary fermions. Conversely, if $m$ is an ordinary fermion, then formal real and imaginary parts of $m$ yield a mathematical pair of Majorana fermions. A chain of electrons in a nano-wire, conceived in this way can give rise to a chain of Majorana fermions with a non-localized pair corresponding to the distant ends of the chain. The non-local nature of this pair is promising for creating topologically protected qubits, and 
there is at this writing an experimental search for evidence for the existence of such end-effect Majorana fermions.
\bigbreak

We now see that it is exactly the Majorana fermion algebra that matches the properties of the knot sets.
Here is an example that shows how the topology comes in. Let $x, y, z$ be three Majorana fermions.
Let $A = yx, B =  zy, C = xz.$ We have already seen that $A, B, C$ represent the quaternions.
Now define 
$$s_{1} = (1 + A) /\sqrt{2}, s_{2} = (1 + B)/ \sqrt{2}, s_{3} = (1 + C) /\sqrt{2}.$$
It is easy to see that $s_{i}$ and $s_{j}$ satisfy the braiding relation for any $i \ne j.$ For example, here is the verification for $i =1, j =2.$
$$s_{1}s_{2}s_{1} = (1/2 \sqrt{2})(1 + A)(1+ B)(1+A)$$
$$= (1/2 \sqrt{2})(1 + A + B + AB)(1+A)$$
$$ =  (1/2 \sqrt{2})(1 + A + B + AB + A + A^2 + BA + ABA)$$
$$=  (1/2 \sqrt{2})(1 + A + B + AB + A  -1 - AB + B)$$
$$ = (1/ \sqrt{2})(A + B).$$
Similarly,
$$s_{2}s_{1}s_{2} = (1/2 \sqrt{2})(1 + B)(1+ A)(1+B)$$
$$= (1/2 \sqrt{2})(1 + A + B + BA)(1+B)$$
$$ =  (1/2 \sqrt{2})(1 + A + B + BA + B + AB + B^2 + BAB)$$
$$=  (1/2 \sqrt{2})(1 + A + B + BA + B  - BA -1  + A)$$
$$ = (1/ \sqrt{2})(A + B).$$
Thus $$s_{1}s_{2}s_{1} = s_{2}s_{1}s_{2},$$
and so a natural braid group representation arises from the Majorana fermions.
This braid group representation is significant for quantum computing as we shall see in Section 7.
For the purpose of this discussion, the braid group representation shows that the Clifford algebraic representation for knot sets is related to topology at more than one level. The relation $x^2 = 1$ for generators makes the individual sets, taken as products of generators, invariant under the Reidemeister moves (up to a global sign). But braiding invariance of certain linear combinations of sets is a relationship with knotting at a second level. This multiple relationship certainly deserves more thought.
We will make one more remark here, and reserve further analysis for a subsequent paper.
\bigbreak

\begin{figure}
     \begin{center}
     \begin{tabular}{c}
     \includegraphics[height=7cm]{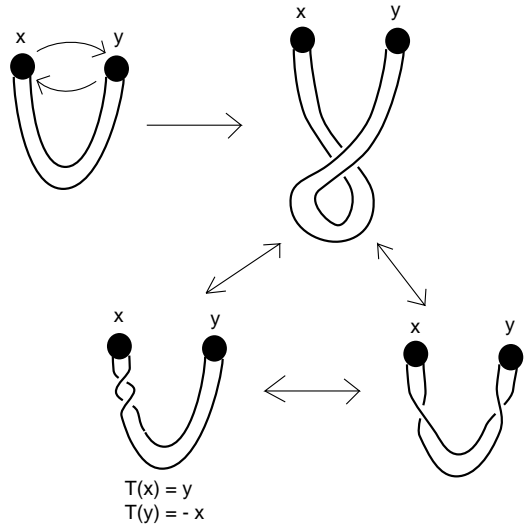}
     \end{tabular}
     \end{center}
     \caption{\bf Braiding Action on a Pair of Fermions}
     \label{ex}
     \end{figure}

These braiding operators can be seen to act on the vector space over the complex numbers that is spanned by the fermions $x, y, z.$ To see how this works, consider
$$s =  \frac{1 + yx}{\sqrt{2}},$$
$$T(p) = sps^{-1} = (\frac{1 + yx}{\sqrt{2}})p(\frac{1 - yx}{\sqrt{2}}),$$ 
and verify that 
$T(x) = y$ and $T(y) = -x.$ Now view Figure~\ref{ex} where we have illustrated a 
topological interpretation for the braiding of two fermions. In the topological interpretation the two fermions are connected by a flexible belt. On interchange, the belt becomes twisted  by $2 \pi.$ 
In the topological interpretation a twist of $2 \pi$ corresponds to a phase change of $-1.$ 
(For more information on this topological interpretation of $2 \pi$ rotation for fermions, see \cite{KP}.)
Without a further choice it is not evident which particle of the pair should receive the phase change. The topology alone tells us only the relative change of phase between the two particles. The Clifford algebra for Majorana fermions makes a specific choice in the matter and in this way fixes the representation of the braiding.
\bigbreak

Finally, we remark that linear combinations of products in the Clifford algebra can be regarded as superpositions of the knot sets. Thus $xy + xz$ is a superposition of the sets with members $\{ x,y \}$ and$\{ x,z \}.$ Superposition of sets suggests that we are creating a species of 
{\it quantum set theory} and indeed Clifford algebra based quantum set theories have been suggested
(see \cite{Fink}) by David Finkelstein and others. It may come as a surprise to a quantum set theorist to find that knot theoretic topology is directly related to this subject.  It is also clear that this Clifford algebraic quantum set theory should be related to our previous constructions for quantum knots 
\cite{QK1,QK2,QK3,QK4,QK5}. 
This requires more investigation, and it suggests that knot theory and the theory of braids occupy a fundamental place in the foundations of quantum mechanics.
\bigbreak

     \bigbreak

\section{Laws of Form}
\begin{figure}
     \begin{center}
     \begin{tabular}{c}
     \includegraphics[height=4cm]{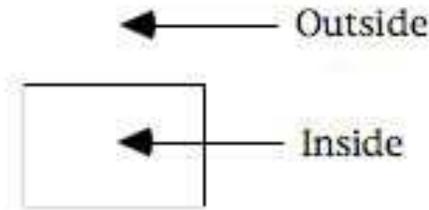}
     \end{tabular}
     \end{center}
     \caption{\bf Inside and Outside}
     \label{outin}
     \end{figure}

In this section we discuss a formalism due the G. Spencer-Brown \cite{LOF} that is often called the ``calculus of indications". This calculus is a study of mathematical foundations with a topological notation based on one symbol, the mark:
$$\M{ } \, .$$
This single symbol represents a distinction between its own inside and outside.
As is evident from Fgure~\ref{outin}, the mark is regarded as a shorthand for a rectangle drawn in the plane and dividing the plane into the regions inside and outside the rectangle.   
\bigbreak

     \bigbreak

The reason we introduce this notation is that in the calculus of indications the mark can interact with itself in two possible ways. The resulting formalism becomes a version of Boolean arithmetic, but fundamentally simpler than the usual Boolean arithmetic of $0$ and $1$ with its 
two binary operations and one unary operation (negation). In the calculus of indications one takes a step in the direction of simplicity, and also a step in the direction of physics. The patterns of this
mark and its self-interaction match those of a {\it Majorana fermion}  as discussed in the previous section. A Majorana fermion is a particle that is its own anti-particle.  \cite{ Majorana}. We will later see, in this paper, that by adding braiding to the calculus of indications we arrive at the Fibonacci model, that can in principle support quantum computing.
\bigbreak

In the previous section we described Majorana fermions in terms of their algebra of creation and annihilation operators. Here we describe the particle directly in terms of its interactions. This is part of a general scheme called ``fusion rules" \cite{MooreRead} that can be applied to discrete particle interacations. A fusion rule represents all of the different particle interactions in the form of a set of equations.
The bare bones of the Majorana fermion consist in a particle $P$ such that $P$ can interact with itself to produce a neutral particle $*$ or produce itself $P.$ Thus the possible interactions are 
$$PP \longrightarrow *$$
and 
$$PP \longrightarrow P.$$
This is the bare minimum that we shall need. 
The fusion rule is $$P^2 = 1 + P.$$ This represents the fact that $P$ can interact with itself to produce the neutral particle (represented as $1$ in the fusion rule) or itself (represented by $P$ in the fusion rule). We shall come back to the combinatorics related to this fusion equation.
\bigbreak

Is there  a {\it linguistic} particle that is its own anti-particle? Certainly we have
$$\sim \sim Q = Q$$ for any proposition $Q$ (in Boolean logic). And so we might 
write 
$$\sim \sim \longrightarrow *$$
where $*$ is a neutral linguistic particle, an identity operator so that
$$*Q = Q$$ for any proposition $Q.$ But in the normal use of negation there is no way that the negation sign combines with itself to produce itself. This appears to ruin the analogy between negation and the 
Majorana fermion. Remarkably, the calculus of indications provides a context in which we can say exactly that a certain logical particle, the mark,  can act as negation {\it and} can interact with itself to produce itself.
\bigbreak

\begin{figure}
     \begin{center}
     \begin{tabular}{c}
     \includegraphics[height=4cm]{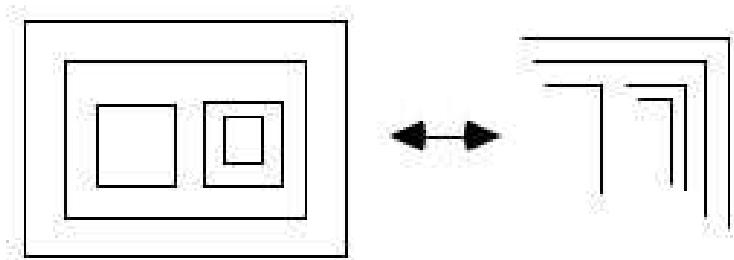}
     \end{tabular}
     \end{center}
     \caption{\bf Boxes and Marks}
     \label{boxmark}
     \end{figure}

In the calculus of indications patterns of non-intersecting marks (i.e. non-intersecting rectangles) are called {\it expressions.} For example in Figure~\ref{boxmark} we see how patterns of boxes correspond to patterns of marks.

     \bigbreak

\begin{figure}
     \begin{center}
     \begin{tabular}{c}
     \includegraphics[height=4cm]{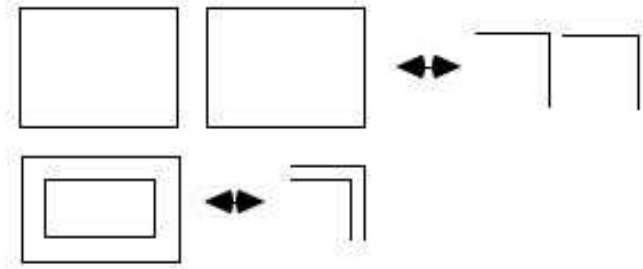}
     \end{tabular}
     \end{center}
     \caption{\bf Translation between Boxes and Marks}
     \label{markbox}
     \end{figure}

In Figure~\ref{boxmark}, we have illustrated both the rectangle and the marked version of the expression.  In an expression you can say definitively of any two marks whether one is or is not inside the other.  The relationship between two marks is either that one is inside the other, or that neither is inside the other.  These two conditions correspond to the two elementary expressions shown in
Figure~\ref{markbox}.

     \bigbreak

The mathematics in Laws of Form begins with two laws of transformation about these two basic expressions. Symbolically, these laws are:
\begin{enumerate}
\item Calling : $$\M{} \, \M{} \, =   \M{}$$  
\item Crossing: $$\M{ \M{ } }  =  \,\,\,\,.$$
\end{enumerate}
The equals sign denotes a replacement step that can be performed on instances of these patterns
(two empty marks that are adjacent or one mark surrounding an empty mark).
In the first of these equations two adjacent marks condense to a single mark, or a single mark expands to form two adjacent marks.  In the second equation  two marks, one inside the other, disappear to form the unmarked state indicated by nothing at all. 
That is, two nested marks can be replaced by an empty word in this formal system.  Alternatively, the unmarked state can be replaced by two nested marks. These equations give rise to a natural calculus, and the mathematics can begin.  For example,  {\it any expression can be reduced uniquely  to either the marked or the unmarked state.}  The he following example illustrates the method:
$$     \M{\M{\M{\M{} \M{}} \M{}} \M{}} \M{}  =   \M{\M{\M{\M{}} \M{}} \M{}}\M{} =   \M{\M{ \M{}} \M{}}\M{} $$
$$ = \M{\M{}}\M{} = \M{} \,\,\,.$$
The general method for reduction is to locate marks that are at the deepest places in the expression
(depth is defined by counting the number of inward crossings of boundaries needed to reach the given mark). Such a deepest mark must be empty and it is either surrounded by another mark, or it is adjacent to an empty mark. In either case a reduction can be performed by either calling or crossing. 
\bigbreak 

Laws of Form begins with the following statement.
``We take as given the idea of a distinction and the idea of an indication, and that it is not possible to make an indication without drawing a distinction. We take therefore the form of distinction for the form."  
Then the author makes the following two statements (laws):
\begin{enumerate}
\item {\it The value of a call made again is the value of the call.}
\item {\it The value of a crossing made again is not the value of the crossing.}
\end{enumerate}
The two symbolic equations above correspond to these statements. First examine the law of calling. It says that the value of a repeated name is the value of the name. In the equation
$$\M{} \, \M{} \, = \M{}$$
one can view either mark as the name of the state indicated by the outside of the other mark.  
In the other equation
$$\M{ \M{ } } = \,\,\,\,.$$
the state indicated by the outside of a mark is the state obtained by crossing from the state indicated on the inside of the mark. Since the marked state is indicated on the inside, the outside must indicate the unmarked state.  The Law of Crossing indicates how opposite forms can fit into one another and vanish into nothing, or how nothing can  produce opposite and distinct forms that fit one another, hand in glove.  The same interpretation yields the equation
$$\M{} \, = \, \M{}$$
where the left-hand side is seen as an instruction to cross from the unmarked state, and the right hand side is seen as an indicator of the marked state. The mark has a double carry of meaning. It can be seen as an operator, transforming the state on its inside to a different state on its outside, and it can be seen as the name of the marked state. That combination of meanings is compatible in this interpretation.  
\bigbreak

From the calculus of indications, one moves to algebra.  Thus 
 $$\M{\M{A}}$$
stands for the two possibilities
  $$\M{\M{\M{}}} \, = \, \M{} \,  \longleftrightarrow  \, A = \M{}$$
$$\M{\M{}} \, = \, \, \, \,  \longleftrightarrow \, A \,  = $$
In all cases we have
$$\M{\M{A}} \, = \, A.$$
 
 By the time we articulate the algebra, the mark can take the role of a unary operator
 $$ A \longrightarrow \M{A}.$$ But it retains its role as an element in the algebra.
Thus begins algebra with respect to this non-numerical arithmetic of forms.  The primary algebra that emerges is a subtle precursor to Boolean algebra.  One can translate back and forth between elementary logic and primary algebra:
\begin{enumerate}
\item $\M{} \longleftrightarrow T$
\item $\M{\M{}} \longleftrightarrow F$
\item $\M{A} \longleftrightarrow \sim A$
\item $AB \longleftrightarrow A \vee B$
\item $\M{\M{A} \M{B}} \longleftrightarrow A \wedge B$
\item $\M{A}B \,\, \longleftrightarrow \,\,A \Rightarrow B$
\end{enumerate}
The calculus of indications and the primary algebra form an efficient system for working with basic symbolic logic.
\bigbreak

By reformulating basic symbolic logic in terms of the calculus of indications, we have a ground in which negation is represented by  the mark {\em and} the mark is also interpreted as a value (a truth value for logic) and these two intepretations are compatible with one another in the formalism. The key to this compatibility is the choice to represent the value ``false"  by a literally unmarked state in the notational plane. With this the empty mark (a mark with nothing on its inside)  can be interpreted as the negation of ``false" and hence represents ``true".
The mark interacts with itself to produce itself (calling) and the mark interacts with itself to produce nothing (crossing). We have expanded the conceptual domain of negation so that it satisfies the mathematical pattern of an abstract Majorana fermion.
\bigbreak

Another way to indicate these two interactions symbolically is to use a box,for the marked state and a blank space for the unmarked state.
Then one has two modes of interaction of a box with itself:
\begin{enumerate}
\item Adjacency: $\fbox{~} ~~ \fbox{~}$
\smallbreak
\noindent and 
\item Nesting: $\fbox{ \fbox{~~} }.$
\end{enumerate}

\noindent With this convention we take the adjacency interaction to yield a single box, and the nesting interaction to produce nothing:

$$\fbox{~} ~~ \fbox{~} = \fbox{~}$$
$$\fbox{ \fbox{~~} } =  $$

\noindent We take the notational opportunity to denote nothing by an asterisk (*). The syntatical rules for operating the asterisk are
Thus the asterisk is a stand-in for no mark at all and it can be erased or placed wherever it is convenient to do so.
Thus $$\fbox{ \fbox{~~} } = *. $$
\bigbreak

At this point the reader can appreciate what has been done if he returns to the usual form of symbolic logic. In that form we that $$\sim \sim X = X$$ for all logical objects (propositions or elements of the logical algebra) $X.$ We can summarize this by writing $$\sim \sim \,\,\,= \,\,\, $$ as a symbolic statement that is outside the logical formalism. Furthermore, one is committed to the interpretation of 
negation as an operator and not as an operand. The calculus of indications provides a formalism where
the mark (the analog of negation in that domain) is both a value and an object, and so can act on itself in more than one way.
\bigbreak

The Majorana particle is its own anti-particle. It is exactly at this point that physics meets logical epistemology. Negation as logical entity is its own anti-particle.  Wittgenstein says (Tractatus \cite{Wittgen} $4.0621$) ``$\cdots$ the sign `$\sim$' corresponds to nothing in reality." And he goes on to say (Tractatus  $5.511$) `` How can all-embracing logic which mirrors the world use such special catches and manipulations? Only because all these are connected into an infinitely fine network, the great mirror." For Wittgenstein in the Tractatus,  the negation sign is part of  the mirror making it possible for thought to reflect reality through combinations of signs. These remarks of Wittgenstein are part of his early picture theory of the relationship of formalism and the world. In our view,  the world and the formalism we use to represent the world are not separate.  The observer and the mark are (formally) identical. A path is opened between logic and physics.
\bigbreak

\begin{figure}
     \begin{center}
     \begin{tabular}{c}
     \includegraphics[height=7cm]{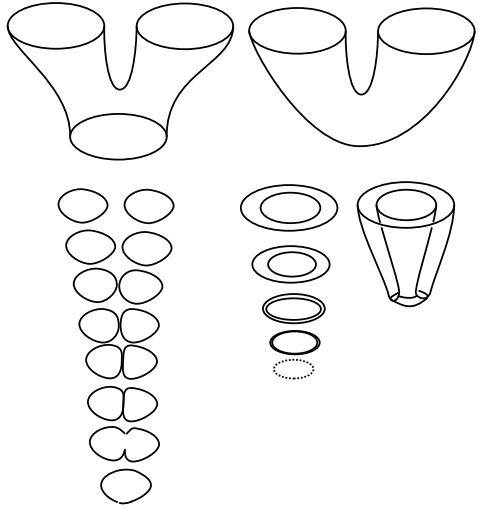}
     \end{tabular}
     \end{center}
     \caption{\bf Calling, Crossing and Cobordism}
     \label{callcross}
     \end{figure}

The visual iconics that create via the boxes of half-boxes of the calculus of indications a model for a logical Majorana fermion can also be seen in terms of cobordisms of surfaces. View Figure~\ref{callcross}. There the boxes have become circles and the interactions of the circles have been displayed as evolutions in an extra dimension, tracing out surfaces in three dimensions. The condensation of two circles to one is a simple cobordism betweem two circles and a single circle. The cancellation of two circles that are concentric can be seen as the right-hand lower cobordism in this figure with a level having a continuum of critical points where the two circles cancel. A simpler cobordism is illustrated above on the right where the two circles are not concentric, but nevertheless are cobordant to the empty circle. Another way of putting this is that two topological closed strings can interact by cobordism to produce a single string or to cancel one another. Thus a simple circle can be a topological model for a Majorana fermion.  

     \bigbreak

In Sections 15 and 16  we detail how the Fibonacci model for anyonic quantum computing \cite{Kitaev,Preskill} can be constructed by using a version of
the two-stranded  bracket polynomial and a generalization of Penrose spin networks. This is a fragment of the Temperly-Lieb recoupling theory \cite{KL}.  
\bigbreak

\subsection{The Square Root of Minus One is an Eigenform and a  Clock}
 So far we have seen that the mark can represent the fusion rules for a Majorana fermion since it can interact with itself to produce either itself or nothing. But we have not yet seen the anti-commuting fermion algebra emerge from this context of making a distinction. Remarkably, this algebra does emerge when one looks at the mark recursively.  
 \bigbreak
 
\begin{figure}
     \begin{center}
     \begin{tabular}{c}
     \includegraphics[width=6cm]{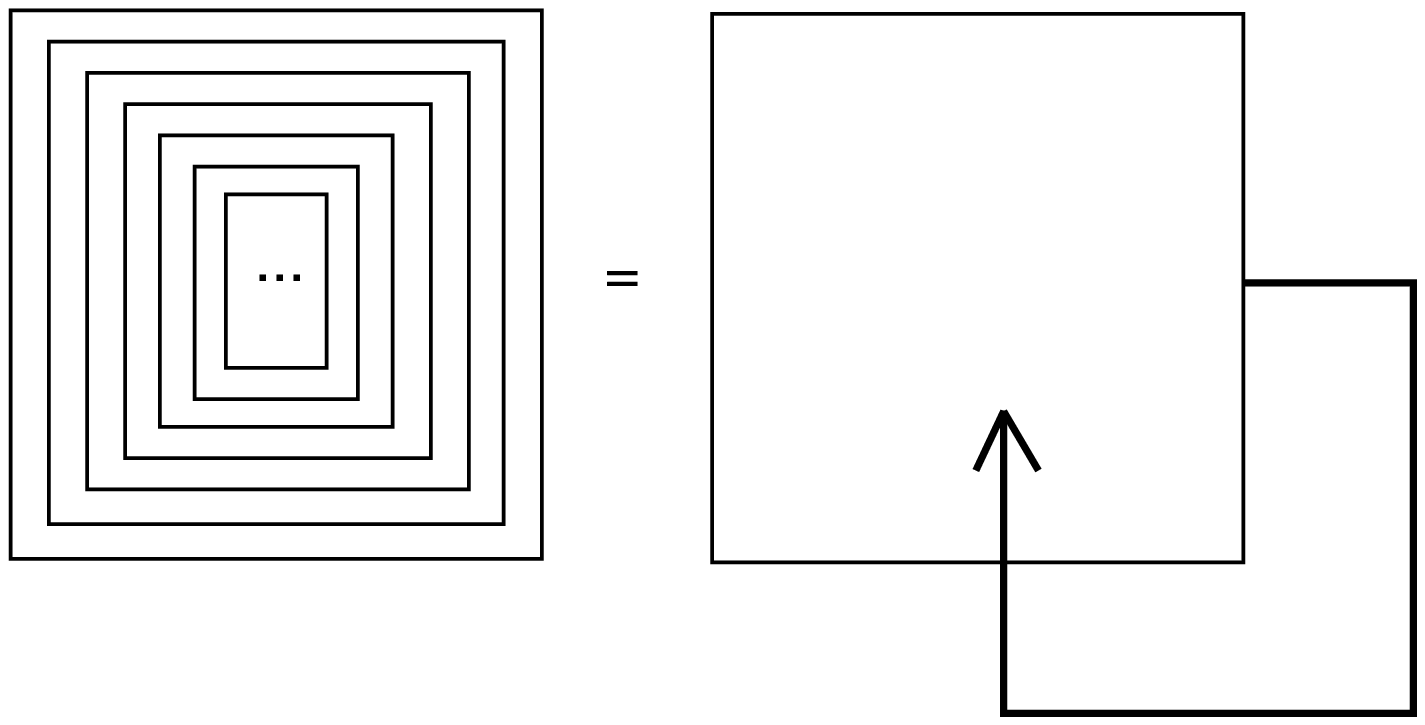}
     \end{tabular}
     \caption{\bf}
     \label{fix}
\end{center}
\end{figure}

 Consider the transformation $$F(X) = \M{X}.$$
 If we iterate it  and take the limit we find 
 $$G = F(F(F(F( \cdots )))) = \M{\M{\M{\M{ ... }}}}$$ 
 an infinite nest of marks satisfying the equation
 $$G = \M{G}.$$
 With $G = F(G),$ I say that $G$ is an {\em eigenform} for the transformation $F.$
 See \cite{} for more about this point of view.
 See Figure~\ref{fix} for an illustration of this nesting with boxes and an arrow that points inside the reentering mark to indicate its appearance inside itself. If one thinks of the mark itself as a Boolean logical value, then extending the language to include the reentering mark $G$ goes beyond the boolean. We will not detail here how this extension can be related to non-standard logics, but refer the reader to \cite{KnotLogic}. Taken at face value the reentering mark cannot be just marked or just unmarked, for by its very definition, if it is marked then it is unmarked and if it is unmarked then it is marked. In this sense the reentering mark has the form of a self-contradicting paradox. There is no paradox since we do not have to permanently assign it to either value. The simplest interpretation of the reentering mark is that it is temporal and that it represents an oscillation between markedness and unmarkedness.  In numerical terms it is a discrete dynamical system oscillating between $+1$ (marked) and $-1$ (not marked).
 \bigbreak
 
 With the reentering mark in mind consider now the transformation on real numbers given by 
 $$T(x) = -1/x.$$ This has the fixed points $i$ and $-i$, the complex numbers whose squares are 
 negative unity. But lets take a point of view more directly associated with the analogy of the recursive mark. Begin by starting with a simple periodic process that is associated directly 
with the classical attempt to solve for $i$ as a solution to a quadratic equation. We take the point
of view that solving $x^2 = ax + b$ is the same (when $x \ne 0$) as solving 
$$ x = a + b/x,$$ and hence is a matter of finding a fixed point. In the case of $i$ we have
$$x^2 = -1$$ and so desire a fixed point $$x = -1/x.$$ There are no real numbers that are fixed points for this operator and so we consider the oscillatory process generated by
$$T(x) = -1/x. $$
The fixed point would satisfy
$$i = -1/i$$
and multiplying, we get that 
$$ii = -1.$$
On the other hand the iteration of $T$ yields
$$ 1, T(1) = -1 , T(T(1)) = +1 , T(T(T(1))) = -1, +1,-1,+1,-1, \cdots .$$
The square root of minus one is a perfect example of an eigenform that occurs in a new and wider domain than the original context in which its recursive process arose. The process has no fixed point in the original domain.
\smallbreak

\begin{figure}
     \begin{center}
     \begin{tabular}{c}
     \includegraphics[width=6cm]{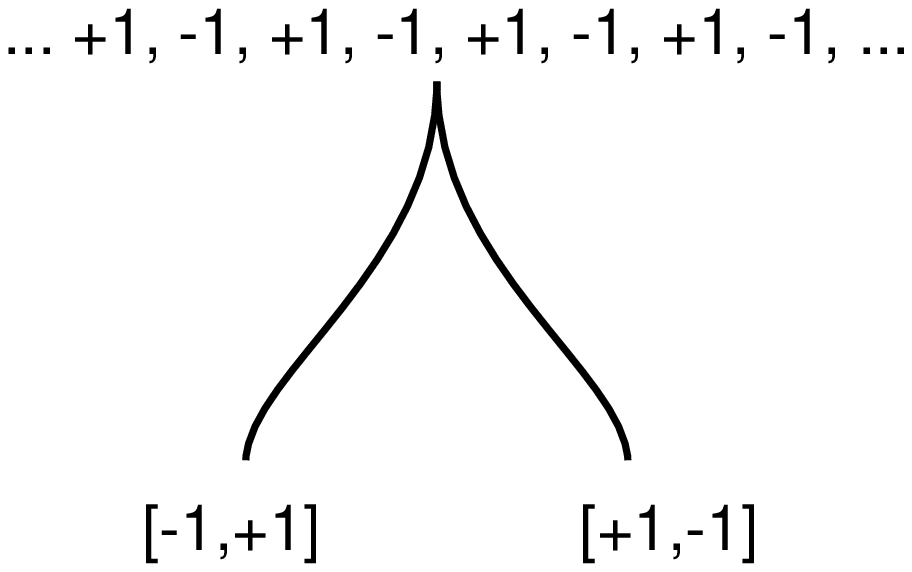}
     \end{tabular}
     \caption{\bf}
     \label{clock}
\end{center}
\end{figure}

Looking at the oscillation between $+1$ and $-1,$ we see that there are naturally two phase-shifted viewpoints.  We denote these two views of the  oscillation by $[+1,-1]$ and$ [-1,+1].$ 
These viewpoints correspond to whether one regards the oscillation at time zero as starting with $+1$ or with $ -1.$ See Figure~\ref{clock}. We shall let the word {\em iterant} stand for an undisclosed alternation or ambiguity between $+1$ and $-1.$ There are two iterant views: $[+1,-1]$ and $[-1,+1] $ for the basic process we are examining. Given an iterant $[a,b],$ we can think of $[b,a]$ as the same process with a shift of one time step. The two iterant views,  $[+1,-1]$ and $[-1,+1] $, will become the square roots of negative unity,  $i$ and $-i.$
\smallbreak

We introduce a {\it temporal shift operator} $ \eta$  such that 
$$[a,b]\eta  = \eta  [b,a]$$ and $$\eta \eta  = 1$$ 
for any iterant $[a,b],$  so that concatenated observations can include a time step of one-half period of the process 
$$\cdots abababab \cdots.$$ 
We combine iterant views term-by-term as in 
$$[a,b][c,d] = [ac,bd]. $$
We now define i by the equation
$$i = [1,-1] \eta . $$
This makes $i $ both a value and an operator that takes into account a step in time.
\smallbreak

We calculate
$$ii = [1,-1] \eta [1,-1] \eta  = [1,-1][-1,1]\eta \eta  = [-1,-1] = -1.$$
Thus we have constructed a square root of minus one by using an iterant viewpoint.  In this view $i$ represents a discrete oscillating temporal process and it is an eigenform for $T(x) = -1/x,$ participating in the algebraic structure of the complex numbers. In fact the corresponding algebra structure of 
linear combinations $[a,b] + [c,d]\eta$ is isomorphic with $2 \times 2$ matrix algebra and iterants can be used to construct $n \times n$ matrix algebra. We treat this generalization elsewhere \cite{NonCom1,NonCom2}.
\smallbreak

Now we can make contact with the algebra of the Majorana fermions. Let $e = [1,-1].$ Then we have
$e^2 = [1,1] = 1$ and $e \eta  =  [1,-1] \eta  =  [-1,1] \eta  = - e \eta.$
Thus we have $$e^2 =1,$$ $$\eta^2 = 1,$$ and $$e \eta = - \eta e.$$
We can regard $e$ and $\eta$ as a fundamental pair of Majorana fermions. This is a formal correspondence, but it is striking how this Marjorana fermion algebra emerges from an analysis of the 
recursive nature of the reentering mark, while the fusion algebra for the Majorana fermion emerges from the distinctive properties of the mark itself. We see how the seeds of the fermion algebra live  in this extended logical context.
\bigbreak 

Note how the development of the algebra works at this point. We have that $$(e \eta)^2 = -1$$
and so regard this as a natural construction of the square root of minus one in terms of the phase synchronization of the clock that is the iteration of the reentering mark. Once we have the square root of minus one it is natural to introduce another one and call this one $i,$ letting it commute with the other operators. Then we have the $(ie \eta)^2 = +1$ and so we have a triple of Majorana fermions: $$a = e, b = \eta, c = ie \eta$$ and we can construct the quaternions
$$I = ba = \eta e, J = cb = ie,  K = ac = i \eta.$$ With the quaternions in place, we have the braiding 
operators $$A = \frac{1}{\sqrt{2}}(1 + I), B = \frac{1}{\sqrt{2}}(1 + J), C = \frac{1}{\sqrt{2}}(1 + K),$$ 
and can continue as we did in Section 4. 
\bigbreak

There is one more comment that is appropriate for this section. Recall from Section 4 that a pair of Majorana fermions can be assembled to form a single standard fermion.  In our case we have the two Marjorana fermions $e$ and $\eta$ and the corresponding standard fermion creation and annihilation operators are then given by the formulas below.
$$\psi = (e + i\eta)/2$$ and
$$\psi^{\dagger} = (e - i\eta)/2.$$
Since $e$ represents a spatial view of the basic discrete oscillation and $\eta$ is the time-shift operator for this oscillation it is of interest to note that the standard fermion built by these two can be regarded as a quantum of spacetime, retrieved from the way that we decomposed the process into space and time. Since all this is initially built in relation to extending the Boolean logic of the mark to a non-boolean recursive context, there is further analysis needed of the relation of the physics and the logic. This will be taken up in a separate paper.
\bigbreak

\subsection{Relativity and the Dirac Equation}
Starting with the algebra structure of $e$ and $\eta$ and adding a commuting square root of $-1$,
$i,$ we have constructed fermion algebra and quaternion algebra. We can now go further and construct the Dirac equation. This may sound circular, in that the fermions arise from solving the Dirac equation, but in fact the algebra underlying this equation has the same properties as the  creation and 
annihilation algebra for fermions, so it is by way of this algebra that we will come to the Dirac equation.
If the speed of light is equal to $1$ (by convention), then energy $E$, momentum $p$ and mass $m$ are
related by the (Einstein) equation $$E^2 = p^2 + m^2.$$ Dirac constructed his equation by looking
for an algebraic square root of $ p^2 + m^2$ so that he could have a linear operator for $E$ that would take the same role as the Hamiltonian in the Schrodinger equation. We will get to this operator by first taking the case where $p$ is a scalar (we use one dimension of space and one dimension of time.
Let $E = \alpha p + \beta m$ where $\alpha$ and $\beta$ are elements of a a possibly non-commutative,
associative algebra. Then $$E^2 = \alpha^2 p^2 + \beta^2 m^2 + pm(\alpha \beta + \beta \alpha).$$
Hence we will satisfiy $E^2 = p^2 +m^2$ if $\alpha^2 = \beta^2 = 1$ and
 $\alpha \beta + \beta \alpha = 0.$ This is our familiar Clifford algebra pattern and we can use the iterant algebra generated by $e$ and $\eta$ if we wish. Then, because the quantum operator for momentum is $-i \partial/\partial x$ and the operator for energy is $i\partial/\partial t,$ we have the Dirac equation $$i\partial \psi /\partial t = -i \alpha \partial \psi /\partial x + \beta m \psi.$$
 Let $${\cal O} = i\partial /\partial t + i \alpha \partial /\partial x - \beta m $$ so that the Dirac equation 
 takes the form $${\cal O} \psi(x,t) = 0.$$ Now note that 
 $${\cal O} e^{i(px - Et)} = (E - \alpha p + \beta m) e^{i(px - Et)}$$ and that  if
 $$ U = (E - \alpha p + \beta m)\beta \alpha = \beta \alpha E + \beta p + \alpha m,$$
 then $$U^2 = -E^2 + p^2 + m^2 = 0, $$ from which it follows that $$\psi = U e^{i(px - Et)}$$ is a (plane wave) solution to the Dirac equation.
\bigbreak

In fact, this calculation suggests that we should multiply the operator ${\cal O}$ by $\beta \alpha$ on the right, obtaining the operator 
$${\cal D} = {\cal O} \beta \alpha =  i\beta \alpha \partial /\partial t + i \beta \partial /\partial x + \alpha m, $$
and the equivalent Dirac equation $${\cal D}\psi = 0.$$ In fact for the specific $\psi$ above we will now 
have ${\cal D} (U  e^{i(px - Et)} ) = U^2  e^{i(px - Et)} = 0.$ This way of reconfiguring the Dirac equation
in relation to nilpotent algebra elements $U$ is due to Peter Rowlands \cite{Rowlands}. 
We will explore this relationship with the Rowlands formulation in a separate paper.
\bigbreak

Return now to the original version of the Dirac equation.
$$i\partial \psi /\partial t = -i \alpha \partial \psi /\partial x + \beta m \psi.$$
 We can rewrite this as
 $$\partial \psi /\partial t =  \alpha \partial \psi /\partial x + i \beta m \psi.$$
 We see that if $i \beta$ is real, then we can write a fully real version of the Dirac equation.
 For example, we can take the equation 
 $$\partial \psi /\partial t =  e \partial \psi /\partial x + e \eta m \psi.$$
 where we represent 
 \[e  = \left( \begin{array}{cc}
-1 & 0 \\
 0  & 1 \\
\end{array} \right)\] and 
 \[\eta  = \left( \begin{array}{cc}
 0 & 1 \\
 1  & 0 \\
\end{array} \right)\]  as matrix versions of the iterants associated with the reentering mark. For the case of one dimension of space and one dimension of time, this is the 
Majorana representation for the Dirac equation (compare \cite{KN1}). Since the equation can have real solutions, these are their own complex conjugates and correspond to particles that are their own anti-particles. As the reader can check, the corresponding Rowland nilpotent $U$ is given by the formula
$$U = -i \eta E + i e \eta p + e m.$$
For effective application to the topics in this paper, one needs to use two dimensions of space and one dimension of time.
This will be explored in another paper. In the present paper  we have given a picture of how, starting with the mark as a logical and recursive particle, one can tell a story that reaches the Dirac equation and its algebra.
\bigbreak

\section{Quantum Mechanics and Quantum Computation}
We shall quickly
indicate the basic principles of quantum mechanics.  The quantum information context 
encapsulates a concise model of quantum theory:
\bigbreak

{\em The initial state of a quantum process is a vector $|v \rangle$ in a complex vector space $H.$
Measurement returns basis elements $\beta$ of $H$ with probability 

$$|\langle \beta \,|v \rangle |^{2}/\langle v \,|v \rangle$$

\noindent where $\langle v \,|w \rangle = v^{\dagger}w$ with $v^{\dagger}$ the conjugate transpose of $v.$
A physical process occurs in steps $|v\rangle \longrightarrow U\,|v \rangle = |Uv \rangle $ where $U$ is a unitary linear transformation.
\bigbreak

Note that since $\langle Uv \,|Uw \rangle = \langle v \,|U^{\dagger}U |w \rangle = \langle v \,|w \rangle = $ when $U$ is unitary, it follows that probability
is preserved in the  course of a quantum process.  }
\bigbreak

One of the details required for any specific quantum problem is the nature of the unitary 
evolution.  This is specified by knowing appropriate information about the classical physics that 
supports the phenomena. This information is used to choose an appropriate Hamiltonian through which the 
unitary operator is constructed via a correspondence principle that replaces classical variables with appropriate quantum
operators. (In the path integral approach one needs a Langrangian to construct the action on which the path
integral is based.) One needs to know certain aspects of classical physics to 
solve any specific quantum problem.  
\bigbreak

A key concept in the quantum information viewpoint is the notion of the superposition of states.
If a quantum system has two  distinct states $|v \rangle$ and $|w \rangle,$ then it has infinitely many states of the form
$a|v \rangle + b|w \rangle$ where $a$ and $b$ are complex numbers taken up to a common multiple. States are ``really" 
in the projective space associated with $H.$ There is only one superposition of a single state $|v \rangle$ with 
itself. On the other hand, it is most convenient to regard the states $|v \rangle$ and $|w \rangle$ as vectors in a vector space.
We than take it as part of the procedure of dealing with states to normalize them to unit length. Once again, the superposition of a state with itself is
again itself.
\bigbreak

Dirac \cite{D} introduced the ``bra -(c)-ket" notation $\langle A\,|B \rangle = A^{\dagger}B$ for the inner product of complex vectors $A,B \in H$.
He also separated the parts of the bracket into the {\em bra} $<A\,|$ and the {\em ket} $|B \rangle.$ Thus

$$\langle A\,|B \rangle = \langle A\,|\,\,|B \rangle$$

\noindent In this interpretation,
the ket $|B \rangle$ is identified with the vector $B \in H$, while the bra $<A\,|$ is regarded as the element dual to $A$ in the 
dual space $H^*$. The dual element to $A$ corresponds to the conjugate transpose $A^{\dagger}$ of the vector $A$, and the inner product is 
expressed in conventional language by the matrix product $A^{\dagger}B$ (which is a scalar since $B$ is a column vector). Having separated the bra and the ket, Dirac can write the
``ket-bra"  $|A \rangle \langle B\,| = AB^{\dagger}.$ In conventional notation, the ket-bra is a matrix, not a scalar, and we have the following formula for
the  square of $P = |A \rangle \langle B\,|:$

$$P^{2} =  |A \rangle \langle B\,| |A \rangle \langle B\,| = A(B^{\dagger}A)B^{\dagger} = (B^{\dagger}A)AB^{\dagger} = \langle B\,|A \rangle P.$$

\noindent The standard example is a ket-bra $P = |A\,\rangle \langle A|$ where $\langle A\,|A \rangle =1$ so that $P^2 = P.$  Then $P$ is a projection
matrix,  projecting to the subspace of $H$ that is spanned by the vector $|A \rangle$. In fact, for any vector $|B \rangle$ we have 

$$P|B \rangle = |A \rangle \langle A\,|\,|B \rangle =  |A \rangle \langle A\,|B \rangle = \langle A\,|B \rangle |A \rangle .$$

\noindent If $\{|C_{1} \rangle, |C_{2} \rangle , \cdots |C_{n} \rangle \}$ is an orthonormal basis for $H$, and $$P_{i} = |C_{i} \,\rangle \langle C_{i}|,$$
\noindent then for any vector $|A \rangle $ we have

$$|A \rangle = \langle C_{1}\,|A \rangle |C_{1} \rangle + \cdots + \langle C_{n}\,|A \rangle |C_{n} \rangle .$$

\noindent Hence 

$$\langle B\,|A \rangle = \langle B\,|C_{1} \rangle \langle C_{1}\,|A \rangle + \cdots + \langle B\,|C_{n} \rangle \langle C_{n}\,|A \rangle $$

One wants the probability of starting in state $|A \rangle $ and ending in state $|B \rangle .$ The 
probability for this event is equal to $|\langle B\,|A \rangle |^{2}$. This can be refined if we have more knowledge. 
If the intermediate states $|C_{i} \rangle $ are a complete set of orthonormal alternatives then we
can assume that 
$\langle C_{i}\,|C_{i} \rangle  = 1$ for each $i$ and that $\Sigma_{i} |C_{i} \rangle \langle C_{i}| = 1.$  This identity now corresponds to the fact that
$1$ is the sum of the probabilities of an arbitrary state being projected into one of these intermediate states.
\bigbreak

If there are intermediate states between the intermediate states this formulation can be continued
until one is summing over all possible paths from $A$ to $B.$ This becomes the path integral expression 
for the amplitude $\langle B|A \rangle .$
\bigbreak

\subsection{What is a Quantum Computer?}

A {\it quantum computer} is, abstractly, a composition $U$ of unitary transformations, together with an initial state and a choice of measurement
basis. One runs the computer by repeatedly initializing it, and then measuring the result of applying the unitary transformation $U$ to the initial state.
The results of these measurements are then analyzed for the desired information that the computer was set to determine. The key to using the computer
is the design of the initial state and the design of the composition of unitary transformations. The reader should consult \cite{N} for more specific
examples of quantum algorithms. 
\bigbreak

Let $H$ be a given finite dimensional vector space over the complex numbers $C.$ Let 
$$\{ W_{0}, W_{1},..., W_{n} \}$$ be an
orthonormal basis for $H$ so that with $|i \rangle := |W_{i} \rangle $ denoting $W_{i}$ and $\langle i|$ denoting the conjugate transpose of $|i \rangle $,
we have
$$\langle i|j \rangle = \delta_{ij}$$
\noindent where $\delta_{ij}$ denotes the Kronecker delta (equal to one when its indices are equal to one another, and equal
to zero otherwise). Given a vector $v$ in $H$ let $|v|^{2} := \langle v|v \rangle .$ Note that $\langle i|v$ is the $i$-th coordinate of $v.$ 
\vspace{3mm}

\noindent An {\em measurement of $v$} returns one of the coordinates $|i \rangle $
of $v$ with probability $|\langle i|v|^{2}.$ This model of measurement is a simple instance of the situation with a quantum
mechanical system that is in a mixed state until it is observed. The result of observation is to put the system into one of
the basis states. 
\vspace{3mm}

When the dimension of the space $H$ is two ($n=1$), a vector in the space is called a {\em qubit}. A qubit represents one
quantum of binary information. On measurement, one obtains either the ket $|0 \rangle $ or the ket $|1 \rangle $. This constitutes the 
binary distinction that is inherent in a qubit.  Note however that the information obtained is probabilistic.  If the qubit is
$$| \psi \rangle = \alpha |0 \rangle + \beta \ |1 \rangle ,$$ \noindent then the ket $|0 \rangle $ is observed with probability $|\alpha|^{2}$, and the ket
$|1 \rangle $ is observed with probability $|\beta|^{2}.$  In speaking of an idealized quantum computer, we do not specify the nature
of measurement process beyond these probability postulates.
\vspace{3mm}
 
In the case of general dimension $n$ of the space $H$, we will call the vectors in $H$
{\em qunits}. It is quite common to use spaces $H$ that are tensor products of two-dimensional spaces (so that all computations 
are expressed in terms of qubits) but this is not necessary in principle. One can start with a given space, and later work out
factorizations into qubit transformations.
\vspace{3mm}

A {\em quantum computation} consists in the application of a unitary
transformation $U$ to an initial qunit $\psi = a_{0}|0 \rangle + ... + a_{n}|n \rangle $  with $|\psi|^{2}=1$, plus an
measurement of
$U\psi.$ A measurement of $U\psi$ returns the ket $|i \rangle $ with probability $|\langle i|U\psi|^{2}$. In particular, if we start the computer
in the state $|i \rangle $, then the probability that it will return the state $|j \rangle $ is $|\langle j|U|i \rangle |^{2}.$

\vspace{3mm} It is the necessity for writing a given computation in terms of unitary transformations, and the probabilistic
nature of the result that characterizes quantum computation. Such computation could be carried out by an idealized quantum
mechanical system. It is hoped that such systems can be physically realized. 
\vspace{3mm}

\section{Braiding Operators and Universal Quantum Gates}
A class of invariants of knots and links called quantum invariants can be constructed by using representations of 
the Artin braid group, and more specifically by using solutions to the Yang-Baxter equation \cite{BA}, 
first discovered in 
relation to $1+1$ dimensional quantum field theory, and $2$ dimensional statistical mechanics. 
Braiding operators feature in constructing representations of the Artin braid group, and in the construction of invariants of knots
and links.
\bigbreak

A key concept in the construction of quantum link invariants is
the association of a Yang-Baxter operator $R$ to each elementary crossing in a
link diagram. The operator $R$ is a linear mapping  
$$R\colon \ V\otimes V \longrightarrow V\otimes V$$ 
defined on the  $2$-fold tensor product of a vector space $V,$ generalizing the permutation of the factors
(i.e., generalizing a swap gate when $V$ represents one qubit). Such transformations are not 
necessarily unitary in 
topological applications. It is useful to understand
when they can be replaced by unitary transformations 
for the purpose of quantum 
computing. Such unitary $R$-matrices can be used to 
make unitary representations of the Artin braid group.
\bigbreak

\begin{figure}
     \begin{center}
     \begin{tabular}{c}
     \includegraphics[height=4cm]{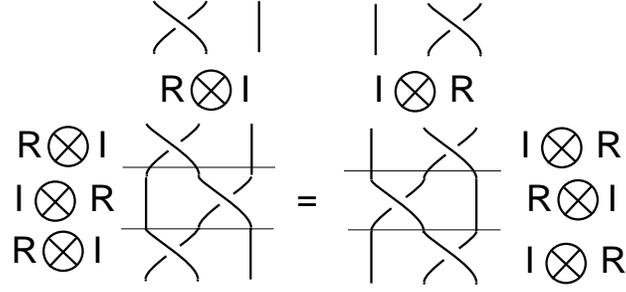}
     \end{tabular}
     \end{center}
     \caption{\bf The Yang-Baxter equation }
     \label{Figure 7 }
     \end{figure}

A solution to the Yang-Baxter equation, as described in the last 
paragraph is a matrix $R,$ regarded as a mapping of a
two-fold tensor product of a vector space
$V \otimes V$ to itself that satisfies the equation 

$$(R \otimes I)(I \otimes R)(R \otimes I) = 
(I \otimes R)(R \otimes I)(I \otimes R).$$ From the point of view of topology, the matrix $R$ 
is regarded as representing an elementary bit of braiding 
represented by one string
crossing over another. In Figure~\ref{Figure 7 } we have illustrated 
the braiding identity that corresponds to the Yang-Baxter equation.
Each braiding picture with its three input lines (below) 
and output lines (above) corresponds to a mapping of the three fold
tensor product of the vector space $V$ to itself, as required 
by the algebraic equation quoted above. The pattern of placement of the 
crossings in the diagram corresponds to the factors 
$R \otimes I$ and $I \otimes R.$ This crucial 
topological move has an algebraic
expression in terms of such a matrix $R.$ Our approach in this section to 
relate topology, quantum computing, and quantum entanglement is through the 
use of the Yang-Baxter equation. In order to accomplish this aim, 
{\em we need to study solutions of the Yang-Baxter equation that are unitary.}
Then the $R$ matrix can be seen {\em either} as a braiding matrix 
{\em or} as a quantum gate in a quantum computer.

     \bigbreak

The problem of finding solutions to the Yang-Baxter equation 
that are unitary turns out to be surprisingly difficult.
Dye \cite{Dye} has classified all such matrices of 
size $4 \times 4.$  A rough summary of her classification is that all $4 \times 4$ unitary 
solutions to the Yang-Baxter equation are similar to one of the following types of matrix:

\[R = \left( \begin{array}{cccc}
1/\sqrt{2} & 0 & 0  & 1/\sqrt{2}\\
0  & 1/\sqrt{2} & -1/\sqrt{2} & 0\\
0 & 1/\sqrt{2} & 1/\sqrt{2} & 0 \\
-1/\sqrt{2} & 0 & 0 & 1/\sqrt{2}\\
\end{array} \right)\] 
\bigbreak

\[R' = \left( \begin{array}{cccc}
a & 0 & 0  & 0\\
0  & 0 & b & 0\\
0 & c & 0 & 0 \\
0 & 0 & 0 & d\\
\end{array} \right)\]

\[R'' = \left( \begin{array}{cccc}
0 & 0 & 0  & a\\
0  & b & 0 & 0\\
0 & 0 & c & 0 \\
d & 0 & 0 & 0\\
\end{array} \right)\] where $a$,$b$,$c$,$d$ are unit complex numbers.

For the purpose of quantum computing, one should regard each matrix as acting on the stamdard basis
$\{ |00\rangle, |01\rangle, |10\rangle, |11\rangle \}$ of $H = V \otimes V,$ where $V$ is a two-dimensional complex vector
space. Then, for example we have
$$R|00\rangle =(1/\sqrt{2})|00\rangle-(1/\sqrt{2})|11\rangle,$$
$$R|01\rangle=(1/\sqrt{2})|01\rangle+(1/\sqrt{2})|10\rangle,$$
$$R|10\rangle=-(1/\sqrt{2})|01\rangle+(1/\sqrt{2})|10\rangle,$$
$$R|11\rangle=(1/\sqrt{2})|00\rangle+(1/\sqrt{2})|11\rangle.$$
The reader should note that $R$ is the familiar change-of-basis matrix from the standard basis to the Bell basis of entangled states.
\bigbreak

\noindent In the case of $R',$ we have
$$R'|00\rangle = a|00\rangle, R'|01\rangle= c|10\rangle,$$ 
$$R'|10\rangle = b|01\rangle, R'|11\rangle = d|11\rangle.$$ Note that $R'$ can be regarded as a diagonal
phase gate $P$, composed with a swap gate $S.$

\[P = \left( \begin{array}{cccc}
a & 0 & 0  & 0\\
0  & b & 0 & 0\\
0 & 0 & c & 0 \\
0 & 0 & 0 & d\\
\end{array} \right)\]

\[S = \left( \begin{array}{cccc}
1 & 0 & 0  & 0\\
0  & 0 & 1 & 0\\
0 & 1 & 0 & 0 \\
0 & 0 & 0 & 1\\
\end{array} \right)\] Compositions of solutions of the (Braiding) Yang-Baxter equation with the swap gate $S$ are
called {\em solutions to the algebraic Yang-Baxter equation}. Thus the diagonal matrix $P$ is a solution to
the algebraic Yang-Baxter equation. 
\bigbreak

\noindent{\bf Remark.} Another avenue related to unitary solutions to the Yang-Baxter equation as quantum gates
comes from using extra physical parameters in this equation (the rapidity parameter) that are related to statistical physics.
In \cite{Yong} we discovered that solutions to the Yang-Baxter equation with the rapidity parameter allow many new unitary solutions.
The significance of these gates for quatnum computing is still under investigation.
\bigbreak

 \subsection{Universal Gates}
A {\em two-qubit gate} $G$ is a unitary  linear mapping $G:V \otimes V \longrightarrow V$ where $V$ is a two complex dimensional
vector space. We say that the gate $G$ is {\em universal for quantum computation} (or just {\em universal}) if $G$ together with 
local unitary transformations (unitary transformations from $V$ to $V$) generates all unitary transformations of the complex vector
space of dimension $2^{n}$ to itself. It is well-known \cite{N} that $CNOT$ is a universal gate. (On the standard basis,
$CNOT$ is the identity when the first qubit is $|0 \rangle $, and it flips the second qbit, leaving the first alone, when the first qubit is $|1 \rangle .$)
\bigbreak

\noindent A gate $G$, as above, is said to be {\em entangling} if there is a vector  
$$| \alpha \beta \rangle = | \alpha \rangle \otimes | \beta \rangle \in V \otimes V$$ such that 
$G | \alpha \beta \rangle$ is not decomposable as a tensor product of two qubits. Under these circumstances, one says that 
$G | \alpha \beta \rangle$ is {\em entangled}.
\bigbreak

\noindent In \cite{BB},  the Brylinskis
give a general criterion of $G$ to be universal. They prove that {\em a two-qubit gate $G$ is universal if and only if it is
entangling.} 
\bigbreak

\noindent {\bf Remark.} A two-qubit pure state $$|\phi \rangle = a|00 \rangle + b|01 \rangle + c|10 \rangle + d|11 \rangle$$
is entangled exactly when $(ad-bc) \ne 0.$ It is easy to use this fact to check when a specific matrix is, or is not, entangling.
\bigbreak

\noindent {\bf Remark.} There are many gates other than $CNOT$ that can be used as universal gates in the presence of local unitary 
transformations. Some of these are themselves topological (unitary solutions to the Yang-Baxter equation, see \cite{BG}) and themselves generate
representations of the Artin braid group. Replacing $CNOT$ by a solution to the Yang-Baxter equation does not place the local unitary transformations as
part of the corresponding representation of the braid group. Thus such substitutions give only a partial solution to creating topological 
quantum computation. In this paper we are concerned with braid group representations that include all aspects of the unitary
group. Accordingly, in the next section we shall first examine how the braid group on three strands can be represented as local unitary
transformations.
\bigbreak

\noindent {\bf Theorem.} Let $D$ denote the phase gate shown below. $D$ is a solution to the algebraic Yang-Baxter equation
(see the earlier discussion in this section). Then $D$ is a universal gate.

\[D = \left( \begin{array}{cccc}
1 & 0 & 0  & 0\\
0 & 1 & 0 & 0\\
0 & 0 & 1 & 0 \\
0 & 0 & 0& -1\\
\end{array} \right)\] 

\noindent {\bf Proof.} It follows at once from the Brylinski Theorem that $D$ is universal. For a more specific proof,
note that $CNOT = QDQ^{-1},$ where
$Q=H\otimes I$, $H$ is the
$2 \times 2$ Hadamard matrix. The conclusion
then follows at once from this identity and the discussion above. 
We illustrate the matrices involved in this
proof below:
\bigbreak

\[H = (1/\sqrt{2})\left( \begin{array}{cc}
1 & 1 \\
1  & -1 \\
\end{array} \right)\]

\[Q = (1/\sqrt{2})\left( \begin{array}{cccc}
1 & 1 & 0  & 0\\
1  & -1 & 0 & 0\\
0 & 0 & 1 & 1 \\
0 & 0 & 1 & -1\\
\end{array} \right)\]

\[D = \left( \begin{array}{cccc}
1 & 0 & 0  & 0\\
0 & 1 & 0 & 0\\
0 & 0 & 1 & 0 \\
0 & 0 & 0& -1\\
\end{array} \right)\]

\[QDQ^{-1} = QDQ = \left( \begin{array}{cccc}
1 & 0 & 0  & 0\\
0 & 1 & 0 & 0\\
0 & 0 & 0 & 1 \\
0 & 0 & 1 & 0\\
\end{array} \right) = CNOT\]  This completes the proof of the Theorem.  //
\bigbreak

\noindent {\bf Remark.} We thank Martin Roetteles \cite{Martin} for pointing out the specific factorization of $CNOT$ used in this
proof. \smallbreak

\noindent {\bf Theorem.} The matrix solutions $R'$ and $R''$ to the Yang-Baxter equation, described above, are universal gates
exactly when $ad-bc \ne 0$ for their internal parameters $a,b,c,d.$ In particular, 
let $R_{0}$ denote the solution $R'$ (above) to the Yang-Baxter equation with 
$a=b=c=1, d=-1.$

\[R' = \left( \begin{array}{cccc}
a & 0 & 0  & 0\\
0  & 0 & b & 0\\
0 & c & 0 & 0 \\
0 & 0 & 0 & d\\
\end{array} \right)\]

\[R_0 = \left( \begin{array}{cccc}
1 & 0 & 0  & 0\\
0  & 0 & 1 & 0\\
0 & 1 & 0 & 0 \\
0 & 0 & 0 & -1\\
\end{array} \right)\]
Then $R_{0}$ is a universal gate.
\bigbreak 

\noindent {\bf Proof.} The first part follows at once from the Brylinski Theorem. 
In fact, letting $H$ be the Hadamard matrix as before, and 
\[\sigma = \left( \begin{array}{cc}
1/\sqrt{2} & i/\sqrt{2} \\
i/\sqrt{2}  & 1/\sqrt{2} \\
\end{array} \right), \, \lambda = \left( \begin{array}{cc}
1/\sqrt{2} & 1/\sqrt{2} \\
i/\sqrt{2}  & -i/\sqrt{2} \\
\end{array} \right)\]
\[\mu = \left( \begin{array}{cc}
(1-i)/2 & (1+i)/2 \\
(1-i)/2  &(-1-i)/2 \\
\end{array} \right).\]
Then $$CNOT = (\lambda \otimes \mu)(R_{0}(I \otimes \sigma)R_{0})(H \otimes H).$$
This gives an 
explicit expression for $CNOT$ in terms of $R_{0}$ and local unitary transformations
(for which we thank Ben Reichardt).  //
\bigbreak

\noindent {\bf Remark.} Let $SWAP$ denote the Yang-Baxter Solution $R'$ with $a=b=c=d=1.$

\[SWAP = \left( \begin{array}{cccc}
1 & 0 & 0  & 0\\
0  & 0 & 1 & 0\\
0 & 1 & 0 & 0 \\
0 & 0 & 0 & 1\\
\end{array} \right)\]
$SWAP$ is the standard swap gate. Note that $SWAP$ is not a universal gate. This also follows from the Brylinski Theorem, 
since $SWAP$ is not entangling. Note also that $R_{0}$ is the composition of the phase gate $D$ with this swap gate.
\bigbreak

\noindent {\bf Theorem.} Let
\[R = \left( \begin{array}{cccc}
1/\sqrt{2} & 0 & 0  & 1/\sqrt{2}\\
0  & 1/\sqrt{2} & -1/\sqrt{2} & 0\\
0 & 1/\sqrt{2} & 1/\sqrt{2} & 0 \\
-1/\sqrt{2} & 0 & 0 & 1/\sqrt{2}\\
\end{array} \right)\] be the unitary solution to the Yang-Baxter equation discussed above. 
Then $R$ is a universal gate. The proof below gives a specific expression for $CNOT$ in terms of $R.$
\bigbreak

\noindent {\bf Proof.} This result follows at once from the Brylinksi Theorem, since $R$ is highly entangling.
For a direct computational proof, it suffices to show that $CNOT$ can be generated from $R$ and local unitary transformations.
Let
\[\alpha = \left( \begin{array}{cc}
1/\sqrt{2} & 1/\sqrt{2}\\
1/\sqrt{2} & -1/\sqrt{2}\\
\end{array} \right)\]

\[\beta = \left( \begin{array}{cc}
-1/\sqrt{2} & 1/\sqrt{2}\\
i/\sqrt{2} & i/\sqrt{2}\\
\end{array} \right)\]

\[\gamma = \left( \begin{array}{cc}
1/\sqrt{2} & i/\sqrt{2}\\
1/\sqrt{2} & -i/\sqrt{2}\\
\end{array} \right)\]

\[\delta = \left( \begin{array}{cc}
-1 & 0\\
0 & -i\\
\end{array} \right)\]  Let $M= \alpha \otimes \beta$ and $N= \gamma \otimes \delta.$ Then it is straightforward to verify that
$$CNOT = MRN.$$ This completes the proof. //
\bigbreak

\noindent {\bf Remark.}  
See \cite{BG} for more information about these calculations.
\bigbreak

\subsection{\bf Majorana Fermions Generate Universal Braiding Gates}
Recall that in Section  4 we showed how to construct braid group representations by using Majorana fermions in the special case of three particles. Here we generalize this construction and show how
the Marjorana fermions give rise to universal topological gates. Let $c_{1}, c_{2}, \cdots c_{n}$ denote
$n$ Majorana fermion creation operators. Thus we assume that $$c_{k}^2 = 1$$ and 
$$c_{i}c_{j} = - c_{j}c_{i}$$ for each $k = 1 \cdots n$ and whenever $i \ne j.$ Then define operators
$$s_{k} = \frac{1}{\sqrt{2}}(1 + c_{k+1} c_{k})$$ for $k = 1 \cdots n-1.$ Then by the same algebra as we
explored in Section 4 it is easy to verify that 
$s_{k+1}s_{k}s_{k+1} = s_{k}s_{k+1}s_{k}$ and that $s_{i}s_{j} = s_{j}s_{i}$ whenever $|i - j| > 1.$
Thus the $s_{i}$ give a representation of the $n$-strand  braid group $B_{n}.$ Furthermore, it is easy to see that a specific representation is given on the complex vector space $V_{n}$ with basis
$\{ c_{1}, c_{2}, \cdots c_{n} \}$ via the linear transformations $T_{k}: V_{n} \longrightarrow V_{n}$
defined by $$T_{k}(v) = s_{k} v s_{k}^{-1}.$$ Note that $ s_{k}^{-1} = \frac{1}{\sqrt{2}}(1 - c_{k+1} c_{k}).$ It is then easy to verify that $$T_{k}(c_{k}) = c_{k+1},$$ $$T_{k}(c_{k+1}) = - c_{k}$$ and that 
$T_{k}$ is the identity otherwise.
\bigbreak

For universality, take $n = 4$ and regard each $T_{k}$ as operating on $V \otimes V$ where $V$ is a single qubit space.  Then the braiding operators $T_{k}$ each satisfy the Yang-Baxter equation and so we have universal gates (in the presence of single qubit unitary operators) from Majorana fermions. If experimental work shows that Majorana fermions can be detected and controlled, then it is possible that quantum computers based on these topological unitary representations will be constructed.
\bigbreak

In the later sections of this paper we will describe the Fibonacci model, which also uses Majorana fermions, and a different subtler representation of the braid groups that is also promising for topological quantum computing.
\bigbreak

\section{A Remark about $EPR,$ Engtanglement and Bell's Inequality}
A state $|\psi\rangle \in H^{\otimes n}$, where $H$ is the qubit space, is said to be {\it entangled} if it cannot be 
written as a tensor product of vectors from non-trivial factors of $H^{\otimes n}.$ Such states turn out to be related to 
subtle nonlocality in quantum physics. It helps to
place this algebraic structure in the context of a gedanken experiment to see
where the physics comes in. Thought experiments of the sort we are about to describe were first devised by Einstein, Podolosky and 
Rosen, referred henceforth as $EPR.$
\bigbreak

\noindent Consider the entangled state $$S = (|0\rangle|1\rangle + |1\rangle|0\rangle)/\sqrt{2}.$$

\noindent
In an EPR thought experiment, we think of two ``parts" of this state that are
separated in space.  We want a notation for these parts and suggest the following:

$$L = (\{|0\rangle\}|1\rangle + \{|1\rangle\}|0\rangle)/\sqrt{2},$$

$$R = (|0\rangle\{|1\rangle\} + |1\rangle\{|0\rangle\})/\sqrt{2}.$$

\noindent In the left
state  $L$, an observer can only observe the left hand factor. In the right state $R$,
an observer can only observe the right hand factor.
These ``states"  $L$ and $R$ together comprise the EPR state $S,$ but they are
accessible individually just as are the two photons in the usual thought
experiement.  One can transport $L$ and $R$ individually and we shall write

$$S = L*R$$

\noindent to denote that they are the ``parts"  (but not tensor factors) of $S.$
\bigbreak

The curious thing about this formalism is that it includes a little bit of
macroscopic physics implicitly, and so it makes it a bit more apparent what EPR
were concerned about.  After all, lots of things that we can do to $L$ or $R$ do not
affect $S.$ For example, transporting $L$ from one place to another, as in the
original experiment where the photons separate.  On the other hand, if Alice has
$L$ and Bob has $R$ and Alice performs a local unitary transformation on ``her" tensor
factor, this applies to both $L$ and $R$ since the transformation is actually being
applied to the state $S.$ This is also a ``spooky action at a distance" whose
consequence does not appear until a measurement is made.
\bigbreak

To go a bit deeper it is worthwhile seeing what entanglement, in the sense of tensor indecomposability,
has to do with the structure of the $EPR$ thought experiment. To this end, we look at the structure of
the Bell inequalities using the Clauser, Horne, Shimony, Holt formalism ($CHSH$) as explained in the book by Nielsen and Chuang \cite{N}. For this
we use the  following observables with eigenvalues $\pm 1.$
$$Q =  \left(
\begin{array}{cc}
       1 & 0\\
       0 & -1
\end{array}\right)_1, 
$$
$$R =  \left(
\begin{array}{cc}
       0 & 1\\
       1 & 0
\end{array}\right)_1, 
$$
$$S =  \left(
\begin{array}{cc}
       -1 & -1\\
       -1 & 1
\end{array}\right)_2/\sqrt{2}, 
$$
$$T =  \left(
\begin{array}{cc}
       1 & -1\\
       -1 & -1
\end{array}\right)_2/\sqrt{2}. 
$$ 

\noindent The subscripts $1$ and $2$ on these matrices indicate that they are to operate on the first and 
second tensor factors, repsectively, of a quantum state of the form
$$\phi = a|00\rangle + b|01\rangle + c|10\rangle + d|11\rangle.$$ To simplify the results of this calculation we shall here assume that 
the coefficients $a,b,c,d$ are real numbers. We calculate the quantity 
$$\Delta = \langle\phi|QS|\phi\rangle + \langle\phi|RS|\phi\rangle + \langle\phi|RT|\phi\rangle - \langle\phi|QT|\phi\rangle,$$ finding that
$$\Delta = (2 - 4(a+d)^2 + 4(ad -bc))/\sqrt{2}.$$ Classical probability calculation with random variables of value
$\pm 1$ gives the value of $QS + RS + RT - QT = \pm 2$ (with each of $Q$, $R$, $S$ and $T$ equal to $\pm 1$).
Hence the classical expectation satisfies the Bell inequality
$$E(QS) + E(RS) + E(RT) - E(QT) \le 2.$$ That quantum expectation is not classical is embodied in the fact that
$\Delta$ can be greater than $2.$ The classic case is that of the Bell state
$$\phi = (|01\rangle - |10\rangle)/\sqrt{2}.$$ Here $$\Delta =  6/\sqrt{2} > 2.$$ In general we see that the following inequality is 
needed in order to violate the Bell inequality
$$(2 - 4(a+d)^2 + 4(ad -bc))/\sqrt{2} > 2.$$ This is equivalent to 
$$(\sqrt{2}-1)/2 < (ad-bc) - (a+d)^2.$$ Since we know that $\phi$ is entangled exactly when $ad-bc$ is non-zero, this
shows that an unentangled state cannot violate the Bell inequality. This formula {\em also} shows that it is possible
for a state to be entangled and yet not violate the Bell inequality. For example, if
$$\phi = (|00\rangle - |01\rangle + |10\rangle + |11\rangle)/2,$$ then $\Delta(\phi)$ satisfies Bell's inequality, but $\phi$ is an entangled 
state. We see from this calculation that entanglement in the sense of tensor indecomposability, and entanglement in the
sense of Bell inequality violation for a given choice of Bell operators are not equivalent concepts. On the other hand,
Benjamin Schumacher has pointed out \cite{S} that any entangled two-qubit state will violate Bell inequalities for an
appropriate choice of operators. This
deepens the context for our question of the relationship between topological entanglement and quantum entanglement. The
Bell inequality violation is an indication of quantum mechanical entanglement. One's intuition suggests that it is
{\em this} sort of entanglement that should have a topological context.  
\bigbreak

\section{The Aravind Hypothesis}
Link diagrams can be used as graphical devices 
and holders of information. In this vein 
Aravind \cite{Ara} proposed that
the entanglement of a link should correspond to 
the entanglement of a state. {\em Measurement of a link 
would be  modeled by deleting one
component of the link.} 
A key example is the 
Borromean rings. See Figure~\ref{boro}.
Deleting any component of the Boromean rings 
yields a remaining pair of unlinked rings. The Borromean 
rings are entangled, but any two of them are unentangled.
In this sense the Borromean rings are analogous to 
the $GHZ$ state $|GHZ\rangle  = (1/\sqrt{2})(|000\rangle  + |111\rangle )$.
Measurement in any factor of the $GHZ$ yields 
an unentangled state. Aravind points out that 
this property is basis dependent. {\em We point out 
that there are states whose entanglement 
after an measurement is a matter of probability (via quantum amplitudes).}
Consider for example the state 

{\bf \[
|\psi \rangle=|001\rangle  + |010\rangle  + |100\rangle .
\]} 

\noindent Measurement in any coordinate yields probabilistically an entangled or an
unentangled state. For example

{\bf \[
|\psi \rangle= |0\rangle(|01\rangle  + |10\rangle)  + |1\rangle |00\rangle .
\] }

\noindent so that projecting to $|1\rangle$ in the first coordinate yields an unentangled state, while projecting to $|0\rangle$ yields an entangled state.
\bigbreak
 
New ways to use link diagrams must be invented to map the properties 
of such states.  One direction is to consider appropriate notions of quantum knots so that one can formlate
superpositions of topological types as in \cite{QK1}. But one needs to go deeper in this consideration.
The relationship of topology and physics needs to be examined carefully.
We take the stance that topological properties of systems are properties
that remain invariant under certain transformations that are identified as ``topological
equivalences". In making quantum physical models, these equivalences should correspond to unitary 
transformations of an appropriate Hilbert space. Accordingly, we have formulated a model for 
{\it quantum knots} \cite{QK2,QK3,QK4,QK5} that meets these requirements.
A quantum knot system represents the \textquotedblleft quantum embodiment"
of a closed knotted physical piece of rope. \ A quantum knot (i.e., an
element $\left\vert K\right\rangle $ lying in an appropriate Hilbert space $H_{n}$, 
as a state of this system, represents the state of such a knotted
closed piece of rope, i.e., the particular spatial configuration of the knot
tied in the rope. Associated with a quantum knot system is a group of
unitary transformations $A_{n}$, called the {\it ambient group},
which represents all possible ways of moving the rope around (without
cutting the rope, and without letting the rope pass through itself.)  Of
course, unlike a classical closed piece of rope, a quantum knot can exhibit
non-classical behavior, such as quantum superposition and quantum
entanglement.  
The {\it knot type} of a quantum knot $\left\vert K\right\rangle $ is
simply the orbit of the quantum knot under the action of the ambient group 
$A_{n}$. This leads to new questions connecting quantum
computing and knot theory. 
\bigbreak

\section{$SU(2)$ Representations of the Artin Braid Group}
The purpose of this section is to determine all the representations of the three strand Artin braid group $B_{3}$ to the special unitary group $SU(2)$ and
concomitantly to the unitary group $U(2).$ One regards the groups $SU(2)$ and $U(2)$ as acting on a single qubit, and so $U(2)$ is usually regarded as the
group of local unitary transformations in a quantum information setting. If one is looking for a coherent way to represent all unitary transformations by
way of braids, then $U(2)$ is the place to start. Here we will show that there are many representations of the three-strand braid group
that generate a dense subset of $U(2).$ Thus it is a fact that local unitary transformations can be "generated by braids" in many ways.
\bigbreak

We begin with the structure of $SU(2).$ A matrix in $SU(2)$ has the form 
$$ M = 
\left( \begin{array}{cc}
z & w \\
-\bar{w} & \bar{z} \\
\end{array} \right),$$ where $z$ and $w$ are complex numbers, and $\bar{z}$ denotes the complex conjugate of $z.$ 
To be in $SU(2)$ it is required that $Det(M)=1$ and that $M^{\dagger} = M^{-1}$ where $Det$ denotes determinant, and $M^{\dagger}$ is the conjugate transpose of $M.$
Thus if
$z = a + bi$ and $w = c + di$ where $a,b,c,d$ are real numbers, and $i^2 = -1,$ then 
$$ M = 
\left( \begin{array}{cc}
a + bi & c + di \\
-c + di & a - bi \\
\end{array} \right)$$  with $a^2 + b^2 + c^2 + d^2 = 1.$ It is convenient to write
$$M =
a\left( \begin{array}{cc}
1 & 0 \\
0 & 1 \\
\end{array} \right) +
b\left( \begin{array}{cc}
i & 0\\
0 & -i \\
\end{array} \right) +
c\left( \begin{array}{cc}
0  & 1 \\
-1  & 0\\
\end{array} \right) +
d\left( \begin{array}{cc}
0 & i \\
i & 0 \\
\end{array} \right),$$ and to abbreviate this decomposition as
$$M = a + bI +cJ + dK$$
where 
$$ 1 \equiv
\left( \begin{array}{cc}
1 & 0 \\
0 & 1 \\
\end{array} \right),
I \equiv
\left( \begin{array}{cc}
i & 0\\
0 & -i \\
\end{array} \right),
J \equiv
\left( \begin{array}{cc}
0  & 1 \\
-1  & 0\\
\end{array} \right),
K \equiv
\left( \begin{array}{cc}
0 & i \\
i & 0 \\
\end{array} \right)$$ so that 
$$I^2 = J^2 = K^2 = IJK = -1$$ and 
$$IJ = K, JK=I, KI = J$$
$$JI = -K, KJ = -I, IK = -J.$$
The algebra of $1,I,J,K$ is called the {\it quaternions} after William Rowan Hamilton who discovered this algebra prior to the discovery of 
matrix algebra. Thus the unit quaternions are identified with $SU(2)$ in this way. We shall use this identification, and some facts about 
the quaternions to find the $SU(2)$ representations of braiding. First we recall some facts about the quaternions.

\begin{enumerate}
\item Note that if $q = a + bI +cJ + dK$ (as above), then $q^{\dagger} = a - bI - cJ - dK$ so that $qq^{\dagger} = a^2 + b^2 + c^2 + d^2 = 1.$
\item A general quaternion has the form $ q = a + bI + cJ + dK$ where the value of $qq^{\dagger} = a^2 + b^2 + c^2 + d^2,$ is not fixed to unity.
The {\it length} of $q$ is by definition $\sqrt{qq^{\dagger}}.$
\item A quaternion of the form $rI + sJ + tK$ for real numbers $r,s,t$ is said to be a {\it pure} quaternion. We identify the set of pure
quaternions with the vector space of triples $(r,s,t)$ of real numbers $R^{3}.$
\item Thus a general quaternion has the form $q = a + bu$ where $u$ is a pure quaternion of unit length and $a$ and $b$ are arbitrary real numbers.
A unit quaternion (element of $SU(2)$) has the addition property that $a^2 + b^2 = 1.$
\item If $u$ is a pure unit length quaternion, then $u^2 = -1.$ Note that the set of pure unit quaternions forms the two-dimensional sphere
$S^{2} = \{ (r,s,t) | r^2 + s^2 + t^2 = 1 \}$ in $R^{3}.$
\item If $u, v$ are pure quaternions, then $$uv = -u \cdot v + u \times v$$ whre $u \cdot v$ is the dot product of the vectors $u$ and $v,$ and 
$u \times v$ is the vector cross product of $u$ and $v.$ In fact, one can take the definition of quaternion multiplication as
$$(a + bu)(c + dv) = ac + bc(u) + ad(v) + bd(-u \cdot v + u \times v),$$ and all the above properties are consequences of this
definition. Note that quaternion multiplication is associative.
\item Let $g = a + bu$ be a unit length quaternion so that $u^2 = -1$ and $a = cos(\theta/2), b=sin(\theta/2)$ for a chosen angle $\theta.$
Define $\phi_{g}:R^{3} \longrightarrow R^{3}$ by the equation $\phi_{g}(P) = gPg^{\dagger},$ for $P$ any point in $R^{3},$ regarded as a pure quaternion.
Then $\phi_{g}$ is an orientation preserving rotation of $R^{3}$ (hence an element of the rotation group $SO(3)$). Specifically, $\phi_{g}$ is a rotation
about the  axis $u$ by the angle $\theta.$ The mapping $$\phi:SU(2) \longrightarrow SO(3)$$ is a two-to-one surjective map from the special unitary group to
the rotation group. In quaternionic form, this result was proved by Hamilton and by Rodrigues in the middle of the nineteeth century.
The specific formula for $\phi_{g}(P)$ as shown below:
$$\phi_{g}(P) = gPg^{-1} = (a^2 - b^2)P + 2ab (P \times u) + 2(P \cdot u)b^{2}u.$$
\end{enumerate}

We want a representation of the three-strand braid group in $SU(2).$ This means that we want a homomorphism $\rho: B_{3} \longrightarrow SU(2),$ and hence
we want elements $g = \rho(s_{1})$ and $h= \rho(s_{2})$ in $SU(2)$ representing the braid group generators $s_{1}$ and $s_{2}.$ Since $s_{1}s_{2}s_{1} =
s_{2}s_{1}s_{2}$ is the generating relation for $B_{3},$ the only requirement on $g$ and $h$ is that $ghg = hgh.$ We rewrite this relation as
$h^{-1}gh = ghg^{-1},$ and analyze its meaning in the unit quaternions.
\bigbreak

Suppose that $g = a + bu$ and $h=c + dv$ where $u$ and $v$ are unit pure quaternions so that $a^2 + b^2 = 1$ and $c^2 + d^2 = 1.$
then $ghg^{-1} = c +d\phi_{g}(v)$ and $h^{-1}gh = a + b\phi_{h^{-1}}(u).$ Thus it follows from the braiding relation that 
$a=c,$ $b= \pm d,$ and that $\phi_{g}(v) = \pm \phi_{h^{-1}}(u).$  However, in the case where there is a minus sign we have
$g = a + bu$ and $h = a - bv = a + b(-v).$ Thus we can now prove the following Theorem. 
\bigbreak

\noindent {\bf Theorem.} Let $u$ and $v$ be pure unit quaternions and $g = a + bu$ and $h=c + dv$ have unit length. Then (without loss of generality), the braid relation $ghg=hgh$ is true if
and only if
$h = a + bv,$ and $\phi_{g}(v) = \phi_{h^{-1}}(u).$ Furthermore, given that $g = a +bu$ and $h = a +bv,$ the condition $\phi_{g}(v) = \phi_{h^{-1}}(u)$
is satisfied if and only if $u \cdot v = \frac{a^2 - b^2}{2 b^2}$ when $u \ne v.$ If $u = v$ then $g = h$ and the braid relation is trivially
satisfied.
\bigbreak

\noindent {\bf Proof.} We have proved the first sentence of the Theorem in the discussion prior to its statement. Therefore assume that
$g = a +bu, h = a +bv,$ and $\phi_{g}(v) = \phi_{h^{-1}}(u).$ 
We have already stated the formula for $\phi_{g}(v)$ in the discussion about quaternions:
$$\phi_{g}(v) = gvg^{-1} = (a^2 - b^2)v + 2ab (v \times u) + 2(v \cdot u)b^{2}u.$$ By the same token, we have
$$\phi_{h^{-1}}(u) = h^{-1}uh = (a^2 - b^2)u + 2ab (u \times -v) + 2(u \cdot (-v))b^{2}(-v)$$
$$= (a^2 - b^2)u + 2ab (v \times u) + 2(v \cdot u)b^{2}(v).$$ Hence we require that
$$(a^2 - b^2)v + 2(v \cdot u)b^{2}u = (a^2 - b^2)u + 2(v \cdot u)b^{2}(v).$$ This equation is equivalent to
$$2(u \cdot v)b^{2} (u - v) = (a^2 - b^2)(u - v).$$
If $u \ne v,$ then this implies that $$u \cdot v = \frac{a^2 - b^2}{2 b^2}.$$
This completes the proof of the Theorem. //
\bigbreak

\noindent{\bf The Majorana Fermion Example.} Note the case of the theorem where
$$g = a +bu, h = a +bv.$$ Suppose that $u \cdot v = 0.$ Then the theorem tells us that we need 
$a^2 - b^2 = 0$ and since $a^2 +b^2 = 1,$ we conclude that $a = 1/\sqrt{2}$ and $b$ likewise.
For definiteness, then we have for the braiding generators (since $I$, $J$ and $K$ are mutually orthogonal) the three operators
$$A = \frac{1}{\sqrt{2}}(1 + I),$$
$$B =\frac{1}{\sqrt{2}}(1 + J),$$
$$C = \frac{1}{\sqrt{2}}(1 + K).$$
Each pair satisfies the braiding relation so that $ABA = BAB, BCB = CBC, ACA =CAC.$ We have already met this braiding triplet in our discussion of the construction of braiding operators from Majorana fermions in  Section 4. This shows (again) how close Hamilton's quaternions are to topology and how braiding is fundamental to the structure of fermionic physics.
\bigbreak

\noindent{\bf The Fibonacci  Example.} Let
$$g = e^{I\theta} = a + bI$$ where $a = cos(\theta)$ and $b = sin(\theta).$
Let $$h = a + b[(c^2 - s^2)I + 2csK]$$ where $c^2 + s^2 = 1$ and $c^2 - s^2 = \frac{a^2 - b^2}{2b^2}.$ Then we can rewrite $g$ and $h$ in matrix form
as the matrices $G$ and $H.$ Instead of writing the explicit form of $H,$ we write $H = FGF^{\dagger}$ where $F$ is an element of $SU(2)$ as shown below.

$$G =
\left( \begin{array}{cc}
e^{i\theta} & 0 \\
0 & e^{-i\theta} \\
\end{array} \right)$$

$$F =
\left( \begin{array}{cc}
ic & is \\
is & -ic \\
\end{array} \right)$$
This representation of braiding where one generator $G$ is a simple matrix of phases, while the other generator $H = FGF^{\dagger}$ is derived from $G$ by
conjugation by a unitary matrix, has the possibility for generalization to representations of braid groups (on greater than three strands) to $SU(n)$ or
$U(n)$ for 
$n$ greater than $2.$ In fact we shall see just such representations constructed later in this paper, by using a version of topological quantum field theory.
The simplest example is given by 
$$g = e^{7 \pi I/10}$$
$$f = I\tau  + K \sqrt{\tau}$$
$$h = f g f^{-1}$$
where $\tau^{2} + \tau = 1.$
Then $g$ and $h$ satisfy $ghg=hgh$ and generate a representation of the three-strand braid group that is dense in $SU(2).$ We shall call this the 
{\it Fibonacci} representation of $B_{3}$ to $SU(2).$
\bigbreak

\noindent {\bf Density.} Consider representations of $B_{3}$ into $SU(2)$ produced by the method of this section. That is consider the subgroup $SU[G,H]$ of 
$SU(2)$ generated by a pair of elements $\{g,h \}$ such that $ghg=hgh.$ We wish to understand when such a representation will be dense in $SU(2).$
We need the following lemma.
\bigbreak

\noindent {\bf Lemma.} $e^{aI} e^{bJ} e^{cI} = cos(b) e^{I(a +c)} + sin(b) e^{I(a-c)} J.$ Hence any element of $SU(2)$ can be written in the form
$e^{aI} e^{bJ} e^{cI}$ for appropriate choices of angles $a,b,c.$ In fact, if $u$ and $v$ are linearly independent unit vectors in $R^{3},$ then
any element of $SU(2)$ can be written in the form $$e^{au} e^{bv} e^{cu}$$ for appropriate choices of the real numbers $a,b,c.$
\bigbreak

\noindent {\bf Proof.} See \cite{AnyonicTop} for the details of this proof.  //
\bigbreak

This Lemma can be used to verify the density of a representation, by finding two elements $A$ and $B$ in the representation such that 
the powers of $A$ are dense in the rotations about its axis, and the powers of $B$ are dense in the rotations about its axis, and such that the 
axes of $A$ and $B$ are linearly independent in $R^{3}.$ Then by the Lemma the set of elements $A^{a+c}B^{b}A^{a-c}$ are dense in $SU(2).$ It follows
for example, that the Fibonacci representation described above is dense in $SU(2),$ and indeed the generic representation of $B_{3}$ into
$SU(2)$ will be dense in $SU(2).$  Our next task is to describe representations of the higher braid groups that will extend some of these unitary
representations of the three-strand braid group. For this we need more topology.
\bigbreak

\section{The Bracket Polynomial and the Jones Polynomial}
We now discuss the Jones polynomial. We shall construct the Jones polynomial by using the bracket state 
summation model \cite{KA87}. The bracket polynomial, invariant under Reidmeister moves II and III, can be normalized to give an invariant of all three
Reidemeister moves. This normalized invariant, with a change of variable, is the Jones polynomial
\cite{JO1,JO2}. The Jones polynomial was originally discovered by a different method than the one given here. 
\bigbreak 

The {\em bracket polynomial} , $<K> \, = \, <K>(A)$,  assigns to each unoriented link diagram $K$ a 
Laurent polynomial in the variable $A$, such that
   
\begin{figure}
     \begin{center}
     \begin{tabular}{c}
     \includegraphics[height=6cm]{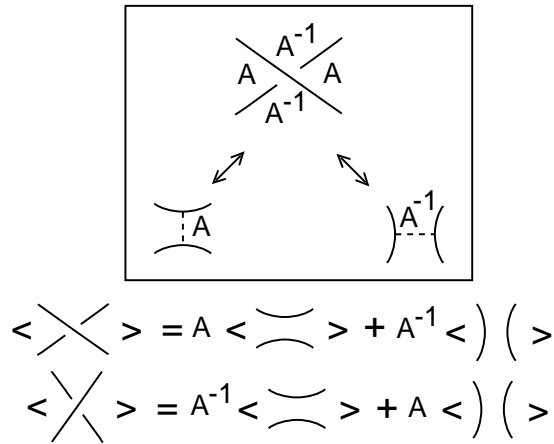}
     \end{tabular}
     \end{center}
     \caption{\bf Bracket Smoothings   }
     \label{Figure 9 }
     \end{figure}

\begin{enumerate}
\item If $K$ and $K'$ are regularly isotopic diagrams, then  $<K> \, = \, <K'>$.
  
\item If  $K \sqcup O$  denotes the disjoint union of $K$ with an extra unknotted and unlinked 
component $O$ (also called `loop' or `simple closed curve' or `Jordan curve'), then 

$$< K \sqcup O> \, = \delta<K>,$$ 
where  $$\delta = -A^{2} - A^{-2}.$$
  
\item $<K>$ satisfies the following formulas 

$$<\mbox{\large $\chi$}> \, = A <\mbox{\large $\asymp$}> + A^{-1} <)(>$$
$$<\overline{\mbox{\large $\chi$}}> \, = A^{-1} <\mbox{\large $\asymp$}> + A <)(>,$$
\end{enumerate}

\noindent where the small diagrams represent parts of larger diagrams that are identical except  at
the site indicated in the bracket. We take the convention that the letter chi, \mbox{\large $\chi$},
denotes a crossing where {\em the curved line is crossing over the straight
segment}. The barred letter denotes the switch of this crossing, where {\em the curved
line is undercrossing the straight segment}.  See Figure~\ref{Figure 9 } for a graphic illustration of this relation, and an
indication of the convention for choosing the labels $A$ and $A^{-1}$ at a given crossing.


     \bigbreak

\noindent It is easy to see that Properties $2$ and $3$ define the calculation of the bracket on
arbitrary link diagrams. The choices of coefficients ($A$ and $A^{-1}$) and the value of $\delta$
make the bracket invariant under the Reidemeister moves II and III. Thus
Property $1$ is a consequence of the other two properties. 
\bigbreak

\begin{figure}
     \begin{center}
     \begin{tabular}{c}
     \includegraphics[height=4cm]{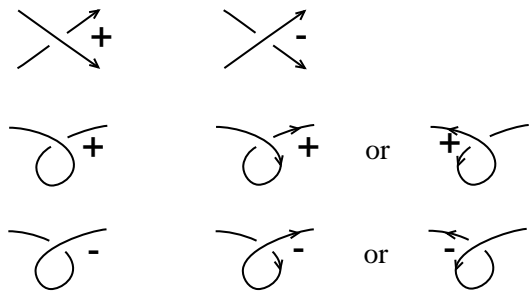}
     \end{tabular}
     \end{center}
     \caption{\bf Crossing Signs and Curls }
     \label{Figure 10 }
     \end{figure}

In computing the bracket, one finds the following behaviour under Reidemeister move I: 
  $$<\mbox{\large $\gamma$}> = -A^{3}<\smile> \hspace {.5in}$$ and 
  $$<\overline{\mbox{\large $\gamma$}}> = -A^{-3}<\smile> \hspace {.5in}$$

\noindent where \mbox{\large $\gamma$}  denotes a curl of positive type as indicated in Figure~\ref{Figure 10 }, 
and  $\overline{\mbox{\large $\gamma$}}$ indicates a curl of negative type, as also seen in this
figure. The type of a curl is the sign of the crossing when we orient it locally. Our convention of
signs is also given in Figure~\ref{Figure 10 }. Note that the type of a curl  does not depend on the orientation
we choose.  The small arcs on the right hand side of these formulas indicate
the removal of the curl from the corresponding diagram.  

\bigbreak
  
\noindent The bracket is invariant under regular isotopy and can be  normalized to an invariant of
ambient isotopy by the definition  
$$f_{K}(A) = (-A^{3})^{-w(K)}<K>(A),$$ where we chose an orientation for $K$, and where $w(K)$ is 
the sum of the crossing signs  of the oriented link $K$. $w(K)$ is called the {\em writhe} of $K$. 
The convention for crossing signs is shown in  Figure~\ref{Figure 10 }.


     \bigbreak

\begin{figure}
     \begin{center}
     \begin{tabular}{c}
     \includegraphics[height=3.5cm]{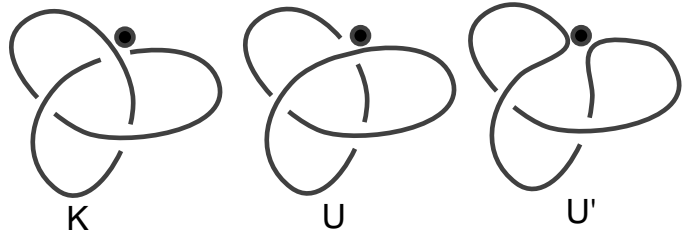}
     \end{tabular}
     \end{center}
     \caption{\bf Trefoil and Two Relatives   }
     \label{Figure 11 }
     \end{figure}

\noindent One useful consequence of these formulas is the following {\em switching formula}
$$A<\mbox{\large $\chi$}> - A^{-1} <\overline{\mbox{\large $\chi$}}> = (A^{2} - A^{-2})<\mbox{\large $\asymp$}>.$$ Note that 
in these conventions the $A$-smoothing of $\mbox{\large $\chi$}$ is $\mbox{\large $\asymp$},$ while the $A$-smoothing of
$\overline{\mbox{\large $\chi$}}$ is $)(.$ Properly interpreted, the switching formula above says that you can switch a crossing and 
smooth it either way and obtain a three diagram relation. This is useful since some computations will simplify quite quickly with the 
proper choices of switching and smoothing. Remember that it is necessary to keep track of the diagrams up to regular isotopy (the 
equivalence relation generated by the second and third Reidemeister moves). Here is an example. View Figure~\ref{Figure 11 }.


     \bigbreak

\noindent Figure~\ref{Figure 11 } shows a trefoil diagram $K$, an unknot diagram $U$ and another unknot diagram $U'.$  Applying the switching formula,
we have $$A^{-1} <K> - A <U> = (A^{-2} - A^{2}) <U'>$$ and 
$<U>= -A^{3}$ and $<U'>=(-A^{-3})^2 = A^{-6}.$  Thus $$A^{-1} <K> - A(-A^{3}) = (A^{-2} - A^{2}) A^{-6}.$$ Hence
$$A^{-1} <K> = -A^4 + A^{-8} - A^{-4}.$$  Thus $$<K> = -A^{5} - A^{-3} + A^{-7}.$$ This is the bracket polynomial of the trefoil diagram $K.$
\bigbreak

\noindent Since the trefoil diagram $K$ has writhe $w(K) = 3,$ we have the normalized polynomial 
$$f_{K}(A) = (-A^{3})^{-3}<K> = -A^{-9}(-A^{5} - A^{-3} + A^{-7}) = A^{-4} + A^{-12} - A^{-16}.$$  
\bigbreak

The bracket model for the Jones polynomial is quite useful both theoretically and in terms
 of practical computations. One of the neatest applications is to simply compute, as we have done, $f_{K}(A)$ for the
trefoil knot $K$ and determine that  $f_{K}(A)$ is not equal to $f_{K}(A^{-1}) = f_{-K}(A).$  This
shows that the trefoil is not ambient isotopic to its mirror image, a fact that is much harder to
prove by classical methods.
\bigbreak

\noindent {\bf The State Summation.} In order to obtain a closed formula for the bracket, we now describe it as a state summation.
Let $K$ be any unoriented link diagram. Define a {\em state}, $S$, of $K$  to be a choice of
smoothing for each  crossing of $K.$ There are two choices for smoothing a given  crossing, and
thus there are $2^{N}$ states of a diagram with $N$ crossings.
 In a state we label each smoothing with $A$ or $A^{-1}$ according to the left-right convention 
discussed in Property $3$ (see Figure~\ref{Figure 9 }). The label is called a {\em vertex weight} of the state.
There are two evaluations related to a state. The first one is the product of the vertex weights,
denoted  

$$<K|S>.$$
The second evaluation is the number of loops in the state $S$, denoted  $$||S||.$$
  
\noindent Define the {\em state summation}, $<K>$, by the formula 

$$<K> \, = \sum_{S} <K|S>\delta^{||S||-1}.$$
It follows from this definition that $<K>$ satisfies the equations
  
$$<\mbox{\large $\chi$}> \, = A <\mbox{\large $\asymp$}> + A^{-1} <)(>,$$
$$<K \sqcup  O> \, = \delta<K>,$$
$$<O> \, =1.$$
  
\noindent The first equation expresses the fact that the entire set of states of a given diagram is
the union, with respect to a given crossing, of those states with an $A$-type smoothing and those
 with an $A^{-1}$-type smoothing at that crossing. The second and the third equation
are clear from the formula defining the state summation. Hence this state summation produces the
bracket polynomial as we have described it at the beginning of the  section. 

\bigbreak

\noindent {\bf Remark.} By a change of variables one obtains the original
Jones polynomial, $V_{K}(t),$  for oriented knots and links from the normalized bracket:

$$V_{K}(t) = f_{K}(t^{-\frac{1}{4}}).$$

\noindent {\bf Remark.} The bracket polynomial provides a connection between  knot theory and physics, in that the state summation 
expression for it exhibits it as a generalized partition function defined on the knot diagram. Partition functions
are ubiquitous in statistical mechanics, where they express the summation over all states of the physical system of 
probability weighting functions for the individual states. Such physical partition functions contain large amounts of 
information about the corresponding physical system. Some of this information is directly present in the properties of the 
function, such as the location of critical points and phase transition. Some of the information can be obtained by differentiating the 
partition function, or performing other mathematical operations on it. 
\bigbreak

There is much more in this connection with statistical mechanics in that the local weights in a partition function are often expressed in
terms of solutions to a matrix equation called the Yang-Baxter equation, that turns out to fit perfectly invariance under the third 
Reidemeister move. As a result, there are many ways to define partition functions of knot diagrams that give rise to invariants of knots and links.
The subject is intertwined with the algebraic structure of Hopf algebras and quantum groups, useful for producing systematic solutions to the Yang-Baxter
equation. In fact Hopf algebras are deeply connected with the problem of constructing invariants of three-dimensional manifolds in relation to
invariants of knots. We have chosen, in this survey paper, to not discuss the details of these approaches, but rather to proceed to Vassiliev invariants
and the relationships with Witten's functional integral. The
reader is referred to \cite{KA87,KA89,KL,Kauffman-Graph,KaufInter,KP,AW,JO1,JO2,KR,RT1,RT2,T,TV} for more information about relationships of knot theory with
statistical mechanics, Hopf algebras and quantum groups. For topology, the key point is that Lie algebras can be used to construct invariants of
knots and links. 
\bigbreak

\subsection{Quantum Computation of the Jones Polynomial}
 
{\em Can the invariants of knots and links such as the Jones 
polynomial be configured as quantum computers?} This is an 
important question because the algorithms to compute the Jones 
polynomial are known to be $NP$-hard, and 
so corresponding quantum algorithms may shed light on the 
relationship of this level of computational complexity with 
quantum computing (See \cite{FLZ}).  Such models can be formulated in terms of 
the Yang-Baxter equation \cite{KA87, KA89,KP, QCJP1, QCJP2}. 
The next paragraph explains how this comes about.

\begin{figure}
     \begin{center}
     \begin{tabular}{c}
     \includegraphics[height=5cm]{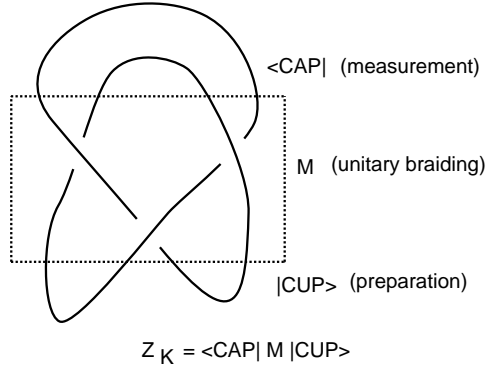}
     \end{tabular}
     \end{center}
     \caption{\bf A Knot Quantum Computer}
     \label{Figure 12 }
     \end{figure}

In Figure~\ref{Figure 12 }, we indicate how topological braiding plus 
maxima (caps) and minima (cups) can be used to configure the 
diagram of a knot or link. This also can be translated into algebra 
by the association of a Yang-Baxter matrix $R$ (not necessarily the $R$ of the previous sections) to each 
crossing and other matrices
to the maxima and minima. There are models 
of very effective invariants of knots and links such as 
the Jones polynomial that can be 
put into this form \cite{QCJP}. In this way of looking at 
things, the knot diagram can be viewed as a picture,
with time as the vertical dimension, of particles arising 
from the vacuum, interacting (in a two-dimensional space) 
and finally annihilating one another.
The invariant takes the form of an amplitude for this process 
that is computed through the association of the Yang-Baxter 
solution $R$ as the 
scattering matrix at the crossings and the minima and maxima 
as creation and annihilation operators. Thus we can write the 
amplitude in the form
\[Z_{K} = \langle CUP|M|CAP\rangle \] where $\langle CUP|$ 
denotes the composition of cups, $M$ is the
composition of elementary braiding matrices, and 
$|CAP\rangle $ is the composition of caps. We regard
$\langle CUP|$ as the preparation of this state, and 
$|CAP\rangle $ as the measurement of this state. In order to
view $Z_{K}$ as a quantum computation, $M$ must be a 
unitary operator. This is the case when 
the $R$-matrices (the solutions to the Yang-Baxter
equation used in the model) are unitary. Each $R$-matrix 
is viewed as a a quantum gate (or
possibly a composition of quantum gates), and the 
vacuum-vacuum diagram for the knot is interpreted
as a quantum computer. This quantum computer will 
probabilistically (via quantum amplitudes) compute the values of the states
in the state sum for $Z_{K}$.  
\bigbreak

     \bigbreak

We should remark, however, that it is not necessary that the invariant be modeled via solutions to the Yang-Baxter equation.
One can use unitary representations of the braid group that are constructed in other ways. In fact, the presently successful
quantum algorithms for computing knot invariants indeed use such representations of the braid group, and we shall see this below.
Nevertheless, it is useful to point out this analogy between the structure of the knot invariants and quantum computation.
\bigbreak

Quantum algorithms for computing the Jones polynomial have been discussed elsewhere. See \cite{QCJP1,BG,Ah1,QCJP2,Ah2,Wo}. Here, as an example, we give a
local unitary representation that can be used to compute the Jones polynomial for closures of 3-braids.   We analyze this representation by making
explicit how the bracket polynomial is computed from it, and showing how the quantum computation devolves to finding the trace of a unitary transformation.
\vspace{3mm}

The idea behind the construction of this representation depends upon the algebra generated by two single qubit density matrices
(ket-bras).
Let
$|v\rangle$ and 
$|w\rangle$ be two qubits in $V,$ a complex vector space of dimension two over the complex numbers. Let
$P = |v\rangle\langle v|$ and $Q=|w\rangle\langle w|$ be the corresponding ket-bras.  Note that
$$P^2 = |v|^{2}P,$$
$$Q^2 = |w|^{2}Q,$$
$$PQP = |\langle v|w \rangle|^{2}P,$$ 
$$QPQ= |\langle v|w\rangle|^{2}Q.$$
$P$ and $Q$ generate a representation of the Temperley-Lieb algebra (See Section 12 of the present paper). One can adjust parameters to make a representation
of the three-strand braid group in the form
$$s_{1} \longmapsto rP + sI,$$
$$s_{2} \longmapsto tQ + uI,$$
where $I$ is the identity mapping on $V$ and $r,s,t,u$ are suitably chosen scalars. In the following we use this method to adjust
such a representation so that it is unitary.  Note also that this is a local unitary representation of $B_{3}$ to $U(2).$ We leave it as an exersise for
the reader to verify that it fits into our general classification of such representations as given in section 10 of the present paper.
\bigbreak

Here is a specific representation depending on two symmetric matrices $U_{1}$ and $U_{2}$ with

$$U_{1} =  \left[
\begin{array}{cc}
     d & 0  \\
     0 & 0
\end{array}
\right] = d|w \rangle \langle w|$$

\noindent and 

$$U_{2} =  \left[
\begin{array}{cc}
     d^{-1} & \sqrt{1-d^{-2}}  \\
         \sqrt{1-d^{-2}}  & d - d^{-1}
\end{array}
\right] = d | v \rangle \langle v |$$
where $w = (1,0),$ and $v = (d^{-1}, \sqrt{1 - d^{-2}}),$ assuming the entries of $v$ are real.
Note that $U_{1}^{2} = dU_{1}$ and $U_{2}^{2} = dU_{1}.$ Moreover, $U_{1}U_{2}U_{1} = U_{1}$ and $U_{2}U_{1}U_{2} = U_{1}.$
 This is an example of a specific
representation of the Temperley-Lieb algebra \cite{KA87, QCJP1}.
\noindent The desired representation of the Artin braid group is given on the two braid generators for the three strand braid group by the
equations:

$$\Phi(s_{1})= AI + A^{-1}U_{1},$$
$$\Phi(s_{2})= AI + A^{-1}U_{2}.$$
Here $I$ denotes the $2 \times 2$ identity matrix.

\noindent For any $A$ with $d = -A^{2}-A^{-2}$ these formulas define a representation of the braid group. With 
$A=e^{i\theta}$, we have $d = -2cos(2\theta)$. We find a specific range of
angles $\theta$ in the following disjoint union of angular intervals
$$\theta \in [0,\pi/6]\sqcup[\pi/3,2\pi/3]\sqcup[5\pi/6,7\pi/6]\sqcup[4\pi/3,5\pi/3]\sqcup[11\pi/6,2\pi]$$
{\it that give unitary representations of
the three-strand braid group.} Thus a specialization of a more general represention of the braid group gives rise to a continuous family
of unitary representations of the braid group.
\vspace{3mm}

\noindent {\bf Lemma.} Note that the traces of these matrices are given by the formulas $tr(U_{1})=tr(U_{2})= d$ while $tr(U_{1}U_{2}) = tr(U_{2}U_{1}) =1.$
If $b$ is any braid, let $I(b)$ denote the sum of the exponents in the braid word that expresses $b$.
For $b$ a three-strand braid, it follows that 
$$\Phi(b) = A^{I(b)}I + \Pi(b)$$

\noindent where $I$ is the $ 2 \times 2$ identity matrix and $\Pi(b)$ is a sum of products in the Temperley-Lieb algebra 
involving $U_{1}$ and $U_{2}.$ 
\bigbreak

We omit the proof of this Lemma. It is a calculation. To see it,
consider an example. Suppose that $b = s_{1}s_{2}^{-1}s_{1}.$ Then
$$\Phi(b) = \Phi(s_{1}s_{2}^{-1}s_{1}) = \Phi(s_{1})\Phi(s_{2}^{-1})\Phi(s_{1}) = $$
$$(AI + A^{-1}U_{1})(A^{-1}I + AU_{2})(AI + A^{-1}U_{1}).$$
The sum of products over the generators $U_{1}$ and $U_{2}$ of the Temperley--Lieb algebra comes from expanding this expression.
\bigbreak

Since the Temperley-Lieb algebra in this dimension is generated by $I$,$U_{1}$, $U_{2}$,
$U_{1}U_{2}$ and $U_{2}U_{1}$, it follows that the value of the bracket polynomial of the closure of the braid $b$, denoted
$<\overline{b}>,$ can be calculated directly from the trace of this representation, except for the part involving the identity matrix. The
result is the equation 
$$<\overline{b}> = A^{I(b)}d^{2} + tr(\Pi(b))$$
\noindent where $\overline{b}$ denotes the standard braid closure of $b$, and the sharp brackets denote the bracket polynomial. 
From this
we see at once that 
$$<\overline{b}> = tr(\Phi(b)) + A^{I(b)}(d^{2} -2).$$

It follows from this calculation that the question of computing the bracket polynomial for the closure of the three-strand
braid $b$ is mathematically equivalent to the problem of computing the trace of the unitary matrix $\Phi(b).$ 
\vspace{3mm}

\noindent {\bf The Hadamard Test}

In order to (quantum) compute the trace of a unitary matrix $U$, one can use the {\it Hadamard test} to obtain the diagonal matrix
elements $\langle \psi|U|\psi \rangle$ of $U.$ The trace is then the sum of these matrix elements as $|\psi \rangle$ runs over an orthonormal basis for 
the vector space. We first obtain $$\frac{1}{2} + \frac{1}{2}Re\langle \psi|U|\psi \rangle$$ as
an expectation by applying the Hadamard gate $H$
$$H|0 \rangle = \frac{1}{\sqrt{2}}(|0\rangle + |1\rangle)$$
$$H|1 \rangle = \frac{1}{\sqrt{2}}(|0\rangle - |1\rangle)$$
to the first qubit of 
$$C_{U} \circ (H \otimes 1) |0 \rangle |\psi \rangle = \frac{1}{\sqrt{2}}(|0\rangle \otimes|\psi \rangle + |1\rangle \otimes U|\psi\rangle.$$
Here $C_{U}$ denotes controlled $U,$ acting as $U$ when the control bit is $|1 \rangle$ and the identity mapping when the control bit is $|0 \rangle.$ We
measure the expectation for the first qubit $|0 \rangle$ of the resulting state
$$\frac{1}{2}(H|0\rangle \otimes|\psi \rangle + H|1\rangle \otimes U|\psi\rangle)
=\frac{1}{2}((|0\rangle + |1\rangle) \otimes|\psi \rangle + (|0\rangle - |1\rangle) \otimes U|\psi\rangle)$$
$$=\frac{1}{2}(|0\rangle \otimes (|\psi \rangle + U|\psi\rangle) + |1\rangle \otimes(|\psi \rangle - U|\psi\rangle)).$$
This expectation is $$\frac{1}{2}(\langle \psi | + \langle \psi| U^{\dagger})(|\psi \rangle + U|\psi\rangle) = \frac{1}{2} + \frac{1}{2}Re\langle \psi|U|\psi
\rangle.$$
\noindent The imaginary
part is  obtained by applying the same procedure to 
$$\frac{1}{\sqrt{2}}(|0\rangle \otimes|\psi \rangle - i|1\rangle \otimes U|\psi\rangle$$
This is the method used in
\cite{Ah1}, and the reader may wish to contemplate its efficiency in the context of this simple model. Note that the Hadamard test enables this quantum 
computation to estimate the trace of any unitary matrix $U$ by repeated trials that estimate individual matrix entries $\langle \psi|U|\psi\rangle.$
We shall return to quantum algorithms for the Jones polynomial and other knot polynomials in a subsequent paper.
\bigbreak

\section{Quantum Topology, Cobordism Categories, Temperley-Lieb Algebra and Topological Quantum Field Theory}
The purpose of this section is to discuss the general idea behind topological quantum field theory, and to illustrate its application to basic
quantum mechanics and quantum mechanical formalism. It is useful in this regard to have available the concept of {\it category}, and we shall begin the 
section by discussing this far-reaching mathematical concept.
\bigbreak

\noindent {\bf Definition.} A {\it category Cat} consists in two related collections:
\begin{enumerate}
\item $Obj(Cat)$, the {\it objects} of $Cat,$ and 
\item $Morph(Cat)$, the {\it morphisms} of $Cat.$
\end{enumerate}
satisfying the following axioms:
\begin{enumerate}
\item Each morphism $f$ is associated to two objects of $Cat$, the {\it domain} of f and the {\it codomain} of f. Letting $A$ denote the domain of $f$ and
$B$ denote the codomain of $f,$ it is customary to denote the morphism $f$ by the arrow notation
$f:A \longrightarrow B.$
\item Given $f:A \longrightarrow B$ and $g:B \longrightarrow C$ where $A$, $B$ and $C$ are objects of $Cat$, then there exists an associated morphism
$g \circ f : A \longrightarrow C$ called the {\it composition} of $f$ and $g$.
\item To each object $A$ of $Cat$ there is a unique {\it identity morphism} $1_{A}:A \longrightarrow A$ such that $1_{A} \circ f = f$ for any
morphism $f$ with codomain $A$, and $g \circ 1_{A} = g$ for any morphism $g$ with domain $A.$
\item Given three morphisms $f:A \longrightarrow B$, $g:B \longrightarrow C$ and $h:C \longrightarrow D$, then composition is associative.
That is $$(h \circ g) \circ f = h \circ (g \circ f).$$
\end{enumerate}

\noindent If $Cat_{1}$ and $Cat_{2}$ are two categories, then a {\it functor} $F:Cat_{1} \longrightarrow Cat_{2}$ consists in functions 
$F_{O}:Obj(Cat_{1}) \longrightarrow Obj(Cat_{2})$ and $F_{M}:Morph(Cat_{1}) \longrightarrow Morph(Cat_{2})$ such that 
identity morphisms and composition of morphisms are preserved under these mappings. That is (writing just $F$ for $F_{O}$ and $F_{M}$),
\begin{enumerate}
\item $F(1_{A}) = 1_{F(A)}$, 
\item $F(f:A \longrightarrow B) = F(f):F(A) \longrightarrow F(B)$,
\item $F(g \circ f) = F(g) \circ F(f)$.
\end{enumerate}

A functor $F:Cat_{1} \longrightarrow Cat_{2}$ is a structure preserving mapping from one category to another.
It is often convenient to think of the image of the functor $F$ as an {\it interpretation} of the first category in terms of the second.
We shall use this terminology below and sometimes refer to an interpretation without specifying all the details of the functor that describes it.
\bigbreak

The notion of category is a broad mathematical concept, encompassing many fields of mathematics. Thus one has the category of sets where the objects
are sets (collections) and the morphisms are mappings between sets. One has the category of topological spaces where the objects are spaces and the morphisms
are continuous mappings of topological spaces. One has the category of groups where the objects are groups and the morphisms are homomorphisms of groups.
Functors are structure preserving mappings from one category to another. For example, the fundamental group is a functor from the category of topological
spaces with base point, to the category of groups. In all the examples mentioned so far, the morphisms in the category are restrictions of mappings in the 
category of sets, but this is not necessarily the case. For example, any group $G$ can be regarded as a category, $Cat(G)$, with one object $*.$
The morphisms from $*$ to itself are the elements of the group and composition is group multiplication. In this example, the object has no internal structure
and all the complexity of the category is in the morphisms. 
\bigbreak

The Artin braid group $B_{n}$ can be regarded as a category whose single object is an ordered row of 
points $[n] = \{1,2,3,...,n \}.$ The morphisms are the braids themselves and composition is the multiplication of the braids. A given ordered row of
points is interpreted as the starting or ending row of  points at the bottom or the top of the braid. In the case of the braid category, the morphisms
have both external and internal structure. Each morphism produces a permutation of the ordered row of points (corresponding to the begiinning and ending
points of the individual braid strands), and weaving of the braid is extra structure beyond the object that is its domain and codomain. Finally, for this
example, we can take all the braid groups $B_{n}$ ($n$ a positive integer) under the wing of a single category, $Cat(B)$, whose objects are all ordered rows
of points
$[n]$, and whose morphisms are of the form $b:[n] \longrightarrow [n]$ where $b$ is a braid in $B_{n}.$ The reader may wish to have
morphisms between objects with different $n$. We will have this shortly in the Temperley-Lieb category and in the category of tangles.
\bigbreak

The {\it $n$-Cobordism Category}, $Cob[n]$, has as its objects smooth manifolds of dimension $n$, and as its morphisms, smooth manifolds $M^{n+1}$ of
dimension $n+1$ with a partition of the boundary, $\partial M^{n+1}$, into two collections of $n$-manifolds that we denote by $L(M^{n+1})$ and $R(M^{n+1}).$
We regard $M^{n+1}$ as a morphism from  $L(M^{n+1})$ to $R(M^{n+1})$ 
$$M^{n+1}: L(M^{n+1}) \longrightarrow R(M^{n+1}).$$
As we shall see, these cobordism categories are highly significant for 
quantum mechanics, and the simplest one, $Cob[0]$ is directly related to the Dirac notation of bras and kets and to the Temperley-Lieb algebara. We shall
concentrate in this section on these cobordism categories, and their relationships with quantum mechanics.
\bigbreak

\noindent One can choose to consider either oriented or non-oriented manifolds, and within unoriented manifolds there are those that are orientable and 
those that are not orientable. In this section we will implicitly discuss only orientable manifolds, but we shall not specify an orientation. In the 
next section, with the standard definition of topological quantum field theory, the manifolds will be oriented. The definitions of the cobordism
categories for oriented manifolds go over mutatis mutandis.
\bigbreak

\begin{figure}
     \begin{center}
     \begin{tabular}{c}
     \includegraphics[height=4cm]{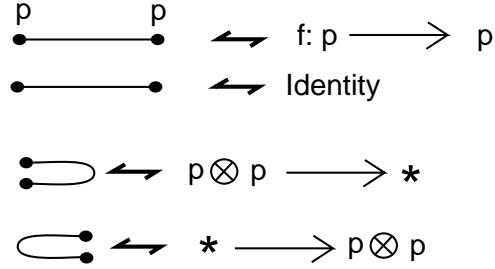}
     \end{tabular}
     \end{center}
     \caption{\bf Elementary Cobordisms  }
     \label{Figure 13 }
     \end{figure}

Lets begin with $Cob[0]$. Zero dimensional manifolds are just collections of points. The simplest zero dimensional manifold is a single point $p$.
We take $p$ to be an object of this category and also $*$,
where $*$ denotes the empty manifold (i.e. the empty set in the category of manifolds). The object $*$ occurs in $Cob[n]$ for every $n$, since
it is possible that either the left set or the right set of a morphism is empty. A line segment $S$ with boundary points $p$ and $q$ is a morphism from $p$
to $q$.
$$S:p \longrightarrow q$$
See Figure~\ref{Figure 13 }. In this figure we have illustrated the morphism from $p$ to $p.$ The simplest convention for this category is to take this morphism to
be the identity. Thus if we look at the subcategory of $Cob[0]$ whose only object is $p$, then the only morphism is the identity morphism. Two points 
occur as the boundary of an interval. The reader will note that $Cob[0]$ and the usual arrow notation for morphisms are very closely related. This is 
a place where notation and mathematical structure share common elements. In general the objects of $Cob[0]$ consist in the empty object $*$
and non-empty rows
of points, symbolized by
$$p \otimes p \otimes \cdots \otimes p \otimes p.$$
Figure~\ref{Figure 13 } also contains a morphism
$$p \otimes p \longrightarrow *$$ and the morphism
$$* \longrightarrow p\otimes p.$$ 
The first represents a cobordism of two points to the empty set (via the bounding curved interval). The second represents a cobordism from the empty set
to two points.


     \bigbreak

\begin{figure}
     \begin{center}
     \begin{tabular}{c}
     \includegraphics[height=6cm]{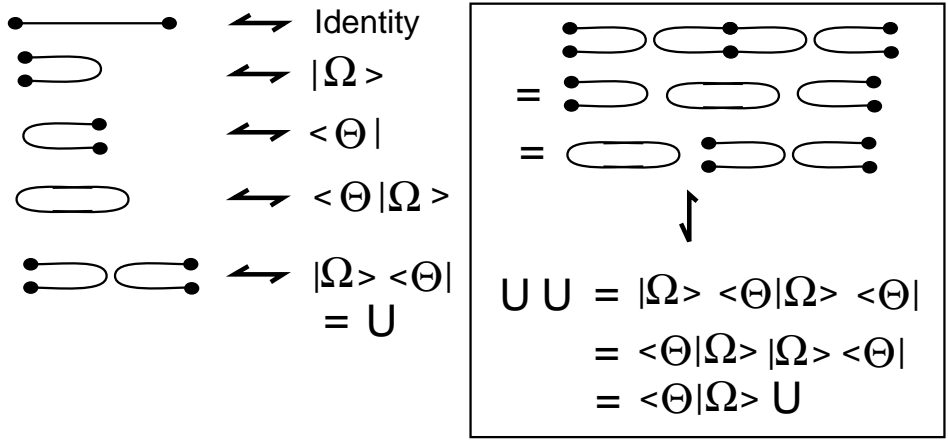}
     \end{tabular}
     \end{center}
     \caption{\bf Bras, Kets and Projectors   }
     \label{Figure 14  }
     \end{figure}

In Figure~\ref{Figure 14 }, we have indicated more morphisms in $Cob[0]$, and we have named the morphisms just discussed as
$$| \Omega \rangle : p \otimes p \longrightarrow *,$$ 
$$\langle \Theta |: * \longrightarrow p\otimes p.$$ 
The point to notice is that the usual conventions for handling Dirac bra-kets are essentially the same as the compostion rules in this
topological category. Thus in Figure~\ref{Figure 14 } we have that
$$\langle \Theta | \circ | \Omega \rangle = \langle \Theta | \Omega \rangle : * \longrightarrow *$$
represents a cobordism from the empty manifold to itself. This cobordism is topologically a circle and, in the Dirac formalism is interpreted as a 
scalar. In order to interpret the notion of scalar we would have to map the cobordism category to the category of vector spaces and linear mappings.
We shall discuss this after describing the similarities with quantum mechanical formalism. Nevertheless, the reader should note that if $V$ is a 
vector space over the complex numbers $\mathcal{C}$, then a linear mapping from $\mathcal{C}$ to $\mathcal{C}$ is determined by the image of $1$, and hence is
characterized by the scalar that is the image of $1$. In this sense a mapping $\mathcal{C} \longrightarrow \mathcal{C}$ can be regarded as a possible image in
vector spaces of the  abstract structure $\langle \Theta | \Omega \rangle : * \longrightarrow *$. It is therefore assumed that in $Cob[0]$ the composition
with the  morphism $\langle \Theta | \Omega \rangle$ commutes with any other morphism. In that way $\langle \Theta | \Omega \rangle$ behaves like a scalar in 
the cobordism category. In general, an $n+1$ manifold without boundary behaves as a scalar in $Cob[n]$, and if a manifold $M^{n+1}$ can be written
as a union of two submanifolds $L^{n+1}$ and $R^{n+1}$ so that that an $n$-manifold $W^{n}$ is their common boundary:
$$M^{n+1} =  L^{n+1} \cup R^{n+1}$$ with
$$ L^{n+1} \cap R^{n+1} = W^{n}$$ then,
we can write $$\langle M^{n+1} \rangle = \langle L^{n+1} \cup R^{n+1} \rangle = \langle L^{n+1} | R^{n+1} \rangle,$$ and $\langle M^{n+1} \rangle$
will be a scalar (morphism that commutes with all other morphisms) in the category $Cob[n]$.
\bigbreak


     \bigbreak


\begin{figure}
     \begin{center}
     \begin{tabular}{c}
     \includegraphics[height=6cm]{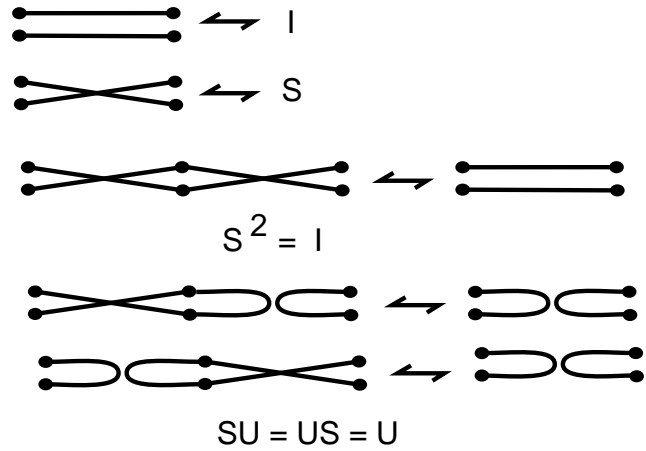}
     \end{tabular}
     \end{center}
     \caption{\bf Permutations  }
     \label{Figure 15 }
     \end{figure} 
     \bigbreak


\begin{figure}
     \begin{center}
     \begin{tabular}{c}
     \includegraphics[height=4cm]{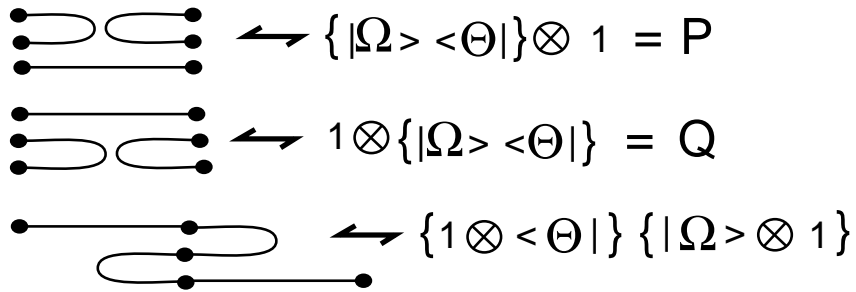}
     \end{tabular}
     \end{center}
     \caption{\bf Projectors in Tensor Lines and Elementary Topology  }
     \label{Figure 16 }
     \end{figure} 
     \bigbreak

Getting back to the contents of Figure~\ref{Figure 14 }, note how the zero dimensional cobordism category has structural parallels to the Dirac ket--bra formalism
$$ U = | \Omega \rangle \langle \Theta |$$
$$ UU = | \Omega \rangle \langle \Theta | \Omega \rangle \langle \Theta |= \langle \Theta | \Omega \rangle | \Omega \rangle \langle \Theta |
= \langle \Theta | \Omega \rangle U.$$ In the cobordism category, the bra--ket and ket--bra formalism is seen as patterns of connection of
the one-manifolds that realize the cobordisms.
\bigbreak

Now view Figure~\ref{Figure 15 }. This Figure illustrates a morphism $S$ in $Cob[0]$ that requires two crossed line segments for its planar representation.
Thus $S$ can be regarded as a non-trivial permutation, and $S^2 = I$ where $I$ denotes the identity morphisms for a two-point row.
From this example, it is clear that $Cob[0]$ contains the structure of all the syymmetric groups and more. In fact, if we take the subcateogry of 
$Cob[0]$ consisting of all morphisms from $[n]$ to $[n]$ for a fixed positive integer $n,$ then this gives the well-known {\it Brauer algebra}  (see
\cite{Benkart}) extending the symmetric group by allowing any connections among the points in the two rows. In this sense, one could call $Cob[0]$ the {\it
Brauer category}. We shall return to this point of view later.
\bigbreak

In this section, we shall be concentrating
on the part of $Cob[0]$ that does not involve permutations. This part can be characterized by those morphisms that can be represented by 
planar diagrams without crosssings between any of the line segments (the one-manifolds). We shall call this crossingless subcategory of $Cob[0]$ the {\em
Temperley-Lieb Category} and denote it by $CatTL.$ In $CatTL$ we have the subcategory $TL[n]$ whose only objects are the row of $n$ points and the empty
object $*$, and whose morphisms can all be represented by configurations that embed in the plane as in the morphisms $P$ and $Q$ in Figure~\ref{Figure 16 }. Note that with
the empty object $*$, the morphism whose diagram is a single loop appears in $TL[n]$ and is taken to commute with all other morphisms.
\bigbreak

The {\em Temperley-Lieb Algebra}, $AlgTL[n]$ is generated by the morphisms in $TL[n]$ that go from $[n]$ to itself. 
Up to multiplication by the loop, the product (composition) of two such morphisms is another flat morphism from $[n]$ to itself.
 For algebraic purposes the loop $*
\longrightarrow *$ is taken to be a scalar  algebraic variable $\delta$ that commutes with all elements in the algebra. Thus the equation 
$$ UU = \langle \Theta | \Omega \rangle U.$$
becomes
$$UU = \delta U$$
in the algebra. In the algebra we are allowed to add morphisms formally and this addition is taken to be commutative. Initially the algebra is taken with 
coefficients in the integers, but a different commutative ring of coefficients can be chosen and the value of the loop may be taken in this ring. For example,
for quantum mechanical applications it is natural to work over the complex numbers. The multiplicative structure of $AlgTL[n]$ can be described by 
generators and relations as follows: Let $I_{n}$ denote the identity morphism from $[n]$ to $[n].$ Let $U_{i}$ denote the morphism from $[n]$ to $[n]$
that connects $k$ with $k$ for $k<i$ and $k>i+1$ from one row to the other, and connects $i$ to $i+1$ in each row. Then the algebra
$AlgTL[n]$ is generated by $\{ I_{n}, U_{1},U_{2},\cdots ,U_{n-1} \}$ with relations
$$U_{i}^{2} = \delta U_{i}$$
$$U_{i}U_{i+1}U_{i} = U_{i}$$
$$U_{i}U_{j} = U_{j}U_{i} \,: \,\, |i-j|>1.$$
These relations are illustrated for three strands in Figure~\ref{Figure 16 }. We leave the commuting relation for the reader to draw in the case where $n$ is 
four or greater. For a proof that these are indeed all the relations, see \cite{KaufDiag}.
\bigbreak

\begin{figure}
     \begin{center}
     \begin{tabular}{c}
     \includegraphics[height=6cm]{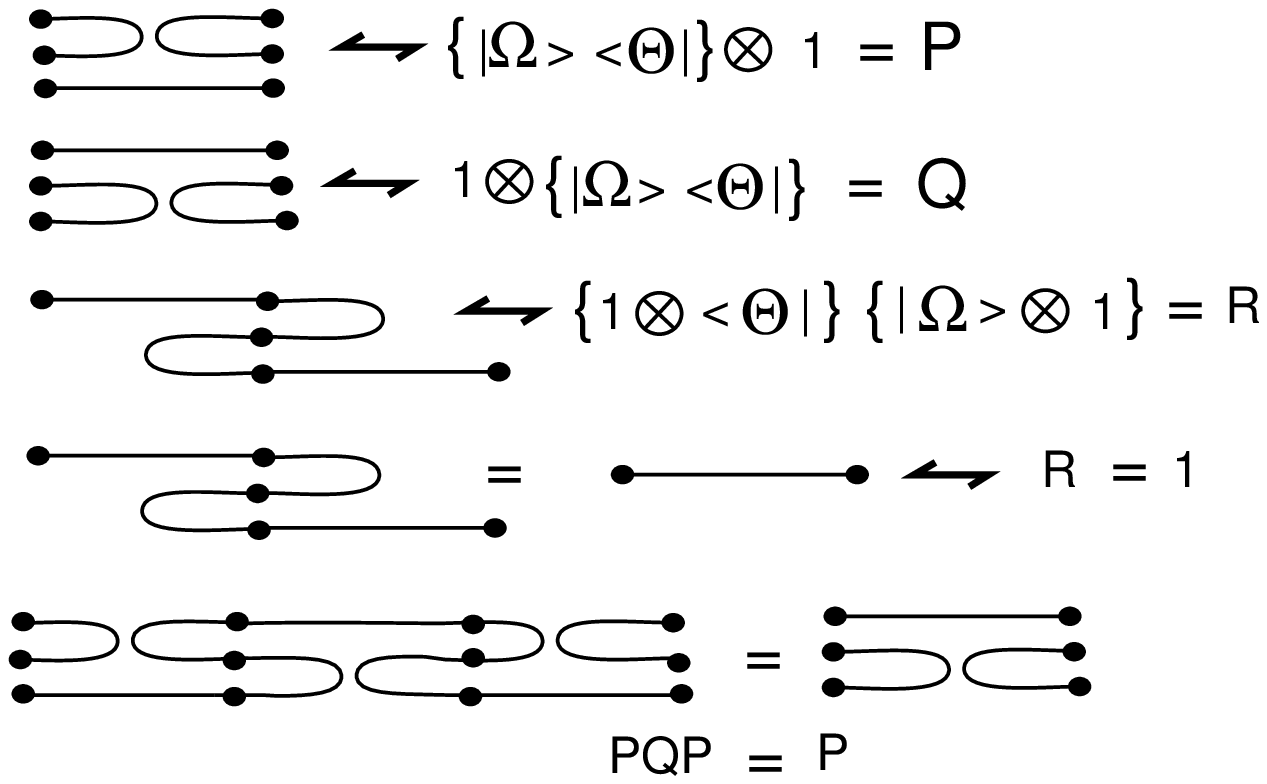}
     \end{tabular}
     \end{center}
     \caption{\bf The Basic Temperley-Lieb Relation   }
     \label{Figure 17 }
     \end{figure}

Figure~\ref{Figure 16 } and Figure~\ref{Figure 17 } indicate how the zero dimensional cobordism category contains structure that goes well beyond the usual Dirac formalism.
By tensoring the ket--bra on one side or another by identity morphisms, we obtain the beginnings of the Temperley-Lieb algebra and the Temperley-Lieb
category. Thus Figure~\ref{Figure 17 } illustrates the morphisms $P$ and $Q$ obtained by such tensoring, and the relation $PQP = P$ which is the same as
$U_{1}U_{2}U_{1} = U_{1}$
\bigbreak

\begin{figure}
     \begin{center}
     \begin{tabular}{c}
     \includegraphics[height=6cm]{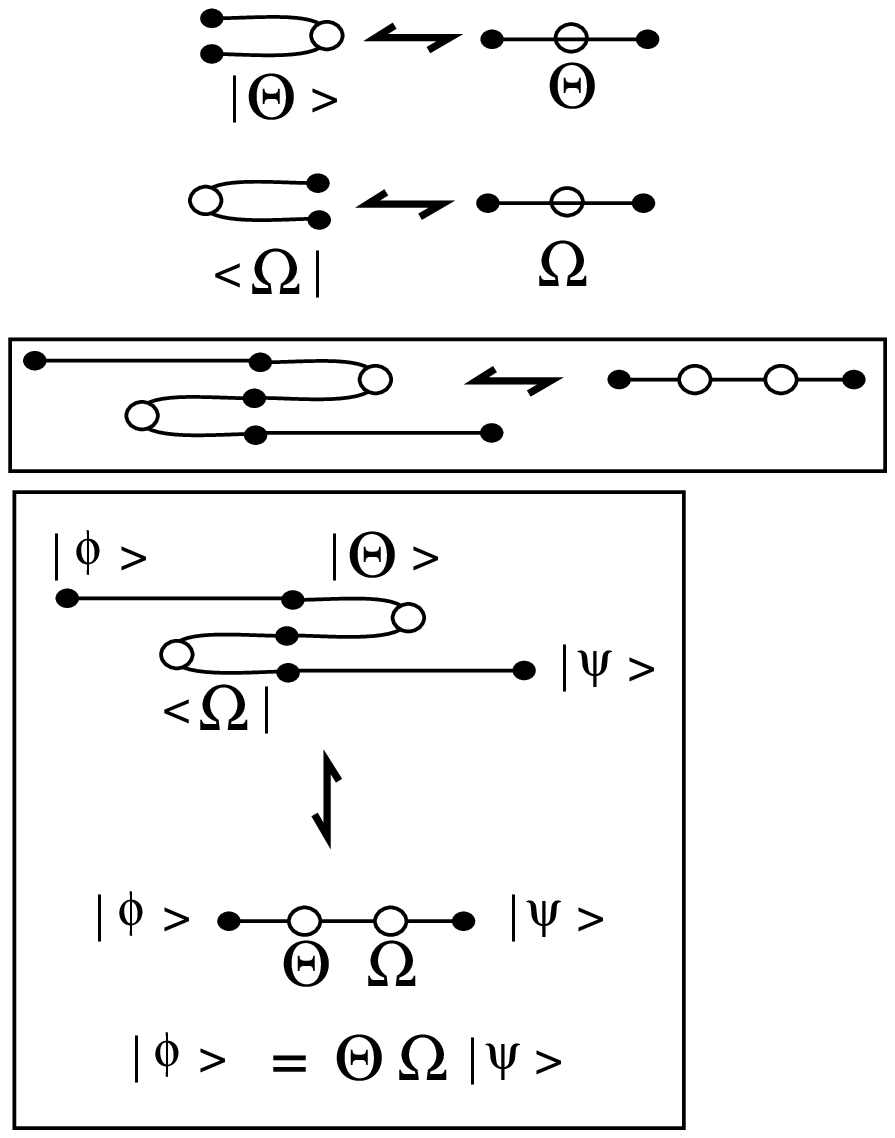}
     \end{tabular}
     \end{center}
     \caption{\bf The Key to Teleportation   }
     \label{Figure 18 }
     \end{figure}

Note the composition at the 
bottom of the Figure~\ref{Figure 17 }. Here we see a composition of the identity tensored with a ket, followed by a bra tensored with the identity.
The diagrammatic for this
association involves ``straightening" the curved structure of the morphism to a straight line. 
In Figure~\ref{Figure 18 } we have elaborated this
situation even further, pointing out that in this category each of the morphisms $\langle \Theta |$ and $| \Omega \rangle$ can be seen, by 
straightening, as mappings from 
the generating object to itself. We have denoted these corresponding morphisms by $\Theta$ and $\Omega$ respectively.
In this way there is a correspondence between
morphisms $p \otimes p \longrightarrow *$ and morphims $p \longrightarrow p.$ 
\bigbreak

In Figure~\ref{Figure 18 } we have illustrated the generalization of the straightening
procedure of Figure~\ref{Figure 17 }. In Figure~\ref{Figure 17 } the straightening occurs because the connection structure in the morphism of $Cob[0]$ does not depend on 
the wandering of curves in diagrams for the morphisms in that category. Nevertheless, one can envisage a more complex interpretation of the 
morphisms where each one-manifold (line segment) has a label, and a multiplicity of morphisms can correspond to a single line segment.
This is exactly what we expect in interpretations. For example, we can interpret the line segment $[1] \longrightarrow [1]$ as a mapping from 
a vector space $V$ to itself. Then $[1] \longrightarrow [1]$ is the diagrammatic abstraction for $ V \longrightarrow V,$ and there are many
instances of linear mappings from $V$ to $V$. 
\bigbreak

At the vector space level there is a duality between mappings 
$V \otimes V \longrightarrow \mathcal{C}$  and linear maps $V \longrightarrow V.$
Specifically, let 
$$\{ | 0 \rangle ,\cdots, | m \rangle \}$$
be a basis for $V.$ Then $\Theta: V \longrightarrow V$ is determined by
$$\Theta |i \rangle = \Theta_{ij} \, |j \rangle$$ (where we have used the Einstein summation convention on the repeated index $j$) 
corresponds to the bra 
$$\langle \Theta |: V \otimes V \longrightarrow \mathcal{C}$$
defined by 
$$\langle \Theta |ij \rangle = \Theta_{ij}.$$
Given $\langle \Theta | :V \otimes V \longrightarrow \mathcal{C},$
we associate $\Theta: V \longrightarrow V$ in this way. 
\bigbreak

Comparing with the diagrammatic for the category $Cob[0]$, we say that $\Theta: V \longrightarrow V$ is obtained by {\it straightening}
the mapping $$\langle \Theta | :V \otimes V \longrightarrow \mathcal{C}.$$ Note that in this interpretation, the bras and kets are defined relative to the 
tensor product of $V$ with itself and $[2]$ is interpreted as $V \otimes V.$ If we interpret $[2]$ as a single vector space $W,$ then 
the usual formalisms of bras and kets still pass over from the cobordism category.
\bigbreak

     \bigbreak


     \bigbreak

Figure~\ref{Figure 18 } illustrates the staightening of $| \Theta \rangle$ and $\langle \Omega |,$ and the straightening of a composition of these
applied to $| \psi \rangle,$ resulting in $| \phi \rangle.$ In the left-hand part of the bottom of Figure~\ref{Figure 18 } we illustrate the preparation
of the tensor product $| \Theta \rangle \otimes | \psi \rangle$ followed by a successful measurement by $\langle \Omega |$ in the second two
tensor factors. The resulting single qubit state, as seen by straightening, is  $| \phi \rangle = \Theta \circ \Omega |\psi \rangle.$ 
\bigbreak

From this, we see that it is possible to reversibly, indeed unitarily, transform a state $| \psi \rangle$ via a combination of preparation and measurement
just so long as the straightenings of the preparation and measurement ($\Theta$ and $\Omega$) are each invertible (unitary). This is the 
key to teleportation \cite{Teleport,C1,C2}. In the standard teleportation procedure one chooses the preparation $\Theta$ to be (up to normalization) the  $2$
dimensional identity matrix so that
$| \theta \rangle = |00\rangle + |11\rangle.$ If the successful measurement $\Omega$ is also the identity, then the transmitted state $| \phi \rangle$
will be equal to $| \psi \rangle.$ In general we will have $| \phi \rangle =  \Omega |\psi \rangle.$ One can then choose a basis of measurements
$|\Omega \rangle,$ each corresponding to a unitary transformation $\Omega$ so that the recipient of the transmission can rotate the result by the 
inverse of $\Omega$ to reconsitute $|\psi \rangle$ if he is given the requisite information. This is the basic design of the teleportation procedure.
\bigbreak  

There is much more to say about the category $Cob[0]$ and its relationship with quantum mechanics. We will stop here, and invite the reader to explore
further. Later in this paper, we shall use these ideas in formulating our representations of the braid group. For now, we point out how things look
as we move upward to $Cob[n]$ for $n > 0.$ In Figure~\ref{Figure 19 } we show typical cobordisms (morphisms) in $Cob[1]$ from two circles to one circle and from 
one circle to two circles. These are often called ``pairs of pants". Their composition is a surface of genus one seen as a morphism from two circles
to two circles. The bottom of the figure indicates a ket-bra in this dimension in the form of a mapping from one circle to one circle as a composition of
a cobordism of a circle to the empty set and a cobordism from the empty set to a circle (circles bounding disks). As we go to higher dimensions the 
structure of cobordisms becomes more interesting and more complicated. It is remarkable that there is so much structure in the lowest dimensions of 
these categories.

\begin{figure}
     \begin{center}
     \begin{tabular}{c}
     \includegraphics[height=6cm]{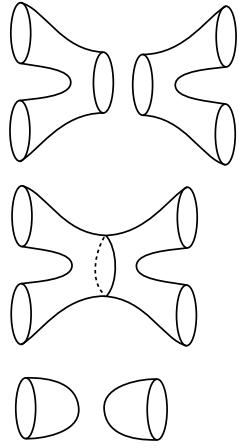}
     \end{tabular}
     \end{center}
     \caption{\bf Corbordisms of $1$-Manifolds are Surfaces   }
     \label{Figure  19}
     \end{figure} 
     \bigbreak

\section{Braiding and Topological Quantum Field Theory}
The purpose of this section is to discuss in a very general way how braiding is related to topological quantum field theory. 
In the section to follow, we will use the Temperley-Lieb recoupling theory to produce specfic unitary representations of the Artin
braid group.
\bigbreak 

The ideas in the subject
of topological quantum field theory (TQFT) are well expressed in the book \cite{Atiyah} by Michael Atiyah and the paper \cite{Witten} by Edward Witten.
Here is Atiyah's definition:
\bigbreak

\noindent {\bf Definition.} A TQFT in dimension $d$ is a functor $Z(\Sigma)$ from the cobordism category $Cob[d]$ to the category  $Vect$ of
vector spaces and linear mappings which assigns
\begin{enumerate}
\item a finite dimensional vector space $Z(\Sigma)$ to each compact, oriented $d$-dimensional manifold $\Sigma,$ 
\item a vector $Z(Y) \in Z(\Sigma)$ for each compact, oriented $(d + 1)$-dimensional manifold $Y$ with boundary $\Sigma.$
\item a linear mapping $Z(Y):Z(\Sigma_{1}) \longrightarrow Z(\Sigma_{2})$ when $Y$ is a $(d + 1)$-manifold that is a cobordism 
between $\Sigma_{1}$ and $\Sigma_{2}$ (whence the boundary of $Y$ is the union of $\Sigma_{1}$ and $-\Sigma_{2}.$
\end{enumerate}

\noindent The functor satisfies the following axioms.

\begin{enumerate}
\item $Z(\Sigma^{\dagger}) = Z(\Sigma)^{\dagger}$ where $\Sigma^{\dagger}$ denotes the manifold $\Sigma$ with the opposite orientation and 
$Z(\Sigma)^{\dagger}$ is the dual vector space.
\item $Z(\Sigma_{1} \cup \Sigma_{2}) = Z(\Sigma_{1}) \otimes Z(\Sigma_{2})$ where $\cup$ denotes disjoint union.
\item If $Y_{1}$ is a cobordism from $\Sigma_{1}$ to $\Sigma_{2},$  $Y_{2}$ is a cobordism from $\Sigma_{2}$ to $\Sigma_{3}$ and 
$Y$ is the composite cobordism $Y = Y_{1} \cup_{\Sigma_{2}} Y_{2},$ then 
$$Z(Y) = Z(Y_{2}) \circ Z(Y_{1}): Z(\Sigma_{1}) \longrightarrow Z(\Sigma_{2})$$ is the composite of the corresponding linear mappings.
\item $Z(\phi) = \mathcal{C}$ ($\mathcal{C}$ denotes the complex numbers) for the empty manifold $\phi.$
\item With $\Sigma \times I$ (where $I$ denotes the unit interval) denoting the identity cobordism from $\Sigma$ to $\Sigma,$ 
$Z(\Sigma \times I)$ is the identity mapping on $Z(\Sigma).$
\end{enumerate}

Note that, in this view a TQFT is basically a functor from the cobordism categories defined in the last section to Vector Spaces
over the complex numbers. We have already seen that in the lowest dimensional case of cobordisms of zero-dimensional manifolds, this gives
rise to a rich structure related to quatum mechanics and quantum information theory. The remarkable fact is that the case of three-dimensions
is also related to quantum theory, and to the lower-dimensional versions of the TQFT. This gives a significant way to think about three-manifold
invariants in terms of lower dimensional patterns of interaction. Here follows a brief description.
\bigbreak

\begin{figure}
     \begin{center}
     \begin{tabular}{c}
     \includegraphics[height=6cm]{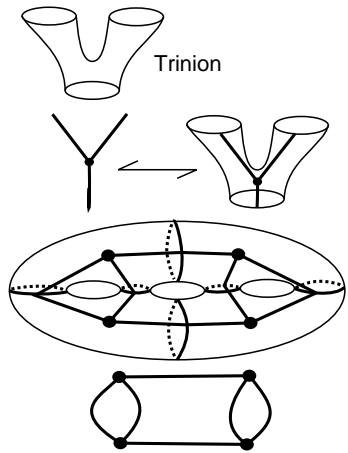}
     \end{tabular}
     \end{center}
     \caption{\bf Decomposition of a Surface into Trinions  }
     \label{trinions}
     \end{figure}

Regard the three-manifold as a union of two handlebodies with boundary an orientable  surface $S_{g}$ of genus $g.$ The surface is divided up into trinions as
illustrated in Figure~\ref{trinions}. A {\it trinion} is a surface with boundary that is topologically equivalent to a sphere with three punctures. The trinion
constitutes, in itself a cobordism in $Cob[1]$ from two circles to a single circle, or from a single circle to two circles, or from three circles to the 
empty set. The {\it pattern} of a trinion is a trivalent graphical vertex, as illustrated in 
Figure~\ref{trinions}. In that figure we show the trivalent vertex
graphical pattern drawn on the surface of the trinion, forming a graphical pattern for this combordism. It should be clear from this figure that any
cobordism in $Cob[1]$ can be diagrammed by a trivalent graph, so that the category of trivalent graphs (as morphisms from ordered sets of points to ordered
sets of points) has an image in the category of cobordisms of  compact one-dimensional manifolds. Given a surface $S$ (possibly with boundary) and a
decomposition of that surface into triions, we associate to it a trivalent graph
$G(S,t)$ where $t$ denotes the particular trinion decomposition.
\bigbreak

In this correspondence, distinct graphs can correspond to topologically identical cobordisms of circles, as
illustrated in Figure~\ref{Figure 22 }. It turns out that the graphical structure is important, and that it is extraordinarily useful to articulate transformations
between the graphs that correspond to the homeomorphisms of the corresponding surfaces. The beginning of this structure is indicated in the bottom part of
Figure~\ref{Figure 22 }. 
\bigbreak

In Figure~\ref{Figure 23 } we illustrate another feature of the relationship betweem surfaces and graphs. At the top of the figure we indicate a 
homeomorphism between a twisted trinion and a standard trinion. The homeomorphism leaves the ends of the trinion (denoted $A$,$B$ and $C$) fixed while undoing
the internal twist. This can be accomplished as an ambient isotopy of the embeddings in three dimensional space that are indicated by this figure.
Below this isotopy we indicate the corresponding graphs. In the graph category there will have to be a transformation between a braided and an unbraided
trivalent vertex that corresponds to this homeomorphism.
\bigbreak

     \bigbreak

\begin{figure}
     \begin{center}
     \begin{tabular}{c}
     \includegraphics[height=4cm]{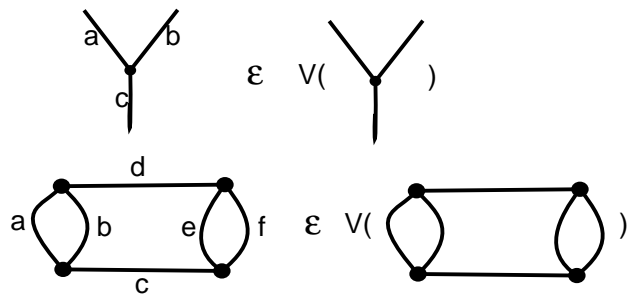}
     \end{tabular}
     \end{center}
     \caption{\bf Trivalent Vectors  }
     \label{Figure  21}
     \end{figure} 
     \bigbreak

\begin{figure}
     \begin{center}
     \begin{tabular}{c}
     \includegraphics[height=6cm]{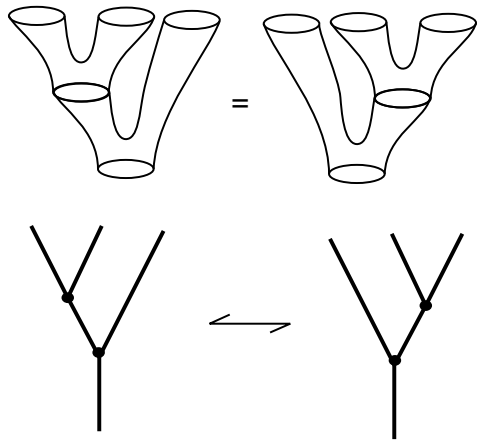}
     \end{tabular}
     \end{center}
     \caption{\bf  Trinion Associativity   }
     \label{Figure 22 }
     \end{figure} 
     \bigbreak

\begin{figure}
     \begin{center}
     \begin{tabular}{c}
     \includegraphics[height=6cm]{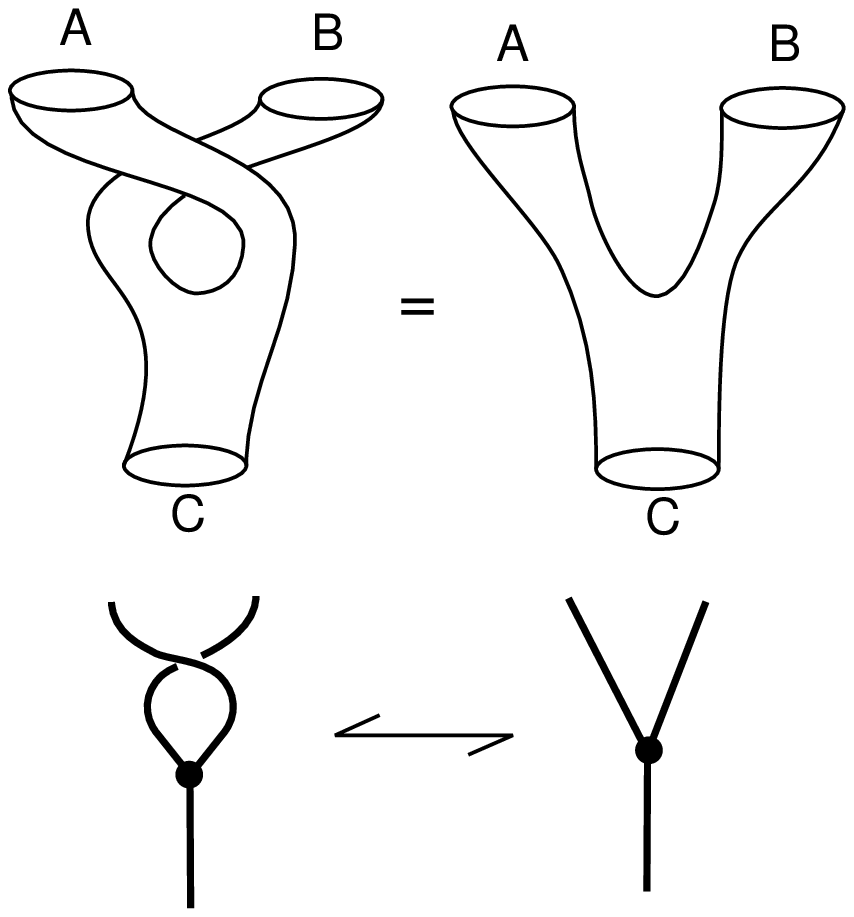}
     \end{tabular}
     \end{center}
     \caption{\bf  Tube Twist   }
     \label{Figure 23 }
     \end{figure} 
     \bigbreak

From the point of view that we shall take in this paper, the key to the mathematical structure of three-dimensional TQFT lies in the trivalent graphs,
including the  braiding of grapical arcs. We can think of these braided graphs as representing idealized Feynman diagrams,
with the trivalent vertex as the basic particle interaction vertex, and the braiding of lines representing an interaction resulting from an exchange of 
particles. In this view one thinks of the particles as moving in a two-dimensional medium, and the diagrams of braiding and trivalent vertex interactions
as indications of the temporal events in the system, with time indicated in the direction of the morphisms in the category. Adding such graphs to the category
of knots and links is an extension of the {\it tangle category} where one has already extended braids to allow any embedding of strands and circles that 
start in $n$ ordered points and end in $m$ ordered points. The tangle category includes the braid category and the Temperley-Lieb category. These are 
both included in the category of braided trivalent graphs. 
\bigbreak

Thinking of the basic trivalent vertex as the form of a particle interaction there will be a set of particle states that can label each arc incident to 
the vertex. In Figure~\ref{Figure 21 } we illustrate the labeling of the trivalent graphs by such particle states. In the next two sections we will see specific rules
for labeling such states. Here it suffices to note that there will be some restrictions on these labels, so that a trivalent vertex has a set of possible
labelings. Similarly, any trivalent graph will have a set of admissible labelings. These are the possible particle processes that this graph can support.
We take the set of admissible labelings of a given graph $G$ as a basis for a vector space $V(G)$ over the complex numbers. This vector space is the space
of {\it processes} associated with the graph $G.$ Given a surface $S$ and a decomposition $t$ of the surface into trinions, we have the associated 
graph $G(S,t)$ and hence a vector space of processes $V(G(S,t))$. It is desirable to have this vector space independent of the particular decomposition 
into trinions. If this can be accomplished, then the set of vector spaces and linear mappings associated to the surfaces can consitute a functor from the
category of cobordisms of  one-manifolds to vector spaces, and hence gives rise to a  one-dimensional topological quantum field theory. To this end we need
some properties of the particle interactions that will be described below. 
\bigbreak

A {\it spin network} is, by definition a lableled trivalent graph in a category of graphs that satisfy the properties outlined in the previous
paragraph. We shall detail the requirements below. 
\bigbreak 

The simplest  case of this idea is C. N. Yang's original interpretation of the Yang-Baxter equation \cite{Yang}.  Yang articulated a
quantum field theory in one dimension of space and one dimension of time in which the $R$-matrix giving the
scattering ampitudes for an interaction of two particles whose (let us say) spins corresponded  to the matrix indices so that
$R^{cd}_{ab}$ is the amplitude for particles of spin $a$ and spin $b$ to interact and produce particles of spin $c$ and $d.$ Since these interactions are
between particles in a line, one takes the convention that the particle with spin
$a$ is to the left of the particle with spin $b,$ and the particle with spin $c$ is to the left of the particle with spin $d.$
If one follows the concatenation of such interactions, then there is an underlying permutation that is obtained
by following strands from the bottom to the top of the diagram (thinking of time as moving up the page). Yang designed the 
Yang-Baxter equation for $R$ so that {\em the amplitudes for a composite process depend only on the underlying permutation corresponding to the
process and not on the individual sequences of interactions.} 
\bigbreak

In taking over the Yang-Baxter equation for topological purposes, we can use the same interpretation, but think of the diagrams with 
their under- and over-crossings as modeling events in a spacetime with two dimensions of space and one dimension of time. The extra
spatial dimension is taken in displacing the woven strands perpendicular to the page, and allows us to use braiding operators $R$ and
$R^{-1}$ as  scattering matrices. Taking this picture to heart, one can add other particle properties to the idealized theory. In
particular one can  add fusion and creation vertices where in fusion two particles interact to become a single particle and in creation
one particle  changes (decays) into two particles. These are the trivalent vertices discussed above. Matrix elements corresponding to trivalent vertices can
represent these interactions. See Figure~\ref{Figure 24 }.
\bigbreak

\begin{figure}
     \begin{center}
     \begin{tabular}{c}
     \includegraphics[height=4cm]{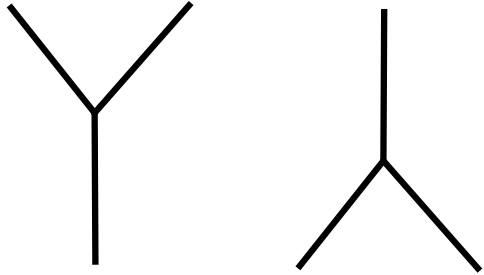}
     \end{tabular}
     \end{center}
     \caption{\bf  Creation and Fusion   }
     \label{Figure  24 }
     \end{figure} 
     \bigbreak

Once one introduces trivalent vertices for fusion and creation, there is the question how these interactions will behave in respect to 
the braiding operators. There will be a matrix expression for the compositions of braiding and fusion or creation as indicated in Figure
25. Here we  will restrict ourselves to showing the diagrammatics with the intent of giving the reader a flavor of these
structures. It is natural to assume that braiding intertwines with creation as shown in Figure~\ref{Figure 27 } (similarly with fusion). This
intertwining identity is clearly the sort of thing that a topologist will love, since it indicates that the diagrams can be interpreted
as embeddings of graphs in three-dimensional space, and it fits with our interpretation of the vertices in terms of trinions. Figure~\ref{Figure 25 } illustrates the
Yang-Baxter equation.  The intertwining identity is an assumption like the
Yang-Baxter equation itself, that simplifies the mathematical structure of the model.
\bigbreak

\begin{figure}
     \begin{center}
     \begin{tabular}{c}
     \includegraphics[height=4cm]{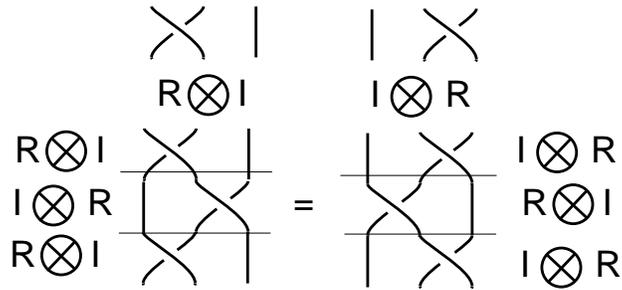}
     \end{tabular}
     \end{center}
     \caption{\bf Yang Baxter Equation   }
     \label{Figure  25 }
     \end{figure} 
     \bigbreak

\begin{figure}
     \begin{center}
     \begin{tabular}{c}
     \includegraphics[height=4cm]{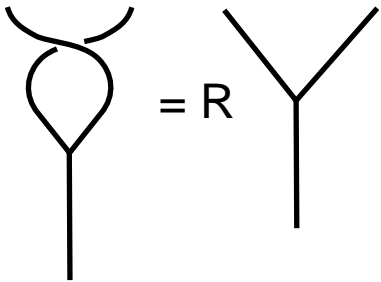}
     \end{tabular}
     \end{center}
     \caption{\bf Braiding  }
     \label{Figure 26 }
     \end{figure} 
     \bigbreak

\begin{figure}
     \begin{center}
     \begin{tabular}{c}
     \includegraphics[height=4cm]{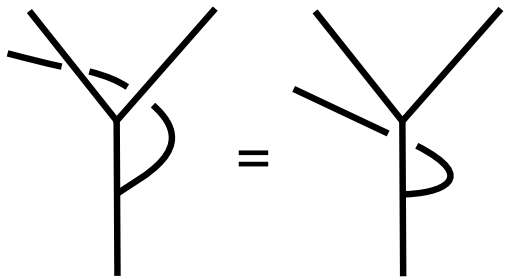}
     \end{tabular}
     \end{center}
     \caption{\bf  Intertwining   }
     \label{Figure 27 }
     \end{figure} 
     \bigbreak

\begin{figure}
     \begin{center}
     \begin{tabular}{c}
     \includegraphics[height=4cm]{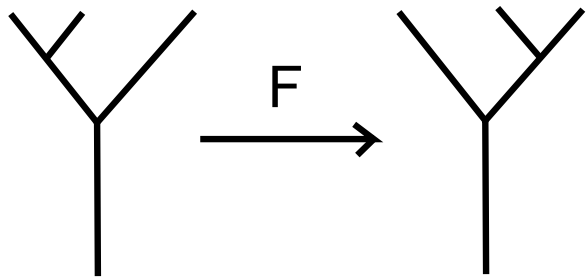}
     \end{tabular}
     \end{center}
     \caption{\bf Recoupling   }
     \label{Figure 28 }
     \end{figure}

It is to be expected that there will be an operator that expresses the recoupling of vertex interactions as shown in Figure~\ref{Figure 28 } and labeled
by $Q.$  This corresponds to the associativity at the level of trinion combinations shown in Figure~\ref{Figure 22 }. The actual formalism of such an operator will
parallel the mathematics of recoupling for angular momentum. See for example 
\cite{KL}. If one just considers the abstract structure of recoupling then one sees that for trees with four branches (each with a single
root) there is a cycle of length five as shown in Figure~\ref{Figure 29 }. One can start with any pattern of three vertex interactions and 
go through a sequence of five recouplings that bring one back to the same tree from which one started. {\em It is a natural simplifying 
axiom to assume that this composition is the identity mapping.} This axiom is called the {\em pentagon identity}. 
\bigbreak

     \bigbreak

\begin{figure}
     \begin{center}
     \begin{tabular}{c}
     \includegraphics[height=6cm]{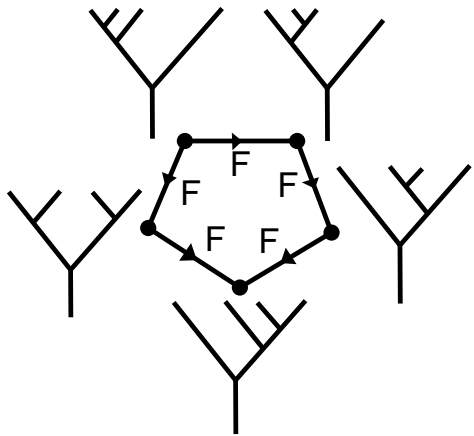}
     \end{tabular}
     \end{center}
     \caption{\bf  Pentagon Identity   }
     \label{Figure  29 }
     \end{figure} 
     \bigbreak

\begin{figure}
     \begin{center}
     \begin{tabular}{c}
     \includegraphics[height=6cm]{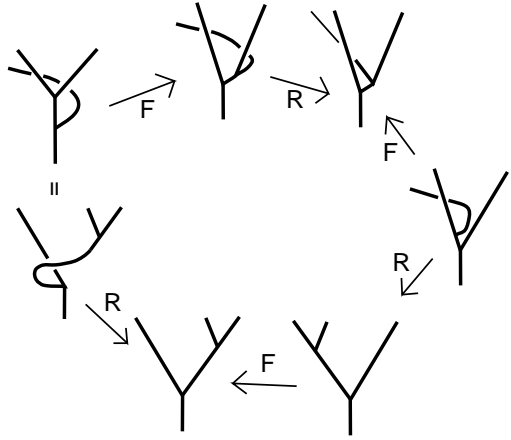}
     \end{tabular}
     \end{center}
     \caption{\bf  Hexagon Identity   }
     \label{Figure 30 }
     \end{figure}

Finally there is a hexagonal cycle of interactions between braiding, recoupling and the intertwining identity as shown in Figure~\ref{Figure 30 }.
One says that the interactions satisfy the {\em hexagon identity} if this composition is the identity.
\bigbreak

     \bigbreak

\noindent {\bf Remark.} It is worth pointing out how these identities are related to the braiding.
The hexagon identity tells us that $$R^{-1}FRF^{-1}RF = I$$ where $I$ is the identity mapping
on the process space for trees with three branches. Letting $$A  = R$$ and $$B = F^{-1}RF,$$ we see that the hexagon identity is equivalent to the statement $$B = R^{-1}F^{-1}R.$$ Thus
$$ABA = R( R^{-1}F^{-1}R)R = F^{-1}R^{2} = (F^{-1}RF)F^{-1}R = (R^{-1}F^{-1}R)F^{-1}R$$
$$= (R^{-1}F^{-1}R)R(R^{-1}F^{-1}R) = BAB.$$ Thus the hexagon relation in this context, implies that
$A$ and $B$ satisfy the braiding relation. The combination of the hexagon and pentagon relations ensures that the braid group representations that are generated are well-defined and fit together as we include smaller numbers of strands in larger numbers of strands. We omit the further details of the verification of this statement. 
\bigbreak 

A {\em graphical three-dimensional topological quantum field theory} is an algebra of interactions that satisfies the Yang-Baxter equation, the
intertwining identity, the pentagon identity  and the hexagon identity. There is not room in this summary to detail the way
that these properties fit into the topology of knots and three-dimensional manifolds, but a sketch is in order. For the case of topological 
quantum field theory related to the group $SU(2)$ there is a construction based entirely on the combinatorial topology of the bracket polynomial
(See Sections 11 to 18  of this paper.). See \cite{KP,KL} for more information on this approach.
\bigbreak

Now return to Figure ~\ref{trinions} where we
illustrate trinions, shown in relation to  a trivalent vertex, and a surface of genus three that is decomposed into four trinions. It
turns out that the vector space
$V(S_g) = V(G(S_{g},t))$ to a surface with a trinion decomposition as $t$ described above, and defined in terms of the graphical topological quantum field
theory, does not depend upon the choice of trinion decomposition. This independence is guaranteed by
the braiding, hexagon and pentagon identities. One can then associate a well-defined vector $|M \rangle$ in $V(S_{g})$ whenenver $M$
is a three manifold whose  boundary is $S_{g}.$  Furthermore, if a closed three-manifold $M^{3}$ is decomposed along a surface $S_{g}$ into 
the union of $M_{-}$ and $M_{+}$
where these parts are otherwise disjoint three-manifolds with boundary $S_{g},$ then the inner product $I(M) = \langle M_{-} | M_{+} \rangle$ is, up to
normalization, an invariant of the three-manifold $M_{3}.$ With the definition of graphical topological quantum field theory given above, knots and links can
be incorporated as well, so that one obtains a source of invariants $I(M^{3},K)$ of knots and links in orientable three-manifolds. Here we see the uses of
the relationships that occur in the higher dimensional cobordism categories, as descirbed in the previous section.
\bigbreak 

\noindent The invariant $I(M^{3},K)$ can be formally compared with the Witten \cite{Witten} integral  $$Z(M^{3},K) = \int DAe^{(ik/4\pi)S(M,A)} W_{K}(A).$$ It
can be shown that up to limits of the heuristics, $Z(M,K)$ and $I(M^{3},K)$ are essentially equivalent for appropriate choice of gauge group and
corresponding spin networks.
\bigbreak

By these graphical reformulations, a three-dimensional $TQFT$ is, at base, a highly simplified theory of point particle interactions in $2+1$
dimensional spacetime. It can be used to articulate invariants of knots and links and invariants of three manifolds. The reader
interested in the
$SU(2)$ case of this structure and its implications for invariants of knots and three manifolds can consult \cite{KL,KP,Kohno,Crane,MS}. One expects that
physical situations involving
$2+1$ spacetime will be approximated by such an idealized theory.  There are also applications to $3 + 1$ quantum gravity \cite{ASR,AL,KaufLiko}.
Aspects of the quantum Hall effect may be related to topological quantum field theory
\cite{Wilczek}. One can study a physics in two dimensional space where the braiding of  particles or collective excitations leads to non-trival
representations of the Artin braid group. Such particles are called {\it Anyons}.  Such $TQFT$ models would describe applicable physics. One can
think about applications of anyons to quantum computing along the lines of the topoological models described here.  
\bigbreak

\begin{figure}
     \begin{center}
     \begin{tabular}{c}
     \includegraphics[height=6cm]{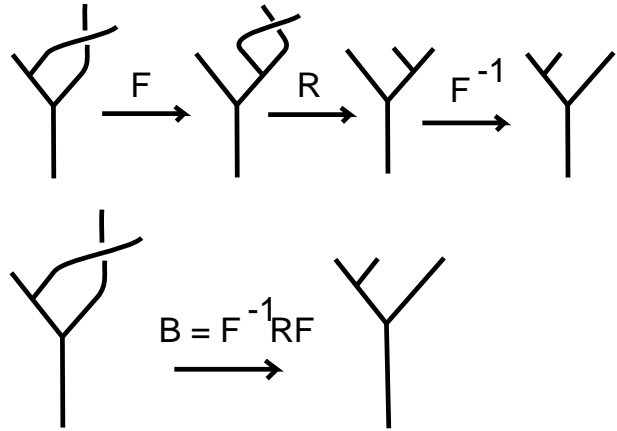}
     \end{tabular}
     \end{center}
     \caption{\bf A More Complex Braiding Operator   }
     \label{Figure  31 }
     \end{figure} 
     \bigbreak

A key point in the application of $TQFT$ to quantum information theory is contained in the 
structure illustrated in Figure~\ref{Figure 31 }. There we show a more complex braiding operator, based on the composition of recoupling with the
elementary braiding at a vertex. (This structure is implicit in the Hexagon identity of Figure~\ref{Figure 30 }.) The new braiding operator is a 
source of unitary representations of braid group in situations (which exist mathematically) where the recoupling transformations are themselves 
unitary. This kind of pattern is utilized in the work of Freedman and collaborators \cite{F,FR98,FLZ,Freedman5,Freedman6}
and in the case of classical angular momentum formalism has been dubbed a ``spin-network quantum simlator" by Rasetti and collaborators
\cite{MR,MR2}. In the next section we show how certain natural deformations \cite{KL} of Penrose spin networks \cite{Penrose} can be used 
to produce these unitary representations of the Artin braid group  and the corresponding models for anyonic topological quantum computation. 
\bigbreak

\section {Spin Networks and Temperley-Lieb Recoupling Theory}
In this section we discuss a combinatorial construction for spin networks that generalizes the original construction of Roger Penrose.
The result of this generalization is a structure that satisfies all the properties of a graphical $TQFT$ as described in the previous section, and 
specializes to classical angular momentum recoupling theory in the limit of its basic variable. The construction is based on the properties of 
the bracket polynomial (as already described in Section 11). A complete description of this theory can be found in the book ``Temperley-Lieb
Recoupling Theory and Invariants of Three-Manifolds" by Kauffman and Lins \cite{KL}.  
\bigbreak

\begin{figure}
     \begin{center}
     \begin{tabular}{c}
     \includegraphics[height=7cm]{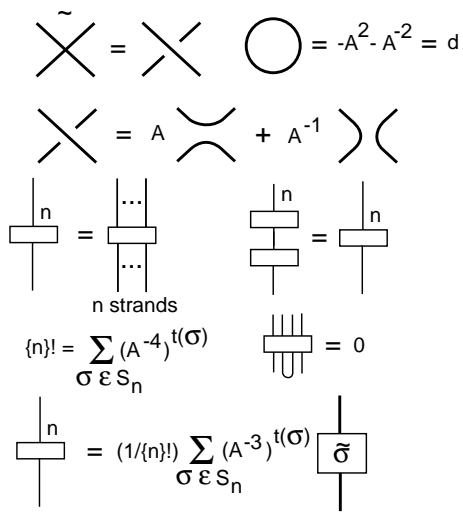}
     \end{tabular}
     \end{center}
     \caption{\bf Basic Projectors  }
     \label{Figure 32 }
     \end{figure}

\begin{figure}
     \begin{center}
     \begin{tabular}{c}
     \includegraphics[height=6cm]{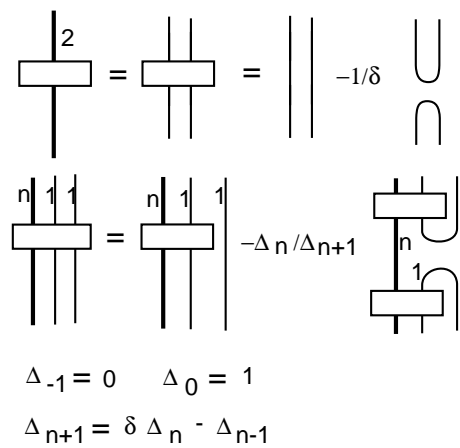}
     \end{tabular}
     \end{center}
     \caption{\bf Two Strand Projector   }
     \label{Figure 33 }
     \end{figure}

The ``$q$-deformed" spin networks that we construct here are based on the bracket polynomial relation. View Figure~\ref{Figure 32 } and Figure~\ref{Figure 33 }.
\bigbreak

     \bigbreak

     \bigbreak

\begin{figure}
     \begin{center}
     \begin{tabular}{c}
     \includegraphics[height=6cm]{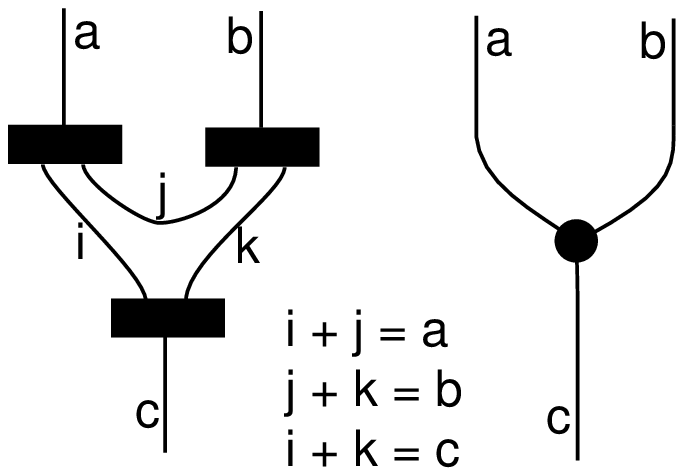}
     \end{tabular}
     \end{center}
     \caption{\bf Vertex   }
     \label{Figure  34 }
     \end{figure} 
     \bigbreak

In Figure~\ref{Figure 32 } we indicate how the basic projector 
\includegraphics[height=.5cm]{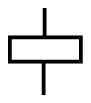} (symmetrizer, Jones-Wenzl projector)
is constructed on the basis of the
bracket polynomial expansion. In this technology a symmetrizer is a sum of tangles on $n$ strands (for a chosen integer $n$). The tangles are made by
summing over braid lifts of  permutations in the symmetric group on $n$ letters, as indicated in Figure~\ref{Figure 32 }. Each elementary braid is then expanded by the
bracket polynomial  relation as indicated in Figure~\ref{Figure 32 } so that the resulting sum  consists of flat tangles without any crossings (these can be viewed as
elements in the Temperley-Lieb algebra). The projectors have the property that the concatenation of a projector with itself is just that projector, and
if you tie two lines on the top or the bottom of a projector together, then the evaluation is zero. This general definition of projectors is very useful for 
this theory. The two-strand projector is shown in Figure~\ref{Figure 33 }. Here the formula for that projector 
is particularly simple. It is the sum of two parallel arcs and two turn-around arcs (with coefficient $-1/d,$  with $d = -A^{2} - A^{-2}$ is the loop
value for the bracket polynomial. Figure~\ref{Figure 33 } also shows the recursion formula for the general projector. This recursion formula is due to Jones and Wenzl and
the projector in this form, developed as a sum in the Temperley--Lieb algebra (see Section 12 of this paper), is usually known as the {\em Jones--Wenzl
projector}.
\bigbreak

The projectors are combinatorial analogs of irreducible representations of a group (the original spin nets were based
on $SU(2)$ and these deformed nets are based on the corresponding quantum group to SU(2)). As such the reader can think of them as ``particles". The
interactions of these particles are governed by how they can be tied together into three-vertices. See Figure~\ref{Figure 34 }.
In Figure~\ref{Figure 34 } we show how to tie three projectors, of $a,b,c$ strands respectively, together to form a three-vertex. In order to accomplish this 
interaction, we must share lines between them as shown in that figure so that there are non-negative integers $i,j,k$ so that
$a = i + j, b = j + k, c = i + k.$ This is equivalent to the condition that $a + b + c$ is even and that the sum of any two of $a,b,c$ is 
greater than or equal to the third. For example $a + b \ge c.$ One can think of the vertex as a possible particle interaction where
$[a]$ and $[b]$ interact to produce $[c].$ That is, any two of the legs of the vertex can be regarded as interacting to produce the third leg.
\bigbreak

\begin{figure}
     \begin{center}
     \begin{tabular}{c}
     \includegraphics[height=6cm]{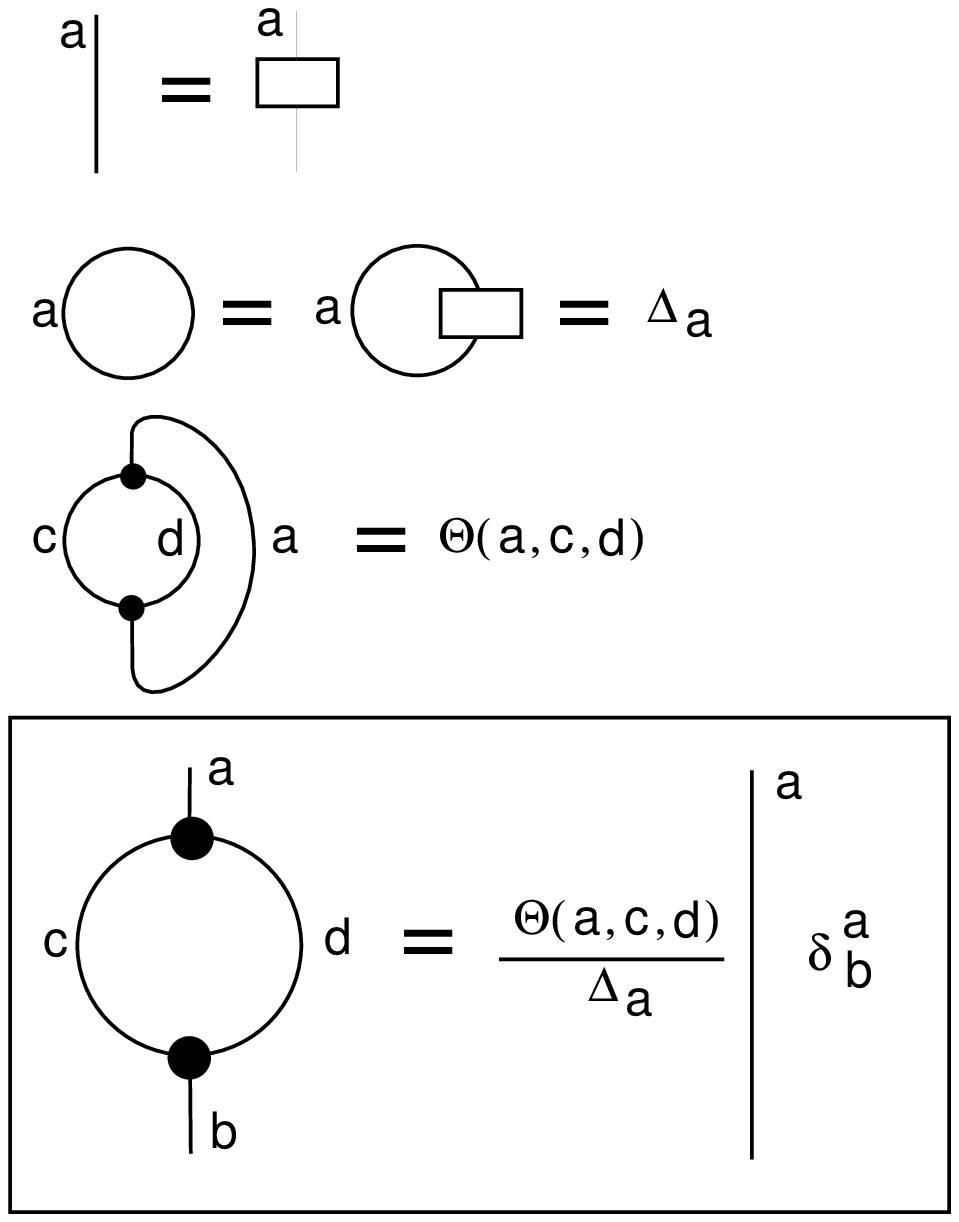}
     \end{tabular}
     \end{center}
     \caption{\bf Orthogonality of Trivalent Vertices   }
     \label{Figure 35 }
     \end{figure}

There is a basic orthogonality of three vertices as shown in Figure~\ref{Figure 35 }. Here if we tie two three-vertices together
so that they form a ``bubble" in the middle, then the resulting network with labels $a$ and $b$ on its free ends
is a multiple of an $a$-line (meaning a line with an $a$-projector on it) or zero (if $a$ is not equal to $b$).
The multiple is compatible with the results of closing the diagram in the equation of Figure~\ref{Figure 35 } so the two free
ends are identified with one another. On closure, as shown in the figure, the left hand side of the equation becomes
a Theta graph and the right hand side becomes a multiple of a ``delta" where $\Delta_{a}$ denotes the bracket 
polynomial evaluation of the $a$-strand loop with a projector on it. The $\Theta(a,b,c)$ denotes the bracket 
evaluation of a theta graph made from three trivalent vertices and labeled with $a, b, c$ on its edges.
\bigbreak

\begin{figure}
     \begin{center}
     \begin{tabular}{c}
     \includegraphics[height=4cm]{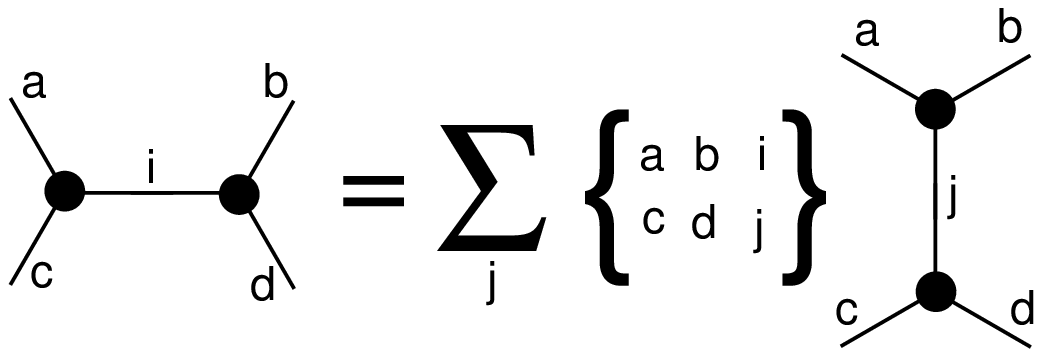}
     \end{tabular}
     \end{center}
     \caption{\bf Recoupling Formula  }
     \label{Figure 36 }
     \end{figure}

There is a recoupling formula in this theory in the form shown in Figure~\ref{Figure 36 }.
Here there are ``$6$-j symbols", recoupling coefficients that can be expressed, as shown in 
Figure~\ref{Figure 36 }, in terms of tetrahedral graph evaluations and theta graph evaluations. The tetrahedral graph is shown in 
Figure~\ref{Figure 37 }. One derives the formulas for 
these coefficients directly from the orthogonality relations for the trivalent vertices by 
closing the left hand side of the recoupling formula and using orthogonality to evaluate the right hand side.
This is illustrated in Figure~\ref{Figure 38 }. The reader should be advised that there are specific calculational formulas for the theta and tetrahedral nets.
These can be found in \cite{KL}. Here we are indicating only the relationships and external logic of these objects.

     \bigbreak

     \bigbreak

\begin{figure}
     \begin{center}
     \begin{tabular}{c}
     \includegraphics[height=4cm]{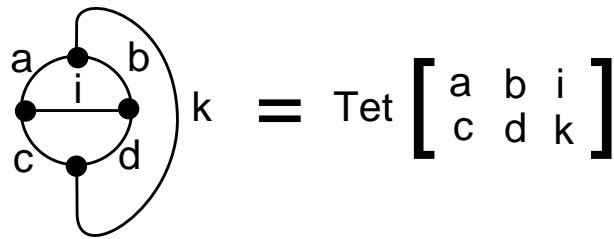}
     \end{tabular}
     \end{center}
     \caption{\bf Tetrahedron Network  }
     \label{Figure  37 }
     \end{figure} 
     \bigbreak

\begin{figure}
     \begin{center}
     \begin{tabular}{c}
     \includegraphics[height=8cm]{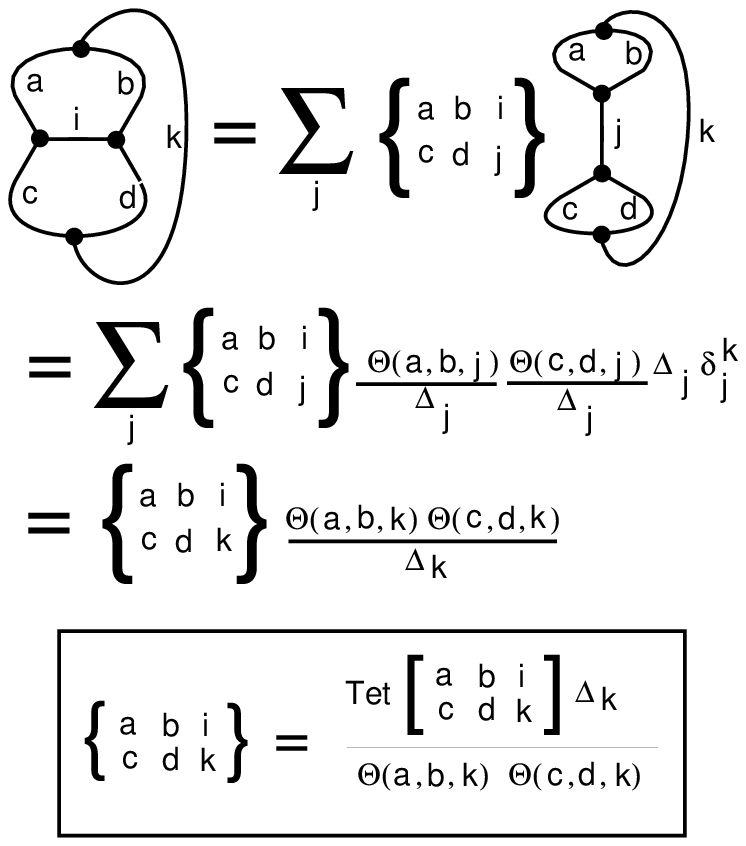}
     \end{tabular}
     \end{center}
     \caption{\bf Tetrahedron Formula for Recoupling Coefficients  }
     \label{Figure  38 }
     \end{figure} 
     \bigbreak

Finally, there is the braiding relation, as illustrated in Figure 36.

\begin{figure}
     \begin{center}
     \begin{tabular}{c}
     \includegraphics[height=6cm]{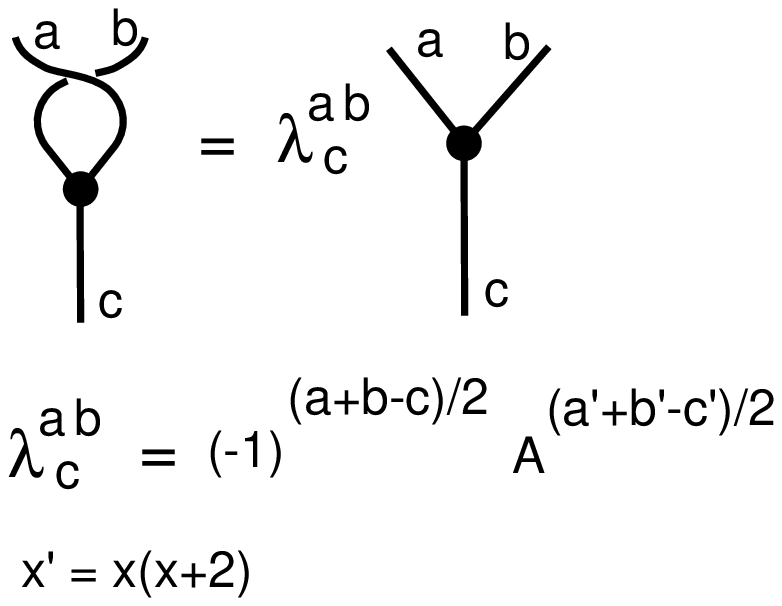}
     \end{tabular}
     \end{center}
     \caption{\bf Local Braiding Formula   }
     \label{Figure 39 }
     \end{figure} 
     \bigbreak

With the braiding relation in place, this $q$-deformed spin network theory satisfies the pentagon, hexagon and braiding naturality identities
needed for a topological quantum field theory. All these identities follow naturally from the basic underlying topological construction of the 
bracket polynomial. One can apply the theory to many different situations.

\subsection{Evaluations}
In this section we discuss the structure of the evaluations for $\Delta_{n}$ and the theta and tetrahedral networks. We refer to 
\cite{KL} for the details behind these formulas. Recall that $\Delta_{n}$ is the bracket evaluation of the closure of the $n$-strand
projector, as illustrated in Figure~\ref{Figure 35 }. For the bracket variable $A,$ one finds that 
$$\Delta_{n} = (-1)^{n}\frac{A^{2n+2} - A^{-2n-2}}{A^{2} - A^{-2}}.$$
One sometimes writes the {\it quantum integer}
$$[n] = (-1)^{n-1}\Delta_{n-1} = \frac{A^{2n} - A^{-2n}}{A^{2} - A^{-2}}.$$
If $$A=e^{i\pi/2r}$$ where $r$ is a positive integer, then 
$$\Delta_{n} = (-1)^{n}\frac{sin((n+1)\pi/r)}{sin(\pi/r)}.$$
Here the corresponding quantum integer is
$$[n] = \frac{sin(n\pi/r)}{sin(\pi/r)}.$$
Note that $[n+1]$ is a positive real number for $n=0,1,2,...r-2$ and that $[r-1]=0.$
\bigbreak

The evaluation of the theta net is expressed in terms of quantum integers by the formula
$$\Theta(a,b,c) = (-1)^{m + n + p}\frac{[m+n+p+1]![n]![m]![p]!}{[m+n]![n+p]![p+m]!}$$
where $$a=m+p, b=m+n, c=n+p.$$ Note that $$(a+b+c)/2 = m + n + p.$$
\bigbreak

When $A=e^{i\pi/2r},$ the recoupling theory becomes finite with the restriction that only three-vertices
(labeled with $a,b,c$) are {\it admissible} when $a + b +c \le 2r-4.$ All the summations in the 
formulas for recoupling are restricted to admissible triples of this form.
\bigbreak

\subsection{Symmetry and Unitarity}
The formula for the recoupling coefficients given in Figure~\ref{Figure 38 } has less symmetry than is actually inherent in the structure of the situation.
By multiplying all the vertices by an appropriate factor, we can reconfigure the formulas in this theory so that the revised recoupling transformation is
orthogonal, in the sense that its transpose is equal to its inverse. This is a very useful fact. It means that when the resulting matrices are real, then
the recoupling transformations are unitary. We shall see particular applications of this viewpoint later in the paper.
\bigbreak

\begin{figure}
     \begin{center}
     \begin{tabular}{c}
     \includegraphics[height=3cm]{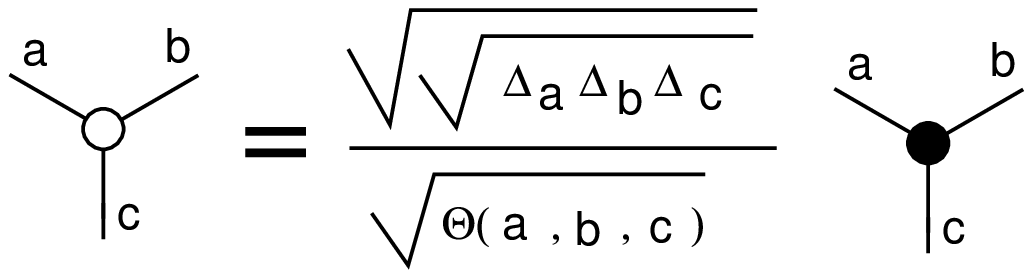}
     \end{tabular}
     \end{center}
     \caption{\bf Modified Three Vertex  }
     \label{Figure 40 }
     \end{figure}

Figure~\ref{Figure 40 } illustrates this modification of the three-vertex. Let $Vert[a,b,c]$ denote the original $3$-vertex of the Temperley-Lieb recoupling theory.
Let $ModVert[a,b,c]$ denote the modified vertex. Then we have the formula
$$ModVert[a,b,c] = \frac{\sqrt{\sqrt{\Delta_{a} \Delta_{b} \Delta_{c}}}}{ \sqrt{\Theta(a,b,c)}}\,\, Vert[a,b,c].$$

\noindent {\bf Lemma.}  For the bracket evaluation at the root of unity $A = e^{i\pi/2r}$ the factor
$$f(a,b,c) = \frac{\sqrt{\sqrt{\Delta_{a} \Delta_{b} \Delta_{c}}}}{ \sqrt{\Theta(a,b,c)}}$$
is real, and can be taken to be a positive real number for $(a,b,c)$ admissible (i.e. $a + b + c \le 2r -4$).
\bigbreak

\noindent {\bf Proof.} By the results from the previous subsection, 
$$\Theta(a,b,c) = (-1)^{(a+b+c)/2}\hat{\Theta}(a,b,c)$$ where $\hat{\Theta}(a,b,c)$ is positive real, and 
$$\Delta_{a} \Delta_{b} \Delta_{c} = (-1)^{(a+b+c)} [a+1][b+1][c+1]$$ where the quantum integers in this formula can be taken to be
positive real. It follows from this that
$$f(a,b,c) = \sqrt{\frac{\sqrt{[a+1][b+1][c+1]}}{\hat{\Theta}(a,b,c)}},$$ showing that this factor can be taken to be positive real. //
\bigbreak

\begin{figure}
     \begin{center}
     \begin{tabular}{c}
     \includegraphics[height=7cm]{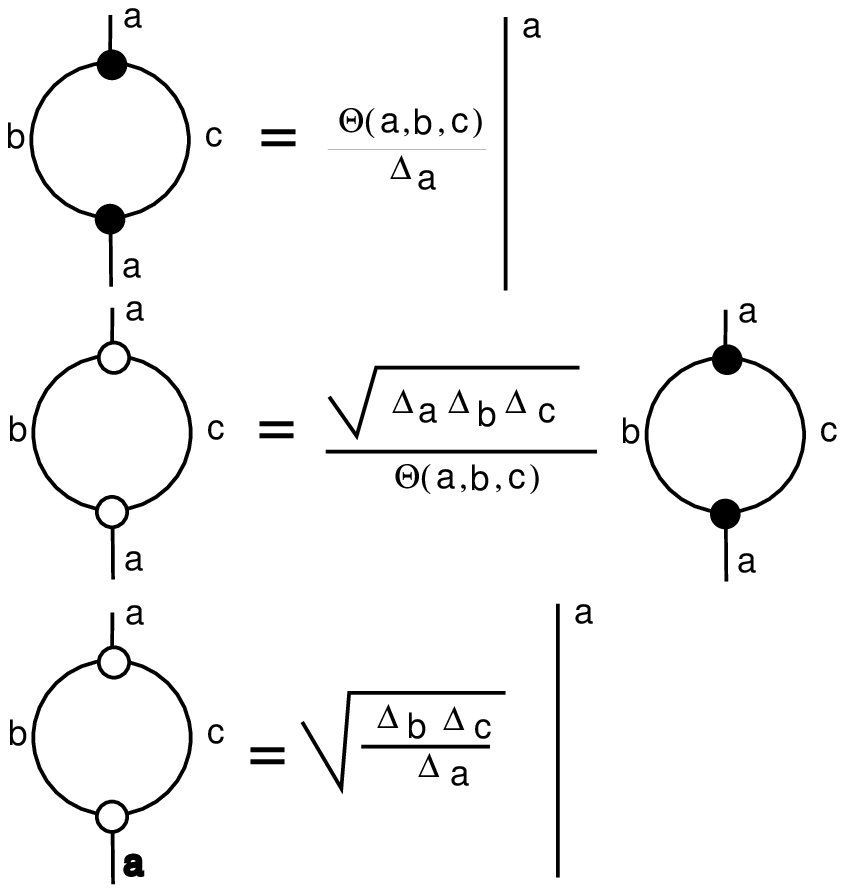}
     \end{tabular}
     \end{center}
     \caption{\bf Modified Bubble Identity   }
     \label{Figure 41 }
     \end{figure}

In Figure~\ref{Figure 41 } we show how this modification of the vertex affects the non-zero term of the orthogonality of trivalent
vertices (compare with Figure~\ref{Figure 35 }). We refer to this as the ``modified bubble identity." The coefficient in the modified bubble identity is
$$\sqrt{ \frac{\Delta_{b}\Delta_{c}}{\Delta_{a}} } = (-1)^{(b+c-a)/2} \sqrt{\frac{[b+1][c+1]}{[a+1]}}$$ 
where $(a,b,c)$ form an admissible triple. In particular $b+c-a$ is even and hence this factor can be taken to be real.
\bigbreak

\begin{figure}
     \begin{center}
     \begin{tabular}{c}
     \includegraphics[height=7cm]{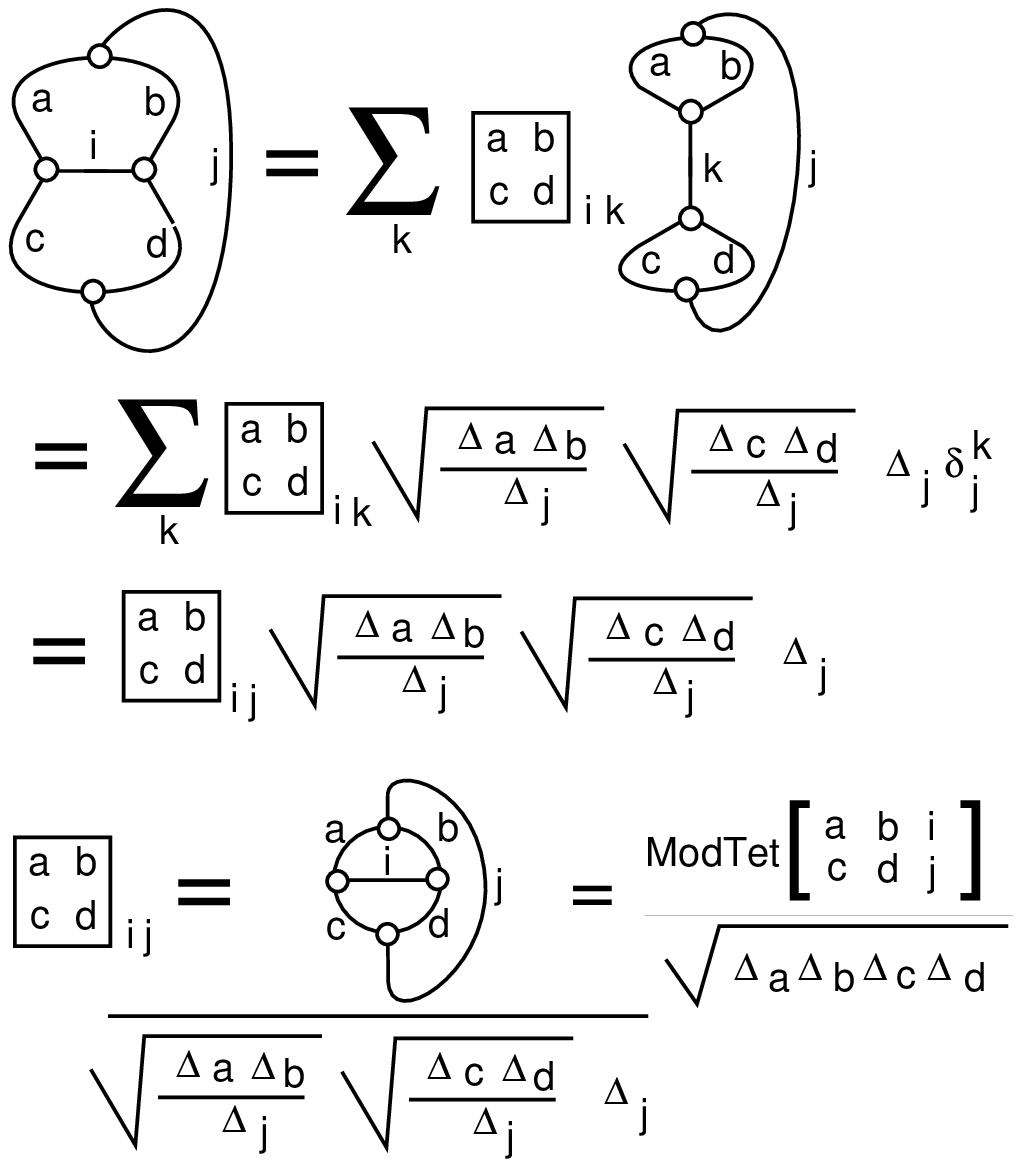}
     \end{tabular}
     \end{center}
     \caption{\bf Derivation of Modified Recoupling Coefficients }
     \label{Figure 42 }
     \end{figure}

\begin{figure}
     \begin{center}
     \begin{tabular}{c}
     \includegraphics[height=3cm]{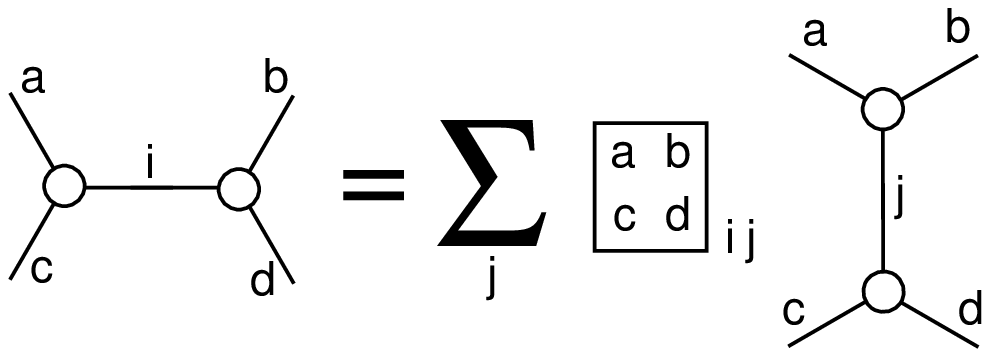}
     \end{tabular}
     \end{center}
     \caption{\bf Modified Recoupling Formula  }
     \label{Figure 43 }
     \end{figure}

We rewrite the recoupling formula in this new basis and emphasize 
that the recoupling coefficients can be seen (for fixed external labels $a,b,c,d$) as a matrix transforming the horizontal ``double-$Y$" basis
to a vertically
disposed double-$Y$ basis. In Figure~\ref{Figure 42 }, Figure~\ref{Figure 43 } and Figure~\ref{Figure 44 } we have shown the form of this transformation,using the matrix notation
$$M[a,b,c,d]_{ij}$$ for the modified recoupling coefficients. In Figure~\ref{Figure 42 } we derive an explicit formula for these matrix elements. The proof of this 
formula follows directly from trivalent--vertex orthogonality (See Figure~\ref{Figure 35 } and Figure~\ref{Figure 38 }.), and is given in Figure~\ref{Figure 42 }. The result shown in Figure~\ref{Figure 42 } and
Figure~\ref{Figure 43 } is the  following formula for the recoupling matrix elements.
$$M[a,b,c,d]_{ij} = ModTet
\left( \begin{array}{ccc}
a &  b & i \\
c & d & j \\
\end{array} \right)/\sqrt{\Delta_{a}\Delta_{b}\Delta_{c}\Delta_{d}}$$
where $\sqrt{\Delta_{a}\Delta_{b}\Delta_{c}\Delta_{d}}$ is short-hand for the product
$$\sqrt{ \frac{\Delta_{a}\Delta_{b}}{\Delta_{j}} }\sqrt{ \frac{\Delta_{c}\Delta_{d}}{\Delta_{j}} } \Delta_{j}$$ 
$$= (-1)^{(a+b-j)/2}(-1)^{(c+d-j)/2} (-1)^{j} \sqrt{ \frac{[a+1][b+1]}{[j+1]}}\sqrt{ \frac{[c+1][d+1]}{[j+1]}} [j+1]$$
$$ = (-1)^{(a+b+c+d)/2}\sqrt{[a+1][b+1][c+1][d+1]}$$
In this form, since
$(a,b,j)$ and $(c,d,j)$ are admissible triples, we see that this coeffient can be taken to be real, and its value is
independent of the choice of $i$ and $j.$
The matrix $M[a,b,c,d]$ is real-valued.
\bigbreak

\noindent It follows from Figure~\ref{Figure 36 } (turn the diagrams by ninety degrees) that 
$$M[a,b,c,d]^{-1} = M[b,d,a,c].$$
In Figure~\ref{Figure 45 } we illustrate the formula
$$M[a,b,c,d]^{T} = M[b,d,a,c].$$ It follows from this formula that 
$$M[a,b,c,d]^{T} = M[a,b,c,d]^{-1}.$$ {\it Hence $M[a,b,c,d]$ is an orthogonal, real-valued matrix.}

     \bigbreak

     \bigbreak

     \bigbreak

     \bigbreak

\begin{figure}
     \begin{center}
     \begin{tabular}{c}
     \includegraphics[height=6cm]{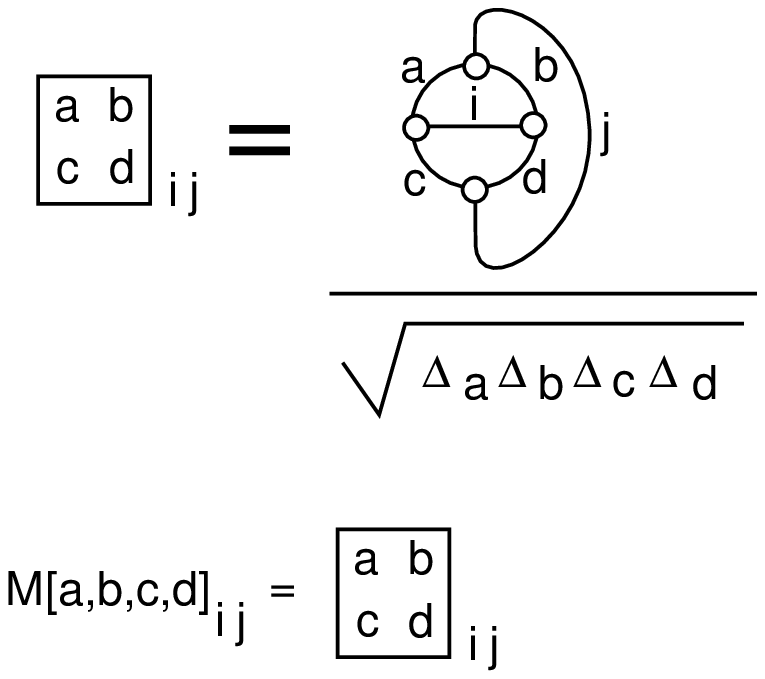}
     \end{tabular}
     \end{center}
     \caption{\bf  Modified Recoupling Matrix  }
     \label{Figure  44 }
     \end{figure} 
     \bigbreak

\begin{figure}
     \begin{center}
     \begin{tabular}{c}
     \includegraphics[height=6cm]{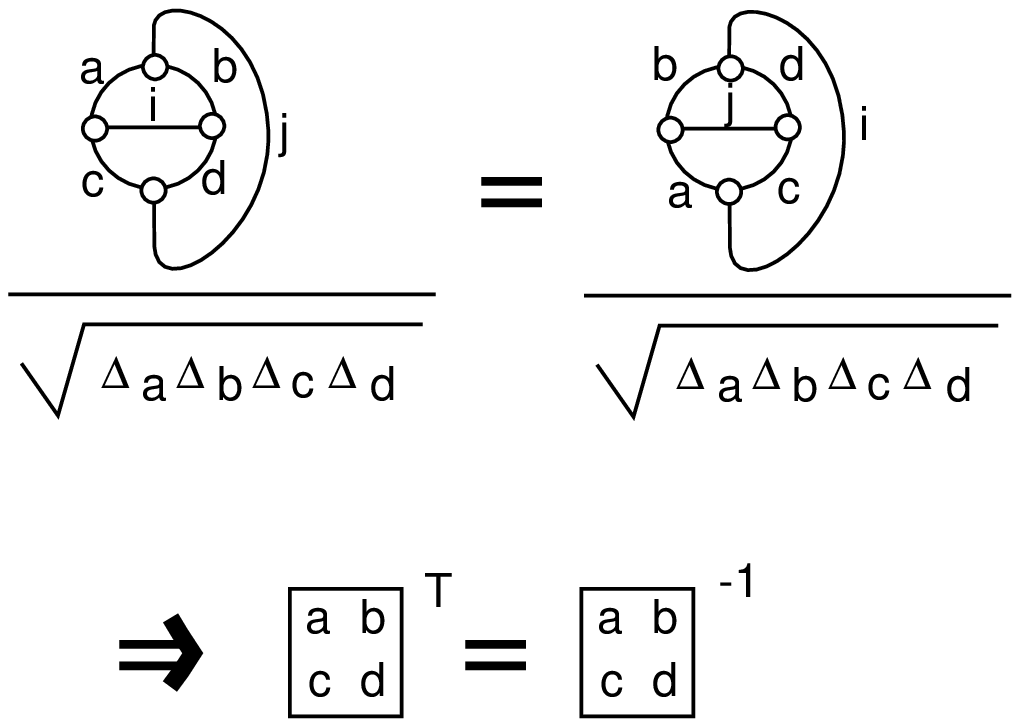}
     \end{tabular}
     \end{center}
     \caption{\bf  Modified Matrix Transpose   }
     \label{Figure  45 }
     \end{figure} 
     \bigbreak

\noindent {\bf Theorem.} In the Temperley-Lieb theory we obtain unitary (in fact real orthogonal) recoupling transformations when the bracket
variable $A$ has the form $A = e^{i\pi/2r}$ for $r$ a positive integer. Thus we obtain families of unitary representations of the Artin braid group
from the recoupling  theory at these roots of unity. 
\bigbreak

\noindent {\bf Proof.} The proof is given the discussion above.  //
\bigbreak

In Section 16 we shall show explictly how these methods work in the case of the Fibonacci model where $A = e^{3i\pi/5}$.  
\bigbreak

\section {Fibonacci Particles}

In this section and the next we detail how the Fibonacci model for anyonic quantum computing \cite{Kitaev,Preskill} can be constructed by using a version of
the two-stranded  bracket polynomial and a generalization of Penrose spin networks. This is a fragment of the Temperly-Lieb recoupling theory \cite{KL}. We
already gave in the preceding sections a general discussion of the theory of spin networks and their relationship with quantum computing.
\bigbreak

The Fibonacci model is a $TQFT$ that is based on a single ``particle" with two states that we shall call the {\it marked state} and the 
{\it unmarked state}. The particle in the marked state can interact with itself either to produce a single particle in the marked state, or 
to produce a single particle in the unmarked state. The particle in the unmarked state has no influence in interactions (an unmarked state interacting
with any state $S$ yields that state $S$). 
One way to indicate these two interactions symbolically is to use a box,for the marked state and a blank space for the unmarked state.
Then one has two modes of interaction of a box with itself:
\begin{enumerate}
\item Adjacency: $\fbox{~} ~~ \fbox{~}$
\smallbreak
\noindent and 
\item Nesting: $\fbox{ \fbox{~~} }.$
\end{enumerate}

\noindent With this convention we take the adjacency interaction to yield a single box, and the nesting interaction to produce nothing:

$$\fbox{~} ~~ \fbox{~} = \fbox{~}$$
$$\fbox{ \fbox{~~} } =  $$

\noindent We take the notational opportunity to denote nothing by an asterisk (*). The syntatical rules for operating the asterisk are
Thus the asterisk is a stand-in for no mark at all and it can be erased or placed wherever it is convenient to do so.
Thus $$\fbox{ \fbox{~~} } = *. $$

\begin{figure}
     \begin{center}
     \begin{tabular}{c}
     \includegraphics[height=4cm]{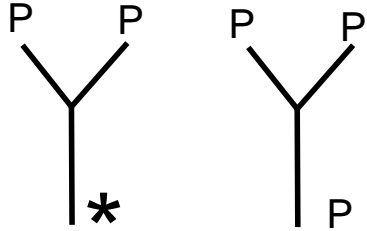}
     \end{tabular}
     \end{center}
     \caption{\bf Fibonacci Particle Interaction   }
     \label{Figure 46 }
     \end{figure} 
     \bigbreak

We shall make a recoupling theory based on this particle, but it is worth noting some of its purely combinatorial properties first.
The arithmetic of combining boxes (standing for acts of distinction) according to these rules has been studied and formalized in 
\cite{LOF} and correlated with Boolean algebra and classical logic. Here {\em within} and {\em next to} are ways to refer to the two sides delineated by
the given distinction. From this point of view, there are two modes of relationship (adjacency and nesting) that arise at once in the presence of a
distinction.  
\bigbreak

\begin{figure}
     \begin{center}
     \begin{tabular}{c}
     \includegraphics[height=6cm]{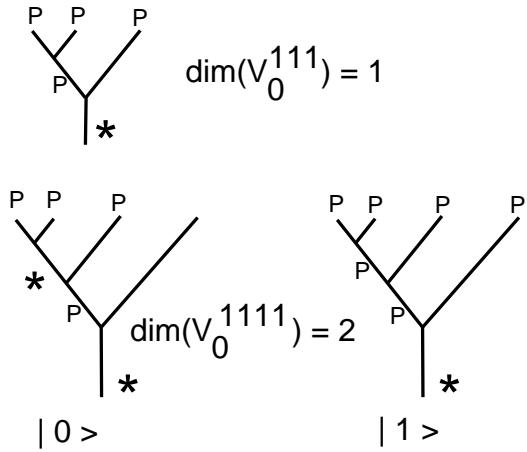}
     \end{tabular}
     \end{center}
     \caption{\bf Fibonacci Trees   }
     \label{Figure 47 }
     \end{figure} 
     \bigbreak

From here on we shall denote the Fibonacii particle by the letter $P.$
Thus the two possible interactions of $P$ with itself are as follows.
\begin{enumerate}
\item $P,P \longrightarrow *$
\item $P,P \longrightarrow P$
\end{enumerate}

\noindent In Figure~\ref{Figure 47 } we indicate in small tree diagrams the two possible interactions of the particle  $P$ with itself.
In the first interaction the particle vanishes, producing the asterix. In the second interaction the particle
 a single copy of $P$ is produced. These are the two basic actions of a single distinction relative to itself, and they
constitute our formalism for this very elementary particle. 
\bigbreak

In Figure~\ref{Figure 47 }, we have indicated the different results of particle
processes where we begin with a left-associated tree structure with three branches, all marked and then four branches all marked.
In each case we demand that the particles interact successively to produce an unmarked particle in the end, at the root of the tree.
More generally one can consider a left-associated tree with $n$ upward branches and one root. Let $T(a_1,a_2, \cdots , a_n : b)$ denote such
a tree with particle labels $a_1, \cdots, a_n$ on the top and root label $b$ at the bottom of the tree. We consider all possible processes
(sequences of particle interactions) that start with the labels at the top of the tree, and end with the labels at the bottom of the tree.
Each such sequence is regarded as a basis vector in a complex vector space 
$$V^{a_1,a_2, \cdots , a_n}_{b}$$
associated with the tree. In the case where all the labels are marked at the top and the bottom label is unmarked, we shall denote this tree
by $$V^{111 \cdots 11}_{0} = V^{(n)}_{0}$$ where $n$ denotes the number of upward branches in the tree. We see from Figure~\ref{Figure 47 } that the dimension 
of $V^{(3)}_{0}$ is $1,$ and that $$dim(V^{(4)}_{0}) = 2.$$ This means that $V^{(4)}_{0}$ is a natural candidate in this context for the two-qubit 
space.
\bigbreak

\begin{figure}
     \begin{center}
     \begin{tabular}{c}
     \includegraphics[height=6cm]{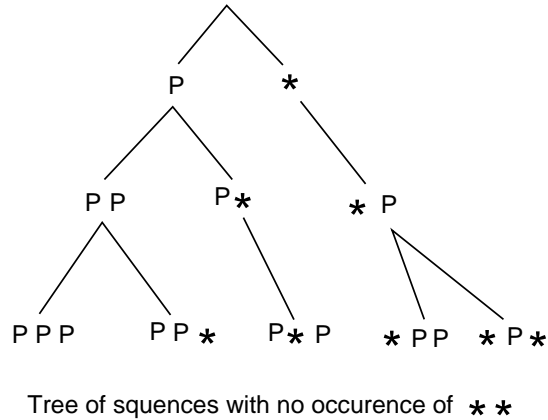}
     \end{tabular}
     \end{center}
     \caption{\bf Fibonacci Sequence   }
     \label{Figure 48 }
     \end{figure}

Given the tree $T(1,1,1,\cdots, 1:0)$ ($n$ marked states at the top, an unmarked state at the bottom), a process basis vector in $V^{(n)}_{0}$
is in direct correspondence with a string of boxes and asterisks ($1$'s and $0$'s) of length $n-2$ with no repeated asterisks and ending in a marked state.
See Figure~\ref{Figure 47 } for an illustration of the simplest cases. It follows from this
that $$dim(V^{(n)}_{0}) = f_{n-2}$$ where $f_k$ denotes the $k$-th Fibonacci number:
$$f_0 = 1, f_1 = 1, f_2 = 2, f_3 = 3, f_4 = 5, f_5= 8, \cdots$$ where $$f_{n+2} = f_{n+1} + f_{n}.$$
The dimension formula for these spaces follows from the fact that there are $f_{n}$ sequences of length $n-1$ of marked and unmarked states with no
repetition of an unmarked state. This fact is illustrated in Figure~\ref{Figure 48 }.
\bigbreak

     \bigbreak

\section{The Fibonacci Recoupling Model}
\begin{figure}
     \begin{center}
     \begin{tabular}{c}
     \includegraphics[height=6cm]{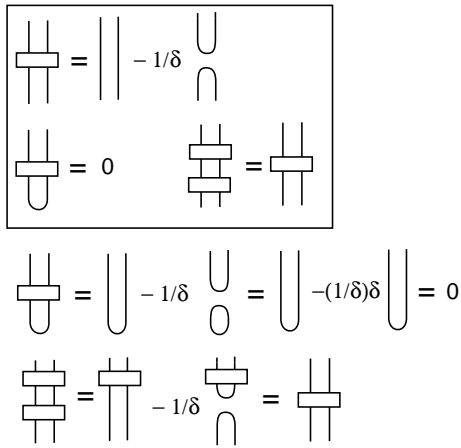}
     \end{tabular}
     \end{center}
     \caption{\bf The 2-Projector   }
     \label{Figure 49 }
     \end{figure}

We now show how to make a model for recoupling the Fibonacci particle by using the Temperley Lieb recoupling theory and the bracket polynomial.
Everything we do in this section will be based on the 2-projector, its properties and evaluations based on the bracket polynomial model for the Jones
polynomial. While we have outlined the general recoupling theory based on the bracket polynomial in earlier sections of this paper,
the present section is  self-contained, using only basic information about the bracket polyonmial, and the essential properties of the 
2-projector as shown in Figure~\ref{Figure 49 }. In this figure we state the definition of the 2-projector, list its two main properties (the operator is idempotent and
a self-attached strand yields a zero evaluation) and give diagrammatic proofs of these properties.

     \bigbreak

\begin{figure}
     \begin{center}
     \begin{tabular}{c}
     \includegraphics[height=6cm]{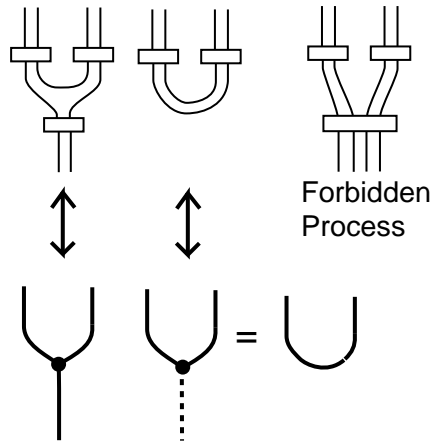}
     \end{tabular}
     \end{center}
     \caption{\bf Fibonacci Particle as 2-Projector  }
     \label{Figure 50 }
     \end{figure}

In Figure~\ref{Figure 50 }, we show the essence of the Temperley-Lieb recoupling model for the Fibonacci particle. The Fibonaccie particle is, in this mathematical
model, identified with the 2-projector itself. As the reader can see from Figure~\ref{Figure 50 }, there are two basic interactions of the 2-projector with itself, 
one giving a 2-projector, the other giving nothing. This is the pattern of self-iteraction of the Fibonacci particle. There is a third possibility,
depicted in Figure~\ref{Figure 50 }, where two 2-projectors interact to produce a 4-projector. We could remark at the outset, that the 4-projector will be zero if we
choose the bracket polynomial variable $A = e^{3 \pi/5}.$ Rather than start there, we will assume that the 4-projector is forbidden and deduce (below)
that the theory has to be at this root of unity.

     \bigbreak

\noindent Note that in Figure~\ref{Figure 50 } we have adopted a single strand notation for the particle interactions, with a solid strand corresponding to the
marked particle, a dotted strand (or nothing) corresponding to the unmarked particle.  A dark vertex indicates either an interaction point, or it
may be used to indicate the single strand is shorthand for two ordinary strands. Remember that these are all shorthand expressions for underlying
bracket polynomial calculations.
\bigbreak

In Figure~\ref{Figure 51}, Figure~\ref{Figure 52 }, Figure~\ref{Figure 53 }, Figure~\ref{Figure 54 }, Figure~\ref{Figure 55 } and Figure~\ref{Figure 56 } we have provided complete diagrammatic calculations of all of the relevant small nets and evaluations that 
are useful in the two-strand theory that is being used here. The reader may wish to skip directly to Figure~\ref{Figure 57 } where we determine the 
form of the recoupling coefficients for this theory. We will discuss the resulting algebra below.
\bigbreak

For the reader who does not want to skip the next collection of figures, here is a guided tour. 
Figure~\ref{Figure 51} illustrates three three basic nets in case of two 
strands. These are the theta, delta and tetrahedron nets. In this figure we have shown the decomposition on the theta and delta nets in terms of
2-projectors. The Tetrahedron net will be similarly decomposed in Figure~\ref{Figure 55 } and Figure~\ref{Figure 56 }. The theta net is denoted $\Theta,$ the delta by $\Delta,$ and the
tetrahedron by $T.$  In Figure~\ref{Figure 52 } we illustrate how a pedant loop has a zero evaluation. In Figure~\ref{Figure 53 } we use the identity in Figure~\ref{Figure 52 } to show how
an interior loop (formed by two trivalent vertices) can be removed and replaced by a factor of $\Theta/\Delta.$ Note how, in this figure, line two proves
that one network is a multiple of the other, while line three determines the value of the multiple by closing both nets.
\bigbreak

\noindent Figure~\ref{Figure 54 } illustrates the explicit calculation of the delta and theta nets. The figure begins with a calculation of the result of closing
a single strand of the 2-projector. The result is a single stand multiplied by $(\delta - 1/\delta)$ where $\delta = -A^2 - A^{-2},$ and $A$ is the bracket
polynomial parameter. We then find that $$\Delta = \delta^{2} - 1$$ and 
$$\Theta = (\delta - 1/\delta)^{2} \delta - \Delta/\delta = (\delta -1/\delta)(\delta^{2} - 2).$$  
\bigbreak

\noindent Figure~\ref{Figure 55 } and Figure~\ref{Figure 56 } illustrate the calculation of the value of the tetrahedral network $T.$ The reader should note the first line of 
Figure~\ref{Figure 55 } where the tetradedral net is translated into a pattern of 2-projectors, and simplified. The rest of these two figures are a diagrammatic
calculation, using the expansion formula for the 2-projector. At the end of Figure~\ref{Figure 56 } we obtain the formula for the tetrahedron
$$T = (\delta - 1/\delta)^{2}(\delta^{2} - 2) - 2\Theta/\delta.$$

\begin{figure}
     \begin{center}
     \begin{tabular}{c}
     \includegraphics[height=6cm]{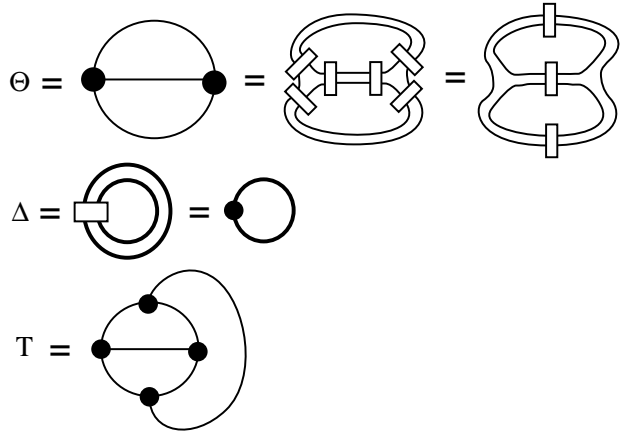}
     \end{tabular}
     \end{center}
     \caption{\bf  Theta, Delta and Tetrahedron  }
     \label{Figure  51}
     \end{figure} 
     \bigbreak

\begin{figure}
     \begin{center}
     \begin{tabular}{c}
     \includegraphics[height=6cm]{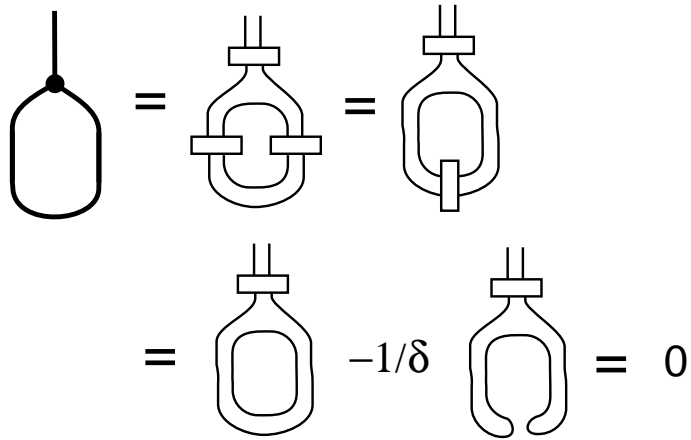}
     \end{tabular}
     \end{center}
     \caption{\bf LoopEvaluation--1   }
     \label{Figure 52 }
     \end{figure} 
     \bigbreak

\begin{figure}
     \begin{center}
     \begin{tabular}{c}
     \includegraphics[height=6cm]{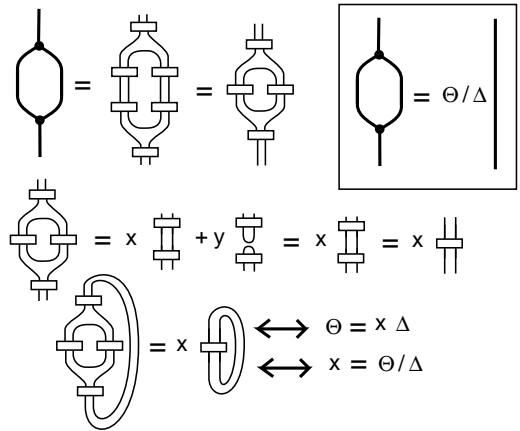}
     \end{tabular}
     \end{center}
     \caption{\bf LoopEvaluation--2   }
     \label{Figure 53 }
     \end{figure} 
     \bigbreak

\begin{figure}
     \begin{center}
     \begin{tabular}{c}
     \includegraphics[height=6cm]{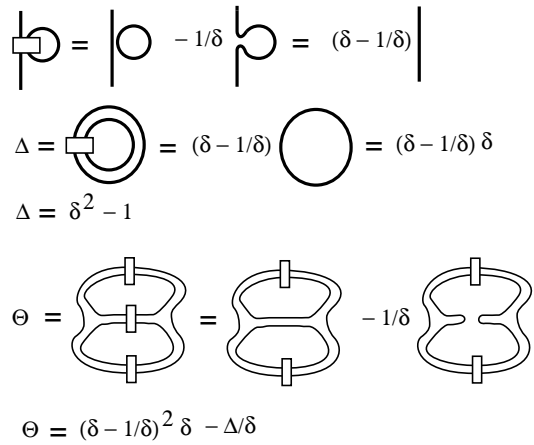}
     \end{tabular}
     \end{center}
     \caption{\bf  Calculate Theta, Delta  }
     \label{Figure 54 }
     \end{figure} 
     \bigbreak

\begin{figure}
     \begin{center}
     \begin{tabular}{c}
     \includegraphics[height=7cm]{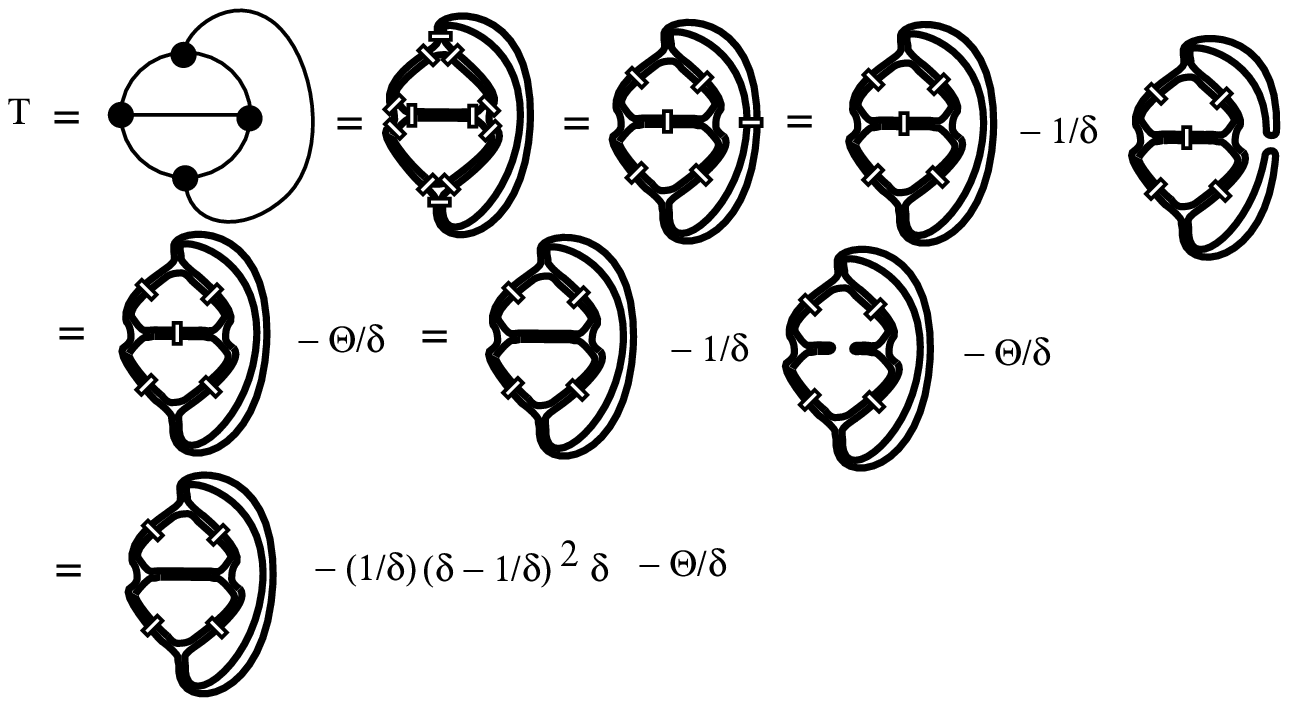}
     \end{tabular}
     \end{center}
     \caption{\bf  Calculate Tetrahedron -- 1   }
     \label{Figure 55 }
     \end{figure} 
     \bigbreak

\begin{figure}
     \begin{center}
     \begin{tabular}{c}
     \includegraphics[height=7cm]{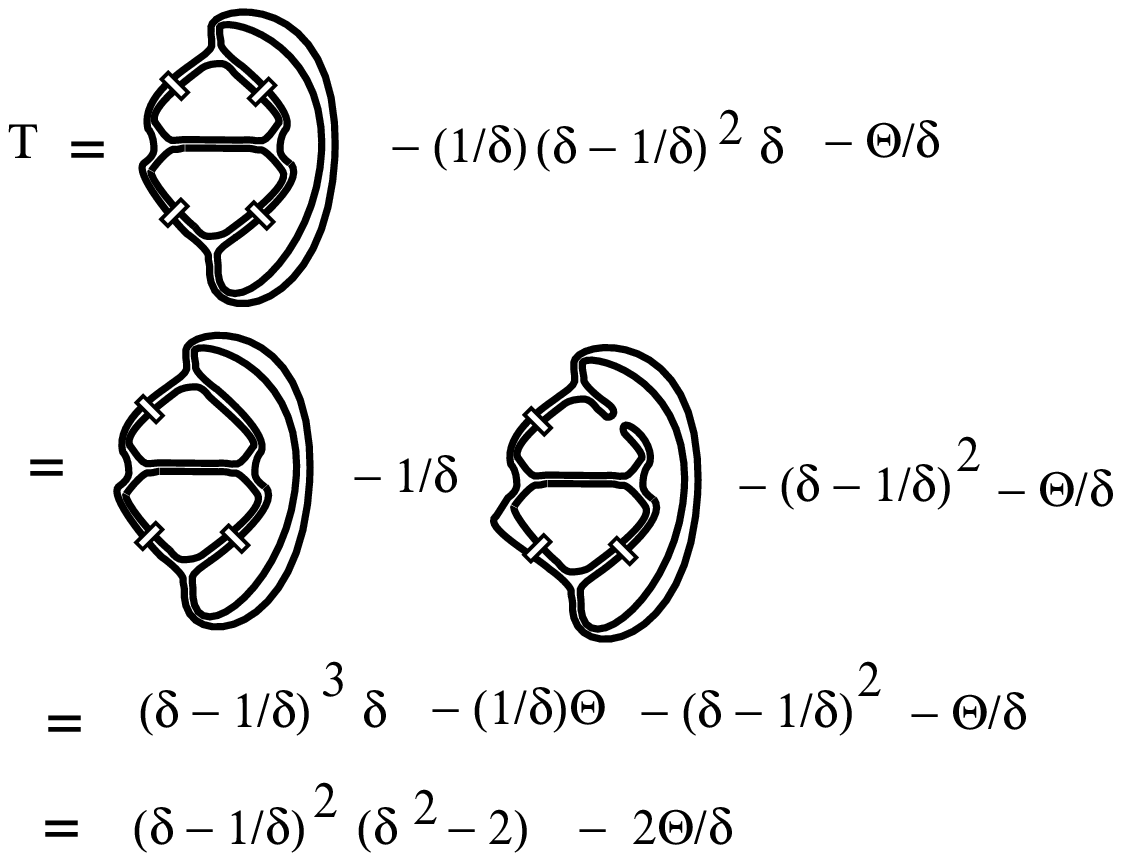}
     \end{tabular}
     \end{center}
     \caption{\bf Calculate Tetrahedron -- 2  }
     \label{Figure 56 }
     \end{figure} 
     \bigbreak

Figure~\ref{Figure 57 } is the key calculation for this model. In this figure we assume that the recoupling formulas involve only $0$ and $2$ strands, with 
$0$ corresponding to the null particle and $2$ corresponding to the 2-projector. ($2 + 2 = 4$ is forbidden as in Figure~\ref{Figure 50 }.) From this assumption we
calculate that the recoupling matrix is given by 
$$ F = 
\left( \begin{array}{cc}
a & b \\
c & d \\
\end{array} \right) =
\left( \begin{array}{cc}
1/\Delta & \Delta/\Theta \\
\Theta/\Delta^{2} & T \Delta/\Theta^{2} \\
\end{array} \right)
$$

\begin{figure}
     \begin{center}
     \begin{tabular}{c}
     \includegraphics[height=7cm]{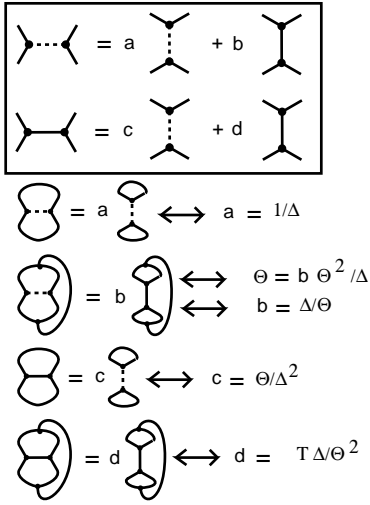}
     \end{tabular}
     \end{center}
     \caption{\bf Recoupling for 2-Projectors  }
     \label{Figure 57 }
     \end{figure} 
     \bigbreak

\begin{figure}
     \begin{center}
     \begin{tabular}{c}
     \includegraphics[height=7cm]{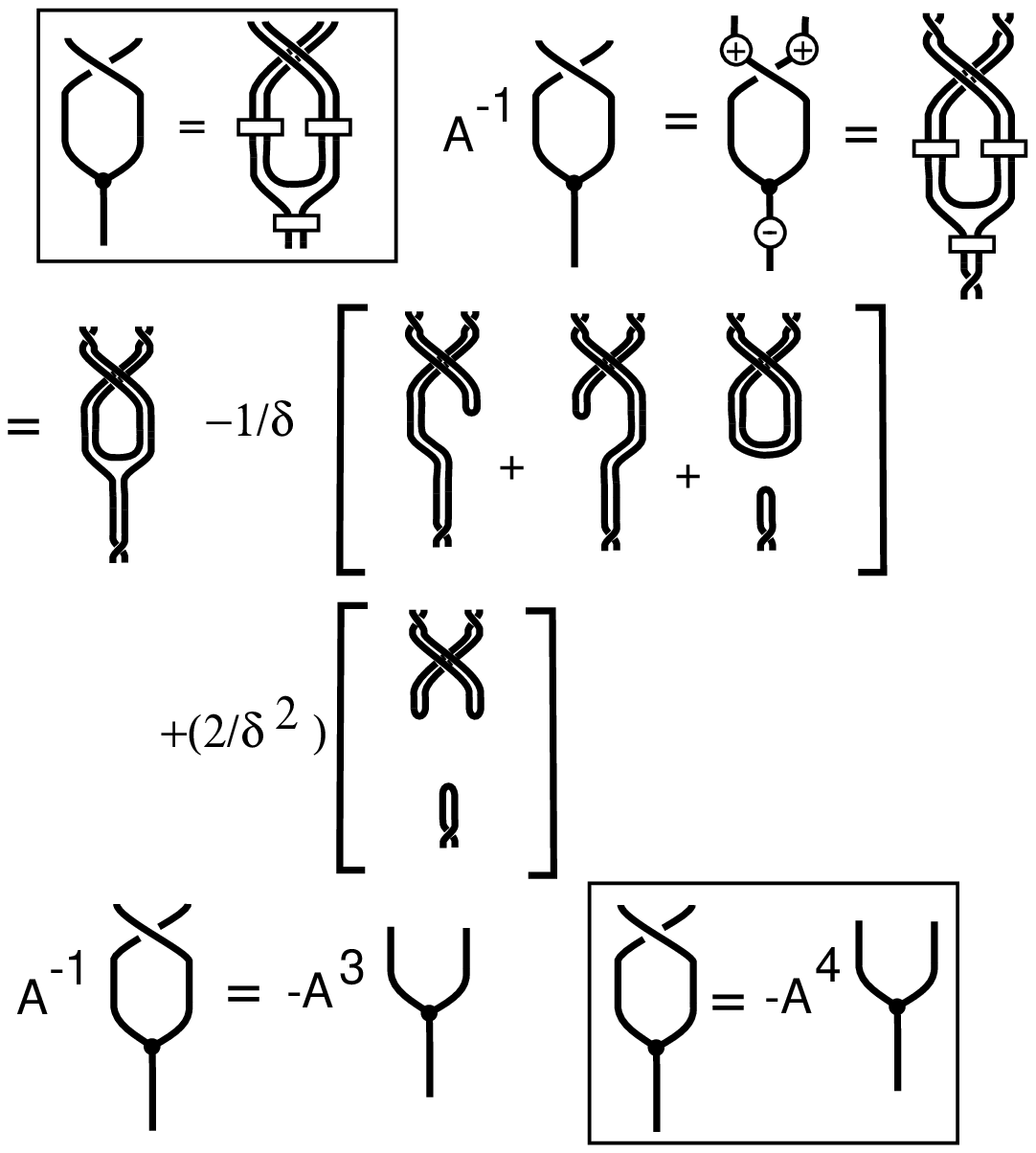}
     \end{tabular}
     \end{center}
     \caption{\bf Braiding at the Three-Vertex }
     \label{Figure 58 }
     \end{figure} 
     \bigbreak

\begin{figure}
     \begin{center}
     \begin{tabular}{c}
     \includegraphics[height=6cm]{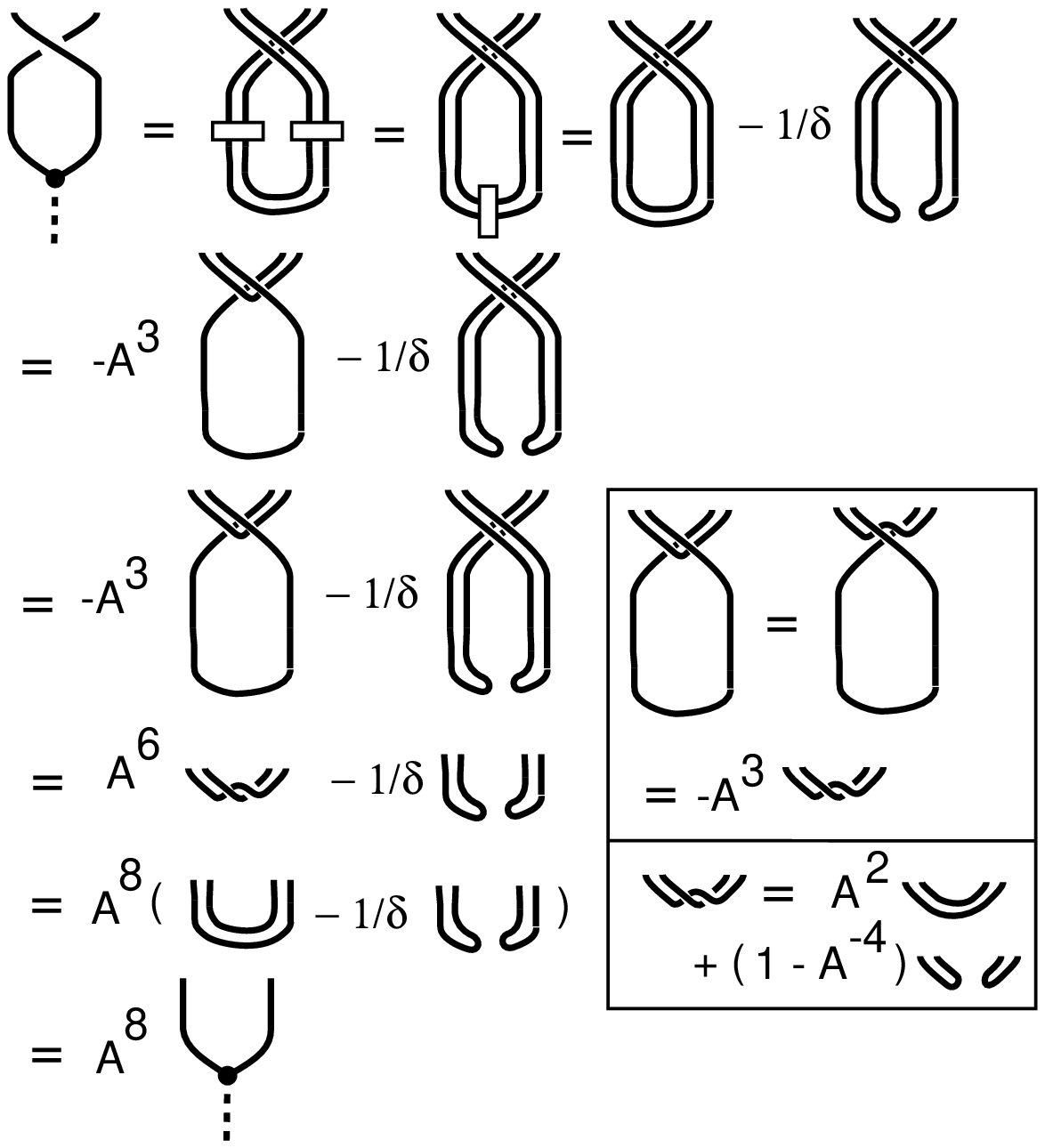}
     \end{tabular}
     \end{center}
     \caption{\bf Braiding at the Null-Three-Vertex  }
     \label{Figure  59 }
     \end{figure} 
     \bigbreak

\noindent Figure~\ref{Figure 58 } and Figure~\ref{Figure 59 } work out the exact formulas for the braiding at a three-vertex in this theory. When the 3-vertex has three marked lines,
then the braiding operator is multiplication by $-A^{4},$ as in Figure 58. When the 3-vertex has two marked lines, then the braiding operator is
multiplication by $A^{8},$ as shown in Figure~\ref{Figure 59 }.
\bigbreak

\noindent Notice that it follows from the symmetry of the diagrammatic recoupling formulas of Figure~\ref{Figure 57 } that 
{\it the square of the recoupling matrix $F$ is equal to 
the identity.} That is, 
$$\left( \begin{array}{cc}
1 & 0 \\
0 & 1 \\
\end{array} \right) = F^{2} =
\left( \begin{array}{cc}
1/\Delta & \Delta/\Theta \\
\Theta/\Delta^{2} & T \Delta/\Theta^{2} \\
\end{array} \right)
\left( \begin{array}{cc}
1/\Delta & \Delta/\Theta \\
\Theta/\Delta^{2} & T \Delta/\Theta^{2} \\
\end{array} \right) =$$ 
$$\left( \begin{array}{cc}
 1/\Delta^{2} + 1/\Delta & 1/\Theta + T\Delta^{2}/\Theta^{3} \\
\Theta/\Delta^{3} + T/(\Delta\Theta) & 1/\Delta + \Delta^{2} T^{2}/\Theta^{4} \\
\end{array} \right).$$
Thus we need the relation
$$1/\Delta + 1/\Delta^{2} = 1.$$
This is equivalent to saying that 
$$\Delta^{2} = 1 + \Delta,$$ a quadratic equation whose solutions are
$$\Delta = (1 \pm \sqrt{5})/2.$$
Furthermore, we know that $$\Delta = \delta^{2} - 1$$ from Figure~\ref{Figure 54 }.
Hence $$\Delta^{2} = \Delta + 1 = \delta^{2}.$$
We shall now specialize to the case where
$$\Delta = \delta = (1 + \sqrt{5})/2,$$
leaving the other cases for the exploration of the reader.
We then take $$A = e^{3\pi i/5}$$ so that 
$$\delta = -A^{2} - A^{-2} = -2cos(6\pi/5) = (1 + \sqrt{5})/2.$$
\bigbreak

Note that $\delta - 1/\delta = 1.$ Thus
$$\Theta = (\delta - 1/\delta)^{2} \delta - \Delta/\delta = \delta - 1.$$
and
$$T = (\delta - 1/\delta)^{2}(\delta^{2} - 2) - 2\Theta/\delta = (\delta^{2} - 2) - 2(\delta - 1)/\delta$$
$$= (\delta - 1)(\delta -2)/\delta = 3\delta - 5.$$
Note that $$T = -\Theta^{2}/\Delta^{2},$$ from which it follows immediately that 
$$F^{2} = I.$$ This proves that we can satisfy this model when $\Delta = \delta = (1 + \sqrt{5})/2.$
\bigbreak

\noindent For this specialization we see that the matrix $F$ becomes
$$ F = 
\left( \begin{array}{cc}
1/\Delta & \Delta/\Theta \\
\Theta/\Delta^{2} & T \Delta/\Theta^{2} \\
\end{array} \right) =
\left( \begin{array}{cc}
1/\Delta & \Delta/\Theta \\
\Theta/\Delta^{2} & (-\Theta^{2}/\Delta^{2}) \Delta/\Theta^{2} \\
\end{array} \right) =
\left( \begin{array}{cc}
1/\Delta & \Delta/\Theta \\
\Theta/\Delta^{2} & -1/\Delta \\
\end{array} \right)$$
This version of $F$ has square equal to the identity  independent of the value of $\Theta,$ so long as $\Delta^{2} = \Delta + 1.$
\bigbreak

\noindent {\bf The Final Adjustment.} Our last version of $F$ suffers from a lack of symmetry. It is not a symmetric matrix, and hence
not unitary. A final adjustment of the model gives this desired symmetry. {\it Consider the result of replacing each trivalent vertex (with three 2-projector
strands) by a multiple by a given quantity $\alpha.$} Since the $\Theta$ has two vertices, it will be multiplied by $\alpha^{2}.$ Similarly,
the tetradhedron $T$ will be multiplied by $\alpha^{4}.$ The $\Delta$ and the $\delta$ will be unchanged. Other properties of the model will remain 
unchanged. The new recoupling matrix, after such an adjustment is made, becomes
$$\left( \begin{array}{cc}
1/\Delta & \Delta/\alpha^{2}\Theta \\
\alpha^{2}\Theta/\Delta^{2} & -1/\Delta \\
\end{array} \right)$$
For symmetry we require $$\Delta/(\alpha^{2}\Theta) = \alpha^{2}\Theta/\Delta^{2}.$$ We take $$\alpha^{2} = \sqrt{\Delta^{3}}/\Theta.$$
With this choice of $\alpha$ we have $$\Delta/(\alpha^{2}\Theta) = \Delta \Theta/(\Theta \sqrt{\Delta^{3}}) = 1/\sqrt{\Delta}.$$
Hence the new symmetric $F$ is given by the equation
$$F =
\left( \begin{array}{cc}
1/\Delta & 1/\sqrt{\Delta} \\
1/\sqrt{\Delta} & -1/\Delta \\
\end{array} \right) =
\left( \begin{array}{cc}
\tau & \sqrt{\tau} \\
\sqrt{\tau} & -\tau \\
\end{array} \right)$$
where $\Delta$ is the golden ratio and $\tau = 1/\Delta$.
This gives the Fibonacci model. Using Figure~\ref{Figure 58 } and Figure~\ref{Figure 59 }, we have that the local braiding matrix for the model is given by the formula
below with $A = e^{3\pi i/5}.$
$$R = 
\left( \begin{array}{cc}
-A^{4} & 0 \\
0 & A^{8} \\
\end{array} \right)=
\left( \begin{array}{cc}
e^{4\pi i/5} & 0 \\
0 & -e^{2\pi i/5} \\
\end{array} \right).$$
\bigbreak

The simplest example of a braid group representation arising from this theory is the representation of the three strand braid group generated by
$S_{1}= R$ and $S_{2} = FRF$ (Remember that $F=F^{T} = F^{-1}.$). The matrices $S_{1}$ and $S_{2}$ are both unitary, and they generate a dense subset of
the unitary group $U(2),$ supplying the first part of the transformations needed for quantum computing.
\bigbreak

\section{Quantum Computation of Colored Jones Polynomials and the Witten-Reshetikhin-Turaev Invariant}
In this section we make some brief comments on the quantum computation of colored Jones polynomials. This material will be expanded in a subsequent
publication.

\begin{figure}
     \begin{center}
     \begin{tabular}{c}
     \includegraphics[height=8cm]{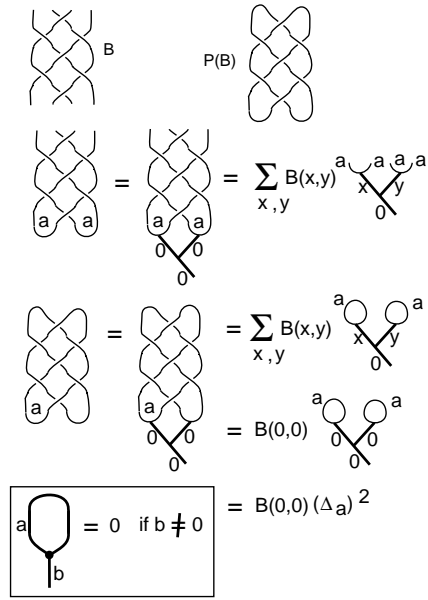}
     \end{tabular}
     \end{center}
     \caption{\bf  Evaluation of the Plat Closure of a Braid }
     \label{Figure 60 }
     \end{figure} 
     \bigbreak

First, consider Figure~\ref{Figure 60 }. In that figure we illustrate the calculation of the evalutation of the {\it ($a$) - colored bracket polynomial} for
the {\it plat closure} $P(B)$  of a braid $B$. 
The reader can infer the definition of the plat closure from Figure~\ref{Figure 60 }. One takes a braid on an even number of strands and closes the top strands with
each other in a row of maxima. Similarly, the bottom strands are closed with a row of minima. It is not hard to see that any knot or link can be represented
as the plat closure  of some braid. Note that in this figure we indicate the action of the braid group on the process spaces corresponding to the small
trees attached below the braids.
\bigbreak

The ($a$) - colored bracket polynonmial of a link $L$, denoted $<L>_{a},$ is the evaluation of that link where each single strand has been replaced by $a$
parallel strands and the insertion of Jones-Wenzl projector (as discussed in Section 14). We then see that we can use our discussion of the Temperley-Lieb
recoupling theory as in sections 14,15 and 16 to compute the value of the colored bracket polynomial for the plat closure $PB.$ As shown in Figure~\ref{Figure 60 }, we regard
the braid as  acting on a process space $V^{a,a,\cdots,a}_{0}$ and take the case of the action on the vector $v$ whose process space coordinates are all
zero. Then the action of the braid takes the form $$Bv(0,\cdots,0) = \Sigma_{x_{1},\cdots,x_{n}} B(x_{1},\cdots,x_{n}) v(x_{1},\cdots,x_{n})$$
where $B(x_{1},\cdots,x_{n})$ denotes the matrix entries for this recoupling transformation and $v(x_{1},\cdots,x_{n})$ runs over a basis for the 
space $V^{a,a,\cdots,a}_{0}.$ Here $n$ is even and equal to the number of braid strands. In the figure we illustrate with $n=4.$ Then, as the figure shows,
when we close the top of the braid action to form $PB,$ we cut the sum down to the evaluation of just one term. In the general case we will get
$$<PB>_{a} = B(0,\cdots,0)\Delta_{a}^{n/2}.$$ The calculation simplifies to this degree because of the vanishing of loops in the recoupling graphs.
The vanishing result is stated in Figure~\ref{Figure 60 }, and it is proved in the case $a =2$ in Figure~\ref{Figure 52 }.
\bigbreak

The {\it colored Jones polynomials} are normalized versions of the colored bracket polymomials, differing just by a normalization factor.
\bigbreak

In order to consider quantumn computation of the colored bracket or colored Jones polynomials, we therefore can consider quantum computation of the 
matrix entries $B(0,\cdots,0).$ These matrix entries in the case of the roots of unity $A = e^{i\pi/2r}$ and for the $a=2$ Fibonacci model with 
$A= e^{3i\pi/5}$ are parts of the diagonal entries of the unitary transformation that represents the braid group on the process space  $V^{a,a,\cdots,a}_{0}.$
{\it We can obtain these  matrix entries by using the Hadamard test as described in section 11.} As a result we get relatively efficient quantum
algorithms for the colored Jones polynonmials at these roots of unity, in essentially the same framework as we described in section 11, but for braids of 
arbitrary size. The computational complexity of these models is essentially the same as the models for the Jones polynomial discussed in \cite{Ah1}.
We reserve discussion of these issues to a subsequent publication.
\bigbreak

\begin{figure}
     \begin{center}
     \begin{tabular}{c}
     \includegraphics[height=6cm]{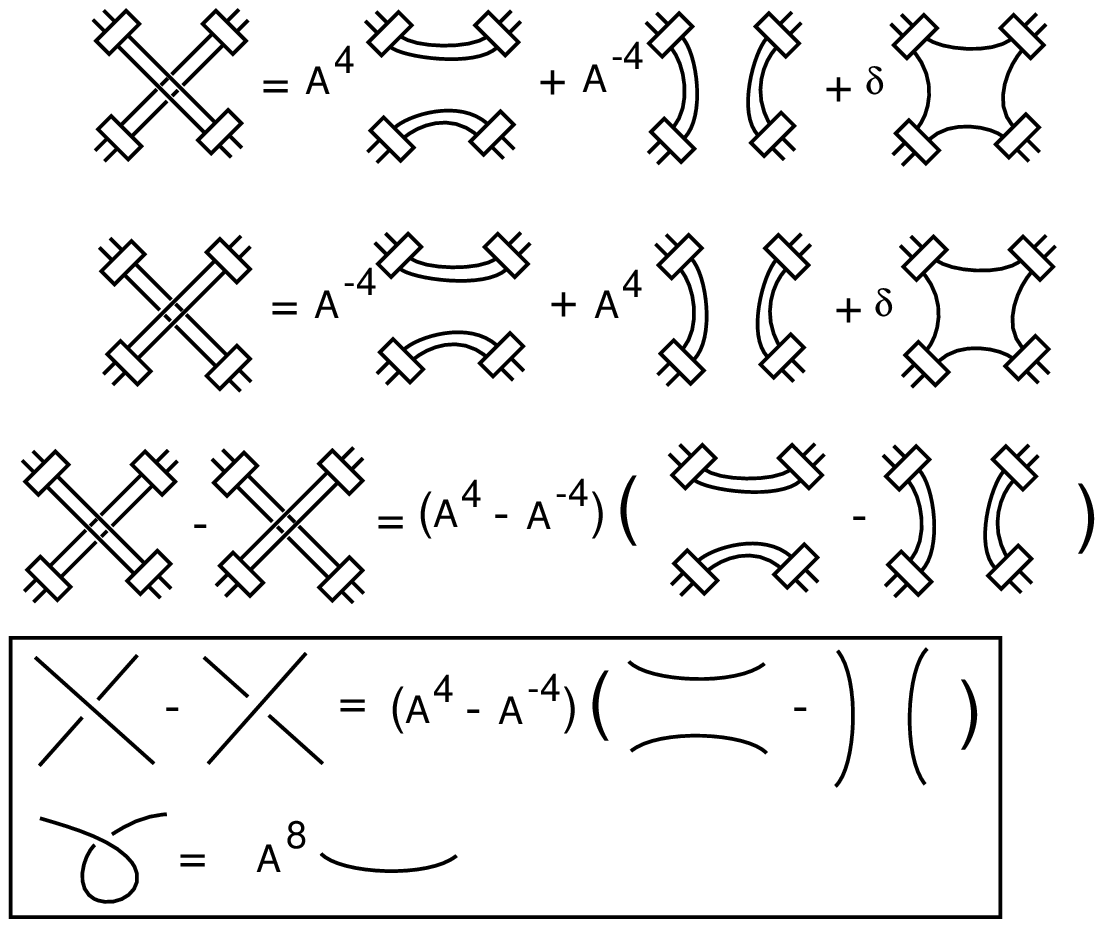}
     \end{tabular}
     \end{center}
     \caption{\bf Dubrovnik Polynomial Specialization at Two Strands   }
     \label{Figure 61 }
     \end{figure} 
     \bigbreak

It is worth remarking here that these algorithms give not only quantum algorithms for computing the colored bracket and Jones polynomials, but also for
computing the Witten-Reshetikhin-Turaev ($WRT$) invariants at the above roots of unity. The reason for this is that the $WRT$ invariant, in unnormalized
form is given as a finite sum of colored bracket polynomials:
$$WRT(L) = \Sigma_{a = 0}^{r-2} \Delta_{a} <L>_{a},$$
and so the same computation as shown in Figure~\ref{Figure 60 } applies to the $WRT.$ This means that we have, in principle, a quantum algorithm for the computation of the
Witten functional integral \cite{Witten} via this knot-theoretic combinatorial topology. It would be very interesting to understand a more direct approach to
such a computation via quantum field theory and functional integration.
\bigbreak

Finally, we note that in the case of the Fibonacci model, the ($2$)-colored bracket polynomial is a special case of the Dubrovnik version of the 
Kauffman polynomial \cite{IRI}. See Figure~\ref{Figure 61 } for diagammatics that resolve this fact. The skein relation for the Dubrovnik polynomial is boxed in this
figure. Above the box, we show how the double strands with projectors reproduce this relation. This observation means that in the Fibonacci model, the
natural underlying knot polynomial is a special evaluation of the Dubrovnik polynomial, and the Fibonacci model  can be used to perform quantum computation
for the values of this invariant.

\section {A Direct Construction of the Fibonacci Model}
In section 10 of this paper, we give elementary constructions for unitary representations of the three
strand braid group in $U(2).$ In section 11 we show how to use unitary representations of the three strand brand group to devise a quantum computation for the Jones polynomial. In this section we return to these considerations, and show how to construct the Fibonacci model by elementary means, without
using the recoupling theory that we have explained in the previous sections of the paper. This final approach is significant in that it shows an even closer relationship of the Fibonacci model with the Temperley Lieb algebra representation associated with the Jones polynomial.
\bigbreak

\begin{figure}
     \begin{center}
     \begin{tabular}{c}
     \includegraphics[height=4cm]{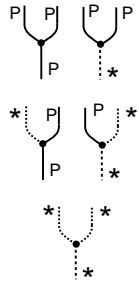}
     \end{tabular}
     \end{center}
     \caption{\bf The Fibonacci Particle $P$}
      \label{Figure 62 }
     \end{figure}

The constructions in this section are based on the combinatorics of the Fibonacci model. While we do not assume the recoupling theory of the previous sections, we essentially reconstruct its patterns for the particular purposes of the Fibonacci model. Recall that in the Fibonacci model we have a (mathematical) particle $P$
that interacts with itself either to produce $P$ or to produce a neutral particle $*$. If $X$ is any particle then $*$ interacts with $X$ to 
produce $X.$ Thus $*$ acts as an identity trasformation. These rules of interaction are illustrated in Figure~\ref{Figure 46 }, Figure~\ref{Figure 47 }, Figure~\ref{Figure 48 } and Figure~\ref{Figure 62 }.

     \bigbreak

\begin{figure}
     \begin{center}
     \begin{tabular}{c}
     \includegraphics[height=4cm]{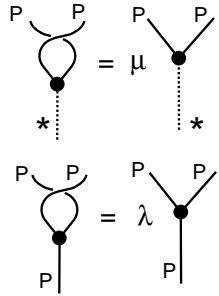}
     \end{tabular}
     \end{center}
     \caption{\bf Local Braiding}
     \label{Figure 63 }
     \end{figure} 
     \bigbreak

The braiding of two particles is measured in relation to their interaction. In Figure~\ref{Figure 63 } we illustrate braiding of $P$ with itself in relation
to the two possible interactions of $P$ with itself. If $P$ interacts to produce $*$, then the braiding gives a phase factor of $\mu.$ If $P$
interacts to produce $P$, then the braiding gives a phase factor of $\lambda.$ We assume at the outset that $\mu$ and $\lambda$ are unit complex
numbers. One should visualize these particles as moving in a plane and the diagrams of interaction are either creations of two particles from one particle,
or fusions of two particles to a single particle (depending on the choice of temporal direction). Thus we have a braiding matrix for these ``local"
particle interactions: 
$$R = 
\left( \begin{array}{cc}
\mu & 0 \\
0 & \lambda \\
\end{array} \right)$$ written with respect to the basis $\{ |* \rangle, |P \rangle \}$ for this space of particle interactions.
\bigbreak

\begin{figure}
     \begin{center}
     \begin{tabular}{c}
     \includegraphics[height=2cm]{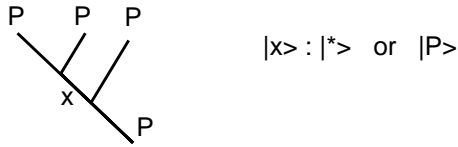}
     \end{tabular}
     \end{center}
     \caption{\bf Three Strands at Dimension Two}
     \label{Figure 64 }
     \end{figure}

\begin{figure}
     \begin{center}
     \begin{tabular}{c}
     \includegraphics[height=4cm]{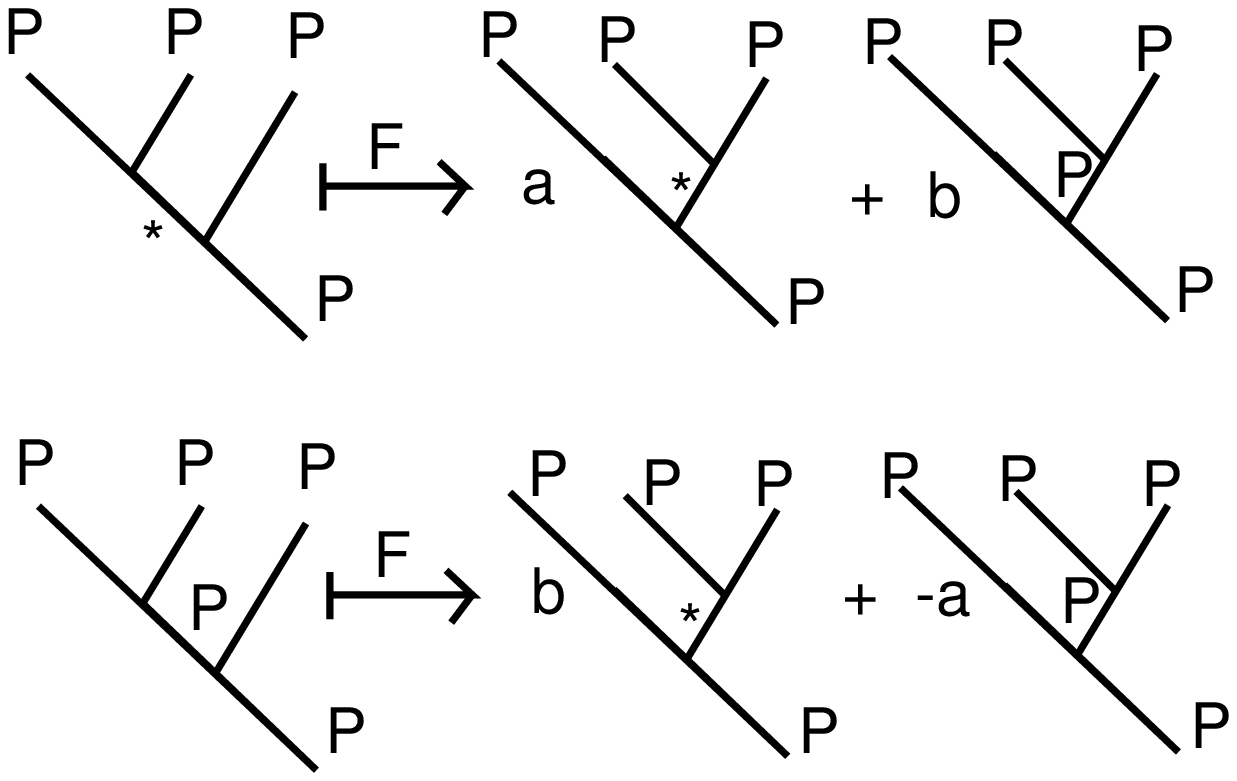}
     \end{tabular}
     \end{center}
     \caption{\bf Recoupling Formula}
     \label{Figure 65 }
     \end{figure}

We want to make this braiding matrix part of a larger representation of the braid group. In particular, we want a representation of the three-strand 
braid group on the process space $V_{3}$ illustrated in Figure~\ref{Figure 6 }. This space starts with three $P$ particles and considers processes associated
in the patttern $(PP)P$ with the stipulation that the end product is $P$. The possible pathways are illustrated in Figure~\ref{Figure 64 }. They correspond
to $(PP)P \longrightarrow (*)P \longrightarrow P$ and $(PP)P \longrightarrow (P)P \longrightarrow P.$ This process space has dimension two and can
support a second braiding generator for the second two strands on the top of the tree. In order to articulate the second braiding we change basis to the
process space corresponding to $P(PP)$ as shown in Figure~\ref{Figure 65 } and 
Figure~\ref{Figure 66 }. The change of basis is shown in Figure~\ref{Figure 66  }  and has matrix $F$ as shown below.
We want a unitary representation $\rho$ of three-strand braids so that $\rho(\sigma_{1}) = R$ and $\rho(\sigma_{2}) = S = F^{-1}RF.$ See Figure~\ref{Figure 66 }.
We take the form of the matrix $F$ as follows.
$$F = 
\left( \begin{array}{cc}
a & b \\
b & -a \\
\end{array} \right)$$ where $a^2 + b^2 = 1$ with $a$ and $b$ real. This form of the matrix for the basis change is determined by the requirement
that $F$ is symmetric with  $F^{2} = I$. The symmetry of the change of basis formula essentially demands that
$F^{2} = I.$  If $F$ is real, symmetric and $F^2 = I$, then $F$ is unitary.  Since
$R$ is unitary we see that $S = FRF$ is also unitary. Thus, if $F$ is constructed in this way then we obtain a unitary representation of $B_{3}.$
\bigbreak

Now we try to simultaneously construct an $F$ {\it and} construct a representation of the Temperley-Lieb algebra.
We begin by noting that 
$$R = 
\left( \begin{array}{cc}
\mu & 0 \\
0 & \lambda \\
\end{array} \right)
=\left( \begin{array}{cc}
\lambda & 0 \\
0 & \lambda \\
\end{array} \right) + 
\left( \begin{array}{cc}
\mu - \lambda & 0 \\
0 & 0 \\
\end{array} \right) =
\left( \begin{array}{cc}
\lambda & 0 \\
0 & \lambda \\
\end{array} \right) + 
\lambda^{-1} \left( \begin{array}{cc}
\delta & 0 \\
0 & 0 \\
\end{array} \right)$$ where $\delta = \lambda (\mu - \lambda).$
Thus $R = \lambda I + \lambda^{-1} U$ where 
$U = 
\left( \begin{array}{cc}
\delta & 0 \\
0 & 0 \\
\end{array} \right)$ so that $U^2 = \delta U.$ For the Temperley-Lieb representation, we want
$\delta = -\lambda^{2} - \lambda^{-2}.$ Hence we need
$- \lambda^{2} - \lambda^{-2} = \lambda (\mu - \lambda),$ which implies that $\mu = - \lambda^{-3}.$
With this restriction on $\mu,$ we have the Temperley-Lieb representation and the corresponding unitary braid group representation 
for $2$-strand braids and the $2$-strand Temperley-Lieb algebra.
\bigbreak

     \bigbreak

     \bigbreak

\begin{figure}
     \begin{center}
     \begin{tabular}{c}
     \includegraphics[height=8cm]{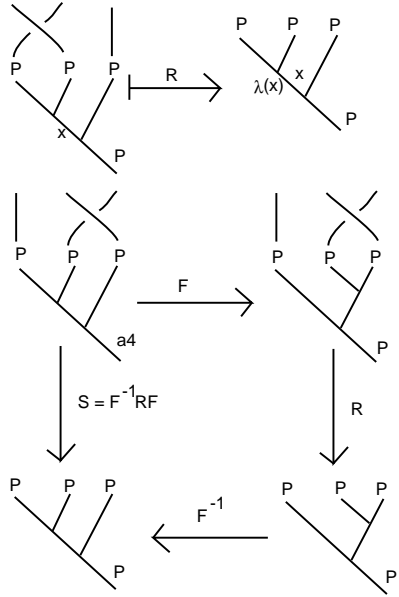}
     \end{tabular}
     \end{center}
     \caption{\bf Change of Basis}
     \label{Figure 66 }
     \end{figure} 
     \bigbreak

Now we can go on to $B_{3}$ and $TL_{3}$ via $S = FRF = \lambda I + \lambda^{-1} V$ with $V = FUF.$ We must examine
$V^{2}$, $UVU$ and $VUV.$ We find that 
$$V^{2} = FUFFUF = FU^{2}F = \delta FUF = \delta V,$$ as desired 	and
$$V = FUF = 
\left( \begin{array}{cc}
a & b \\
b & -a \\
\end{array} \right)
\left( \begin{array}{cc}
\delta & 0 \\
0 & 0 \\
\end{array} \right)
\left( \begin{array}{cc}
a & b \\
b & -a \\
\end{array} \right) =
\delta \left( \begin{array}{cc}
a^2 & ab \\
ab & b^2 \\
\end{array} \right).$$
Thus $V^2 = V$ and since $V = \delta |v\rangle \langle v|$ and $U =\delta |w\rangle \langle w|$ with $w = (1,0)^T$ and $v = Fw = (a,b)^T$
($T$ denotes transpose), we see that 
$$VUV = \delta^3 |v\rangle \langle v|w\rangle \langle w|v\rangle \langle v| = \delta^3 a^2 |v\rangle  \langle v|= \delta^2 a^2 V.$$
Similarly $UVU = \delta^2 a^2 U.$ Thus, we need $\delta^2 a^2 = 1$ and so we shall take $a = \delta^{-1}.$ With this choice, we have 
a representation of the Temperley-Lieb algebra $TL_{3}$ so that $\sigma_{1} = A I + A^{-1} U$ and 
$\sigma_{2} = A I + A^{-1} V$ gives a unitary representation of the braid group when $A = \lambda = e^{i \theta}$ and 
$b = \sqrt{1 - \delta^{-2}}$ is real. This last reality condition is equivalent to the inequality
$$cos^{2}(2 \theta) \ge \frac{1}{4},$$ which is satisfied for infinitely many values of $\theta$ in the ranges 
$$[0,\pi/6] \cup [\pi/3, 2 \pi/3] \cup [5 \pi/6, 7 \pi /6] \cup [4 \pi/3, 5\pi/3].$$
\bigbreak

\noindent With these choices we have 
$$F = 
\left( \begin{array}{cc}
a &  b\\
b & -a \\
\end{array} \right)=
\left( \begin{array}{cc}
1/\delta &  \sqrt{1 - \delta^{-2}}\\
\sqrt{1 - \delta^{-2}} & -1/\delta \\
\end{array} \right)$$ real and unitary, and for the Temperley-Lieb algebra, 
$$U=
\left( \begin{array}{cc}
\delta &  0\\
0 & 0\\
\end{array} \right), 
V = 
\delta \left( \begin{array}{cc}
a^2 &  ab\\
ab & b^2 \\
\end{array} \right) =
\left( \begin{array}{cc}
a &  b\\
b & \delta b^2 \\
\end{array} \right).$$
\bigbreak

\begin{figure}
     \begin{center}
     \begin{tabular}{c}
     \includegraphics[height=6cm]{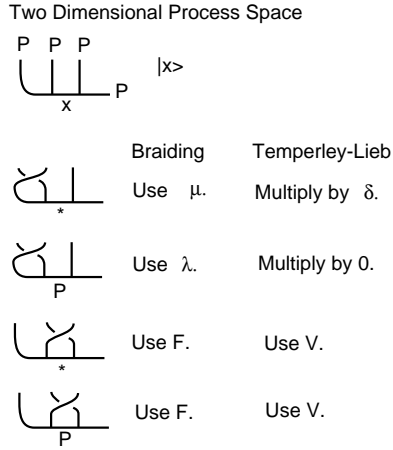}
     \end{tabular}
     \end{center}
     \caption{\bf Algebra for a Two Dimensional Process Space}
     \label{Figure 67 }
     \end{figure}

Now examine Figure~\ref{Figure 67 }. Here we illustrate the action of the braiding and the Temperley-Lieb Algebra on the first Fibonacci 
process space with basis $\{ |*\rangle, |P\rangle \}.$ Here we have $\sigma_{1} = R, \sigma_{2} = FRF$ and $U_{1} =U, U_{2} = V$ as 
described above. Thus we have a representation of the braid group on three strands and a representation of the Temperley-Lieb algebra
on three strands with no further restrictions on $\delta.$
\bigbreak

     \bigbreak

\begin{figure}
     \begin{center}
     \begin{tabular}{c}
     \includegraphics[height=3cm]{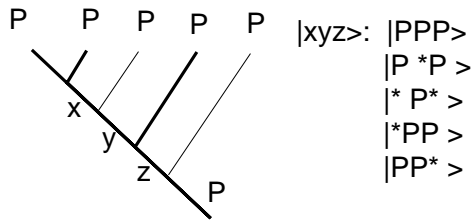}
     \end{tabular}
     \end{center}
     \caption{\bf A Five Dimensional Process Space}
     \label{Figure 68 }
     \end{figure} 
     \bigbreak

\begin{figure}
     \begin{center}
     \begin{tabular}{c}
     \includegraphics[height=7cm]{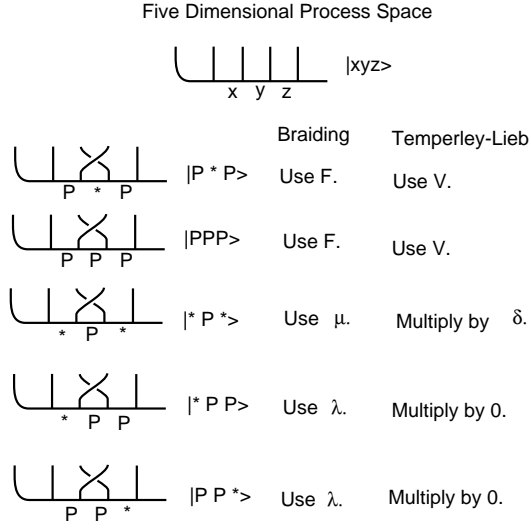}
     \end{tabular}
     \end{center}
     \caption{\bf  Algebra for a Five Dimensional Process Space}
     \label{Figure 69 }
     \end{figure} 
     \bigbreak

So far, we have arrived at exactly the $3$-strand braid representations that we used in our papers \cite{QCJP1,QCJP2} giving a quantum algorithm for the
Jones polynomial for three-strand braids. In this paper we are working in the context of the Fibonacci process spaces and so we wish to see how to make a 
representation of the Temperley-Lieb algebra to this model as a whole, not restricting ourselves to only three strands. The generic case to
consider is the action of the Temperley-Lieb algebra on process spaces of higher dimension as shown in Figure~\ref{Figure 68 } and Figure~\ref{Figure 69 }. In  Figure~\ref{Figure 69 } we
have illustrated the triplets from the previous figure as part of a possibly larger tree and have drawn the strings horizontally rather than diagonally. In
this figure we have listed the effects of braiding the vertical strands $3$ and $4$. We see from this figure that the action of the Temperley-Lieb algebra
must be as follows:
$$U_{3}|P*P\rangle = a|P*P\rangle + b|PPP\rangle,$$
$$U_{3}|PPP\rangle = b|P*P\rangle + \delta b^{2} |PPP\rangle,$$
$$U_{3}|*P*\rangle = \delta |*P*\rangle,$$
$$U_{3}|*PP\rangle = 0,$$
$$U_{3}|PP*\rangle = 0.$$
Here we have denoted this action as $U_{3}$ because it connotes the action on the third and fourth vertical strands in the sequences shown in Figure~\ref{Figure 69 }.
Note that in a larger sequence we can recognize $U_{j}$ by examining the triplet surrounding the $j-1$-th element in the sequence, just as the pattern 
above is governed by the elements surrounding the second element in the sequence. For simplicity, we have only indicated three elements in the sequences
above. Note that in a sequence for the Fibonacci process there are never two consecutive appearances of the neutral element $*.$ 
\bigbreak

We shall refer to a sequence of $*$ and $P$ as a {\it Fibonacci sequence} if it contains no consecutive appearances of $*$. Thus
$|PP*P*P*P\rangle$ is a Fibonacci sequence. In working with this representation of the braid group and
Temperley-Lieb algebra, it is convenient to assume that the ends of the sequence are flanked by $P$ as in Figure~\ref{Figure 68 } and Figure~\ref{Figure 69 } for sequences of length
$3.$ It is convenient to leave out the flanking $P$'s when notating the sequence.
\bigbreak

Using these formulas we can determine conditions on $\delta$ such that this is a representation of the Temperley-Lieb algebra for all Fibonacci sequences.
Consider the following calculation:
$$U_{4}U_{3}U_{4}|PPPP\rangle = U_{3}U_{2}(b|PP*P\rangle + \delta b^{2}|PPPP\rangle)$$
$$=  U_{4}(bU_{3}|PP*P\rangle + \delta b^{2}U_{3}|PPPP\rangle)$$
$$=  U_{4}(0 + \delta b^{2}(b|P*PP\rangle + \delta b^{2}|PPPP\rangle)$$
$$=  \delta b^{2}(bU_{4}|P*PP\rangle + \delta b^{2}U_{4}|PPPP\rangle)$$
$$=  \delta^{2} b^{4}U_{4}|PPPP\rangle.$$
Thus we see that in order for $U_{4}U_{3}U_{4} = U_{4},$ we need that $\delta^{2} b^{4} = 1.$ 
\bigbreak

{\it It is easy to see that $\delta^{2} b^{4} = 1$ is the only remaining
condition needed to make sure that the action of the Temperley-Lieb algebra extends to all Fibonacci Model sequences.} 
\bigbreak

\noindent Note that 
$\delta^{2} b^{4} = \delta^{2}(1 - \delta^{-2})^2 = (\delta - 1/\delta)^{2}.$ Thus we require that 
$$\delta - 1/\delta = \pm 1.$$
When $\delta - 1/\delta = 1,$ we have the solutions $\delta = \frac{1 \pm \sqrt{5}}{2}.$ However, for the reality of $F$ we require that
$1 - \delta^{-2} \ge 0,$ ruling out the choice $\delta = \frac{1 - \sqrt{5}}{2}.$ When $\delta - 1/\delta = -1,$ we have
the solutions $\delta = \frac{-1 \pm \sqrt{5}}{2}.$ This leaves only $\delta = \pm \phi$ where $\phi = \frac{1 + \sqrt{5}}{2}$ (the Golden
Ratio) as possible values for $\delta$ that satisfy the reality condition for $F.$ Thus, up to a sign we have arrived at the well-known value of
$\delta = \phi$ (the Fibonacci model) as essentially the only way to have an extension of this form of the representation of the Temperley-Lieb
algebra for $n$ strands. Let's state this positively as a Theorem.
\bigbreak

\noindent {\bf Fibonacci Theorem.} Let $V_{n+2}$ be the complex vector space with basis $\{ |x_{1}x_{2}\cdots x_{n} \rangle \}$ where each $x_{i}$ equals
either $P$ or $*$ and there do {\it not} occur two consecutive appearances  of $*$ in the sequence $\{ x_{1},\cdots x_{n} \}.$ We refer to this basis for
$V_{n}$ as the set of {\it Fibonacci sequences} of length $n.$ Then the dimension of $V_{n}$ is equal to $f_{n+1}$ where
$f_{n}$ is the $n$-th Fibonacci number: $f_{0} = f_{1} = 1$ and $f_{n+1} = f_{n} + f_{n-1}.$ Let $\delta = \pm \phi$
where $\phi = \frac{1 + \sqrt{5}}{2}.$ Let $a = 1/\delta$ and $b = \sqrt{1 - a^2}.$ Then the Temperley-Lieb algebra on $n+2$ strands with loop value
$\delta$ acts on $V_{n}$ via the formulas given below. 
First we give the left-end actions.
$$U_{1}|*x_{2}x_{3} \cdots x_{n} \rangle = \delta |*x_{2}x_{3} \cdots x_{n} \rangle,$$
$$U_{1}|Px_{2}x_{3} \cdots x_{n} \rangle = 0,$$
$$U_{2}|*Px_{3} \cdots x_{n} \rangle = a|*P x_{3} \cdots x_{n} \rangle + b|PP x_{3} \cdots x_{n} \rangle, $$
$$U_{2}|P* x_{3} \cdots x_{n} \rangle = 0,$$
$$U_{2}|PP x_{3} \cdots x_{n} \rangle = b|*P x_{3} \cdots x_{n} \rangle + \delta b^{2}|PP x_{3} \cdots x_{n} \rangle.$$
Then we give the general action for the middle strands.
$$U_{i}|x_{1} \cdots x_{i-3} P*P x_{i+1} \cdots x_{n} \rangle = a|x_{1} \cdots x_{i-3} P*P  x_{i+1} \cdots x_{n} \rangle$$
$$+ b|x_{1} \cdots x_{i-3} PPP x_{i+1} \cdots x_{n} \rangle,$$
$$U_{i}|x_{1} \cdots x_{i-3} PPP x_{i+1} \cdots x_{n} \rangle = b|x_{1} \cdots x_{i-3} P*P x_{i+1} \cdots x_{n} \rangle$$ 
$$+ \delta b^{2}|x_{1} \cdots x_{i-3} PPP x_{i+1} \cdots x_{n} \rangle,$$
$$U_{i}|x_{1} \cdots x_{i-3} *P* x_{i+1} \cdots x_{n} \rangle = \delta |x_{1} \cdots x_{i-3} *P* x_{i+1} \cdots x_{n} \rangle,$$
$$U_{i}|x_{1} \cdots x_{i-3} *PP x_{i+1} \cdots x_{n} \rangle = 0,$$
$$U_{i}|x_{1} \cdots x_{i-3} PP* x_{i+1} \cdots x_{n} \rangle = 0.$$
Finally, we give the right-end action.
$$U_{n+1}|x_{1} \cdots x_{n-2} *P \rangle = 0,$$
$$U_{n+1}|x_{1} \cdots x_{n-2} P* \rangle =0,$$
$$U_{n+1}|x_{1} \cdots x_{n-2} PP \rangle =b|x_{1} \cdots x_{n-2} P* \rangle + \delta b^{2}|x_{1} \cdots x_{n-2} PP \rangle.$$
\bigbreak

\noindent {\it Remark.} Note that the left and right end Temperley-Lieb actions depend on the same basic pattern as the middle action.
The Fibonacci sequences $|x_{1} x_{2} \cdots x_{n} \rangle$ should be regarded as flanked left and right by $P$'s just as in the special
cases discussed prior to the proof of the Fibonacci Theorem.
\bigbreak

\noindent {\bf Corollary.} With the hypotheses of Theorem $2$, we have a unitary representation of the Artin Braid group $B_{n+2}$ to
$TL_{n+2}$, $\rho: B_{n+2} \longrightarrow TL_{n+2}$ given by the formulas
$$\rho(\sigma_{i}) = A I + A^{-1} U_{i},$$
$$\rho(\sigma_{i}^{-1}) = A^{-1} I + A U_{i},$$ where $A = e^{3 \pi i/5}$
where the $U_{i}$ connote the representation of the Temperley-Lieb algebra on the space $V_{n+2}$ of Fibonacci sequences as described in the Theorem 
above.
\bigbreak

\noindent {\bf Remark.} The Theorem and Corollary give the original parameters of 
the Fibonacci model and shows that this model admits a unitary representation of the braid group via a Jones representation of the Temperley-Lieb
algebra.
\bigbreak

In the original Fibonacci model \cite{SpinTop}, there is a basic non-trivial recoupling matrix $F.$ 
$$F =
\left( \begin{array}{cc}
1/\delta & 1/\sqrt{\delta} \\
1/\sqrt{\delta} & -1/\delta \\
\end{array} \right) =
\left( \begin{array}{cc}
\tau & \sqrt{\tau} \\
\sqrt{\tau} & -\tau \\
\end{array} \right)$$
where $\delta = \frac{1 + \sqrt{5}}{2}$ is the golden ratio and $\tau = 1/\delta$.
The local braiding matrix is given by the formula
below with $A = e^{3\pi i/5}.$
$$R = 
\left( \begin{array}{cc}
A^{8} & 0 \\
0 & -A^{4} \\
\end{array} \right)=
\left( \begin{array}{cc}
e^{4\pi i/5} & 0 \\
0 & -e^{2\pi i/5} \\
\end{array} \right).$$
\bigbreak

This is exactly what we get from our method by using $\delta = \frac{1 + \sqrt{5}}{2}$ and $A = e^{3 \pi i/5}.$
Just as we have explained earlier in this paper, the simplest example of a braid group representation arising from this theory is the representation of the
three strand braid group generated by
$\sigma_{1}= R$ and $\sigma_{2} = FRF$ (Remember that $F=F^{T} = F^{-1}.$). The matrices $\sigma_{1}$ and $\sigma_{2}$ are both unitary, and they generate
a dense subset of $U(2),$ supplying the local unitary transformations needed for quantum computing. The full braid group representation on the Fibonacci
sequences is  computationally universal for quantum computation. In our earlier paper \cite{SpinTop} and in the previous sections of the present work, we gave a construction for the Fibonacci model
based on Temperely-Lieb recoupling theory. In this section, we have reconstructed the Fibonacci model on the more elementary grounds of the representation
of the Temperley-Lieb algebra summarized in the statement of the Fibonacci Theorem and its Corollary.
\bigbreak

\end{document}